\documentclass[12pt]{article}
\usepackage{jheppub}
\pdfoutput=1
\usepackage{epstopdf}

\usepackage{jheppub}
\usepackage{mathrsfs}
\usepackage{psfrag}
\usepackage{color}

\usepackage[table]{xcolor}

\usepackage{amsmath,bbm,array,amsfonts,graphicx,wrapfig,arydshln,lscape,float,slashbox,multirow,longtable,rotating,subfigure,makecell}
\usepackage{url}

\newcommand{\be}{\begin{equation}}
\newcommand{\ee}{\end{equation}}
\newcommand{\beq}{\begin{equation}}
\newcommand{\beql}[1]{\begin{equation}\label{#1}}
\newcommand{\eeq}{\end{equation}}
\newcommand{\ba}{\begin{array}}
\newcommand{\ea}{\end{array}}
\newcommand{\bea}{\begin{eqnarray}}
\newcommand{\beal}[1]{\begin{eqnarray}\label{#1}}
\newcommand{\eea}{\end{eqnarray}}
\newcommand{\ben}{\begin{enumerate}}
\newcommand{\een}{\end{enumerate}}
\newcommand{\bean}{\begin{eqnarray*}}
\newcommand{\eean}{\end{eqnarray*}}
\newcommand{\eref}[1]{(\ref{#1})}
\newcommand{\sref}[1]{\S\ref{#1}}
\newcommand{\tref}[1]{Table~\ref{#1}}

\newcommand{\fref}[1]{Figure \ref{#1}}
\newcommand{\btab}[1]{\begin{tabular}{#1}}
\newcommand{\etab}{\end{tabular}}

\newcommand{\comment}[1]{}

\newcommand{\qed}{\nobreak \ifvmode \relax \else
      \ifdim\lastskip<1.5em \hskip-\lastskip
      \hskip1.5em plus0em minus0.5em \fi \nobreak
      \vrule height0.75em width0.5em depth0.25em\fi}

\newcolumntype{C}[1]{>{\centering\arraybackslash}m{#1}}

\newcommand{\pl}{Pl\"ucker }

\title{Anatomy of the Amplituhedron}

\author{Sebasti\'an Franco,$^1$}
\author{Daniele Galloni,$^1$}
\author{Alberto Mariotti,$^1$}
\author{Jaroslav Trnka$^2$}

\affiliation{$^1$ Institute for Particle Physics Phenomenology, Department of Physics \\
Durham University, Durham DH1 3LE, United Kingdom
}
\affiliation{$^2$ Walter Burke Institute for Theoretical Physics \\
California Institute of Technology, Pasadena, CA 91125, USA
}

\emailAdd{sebastian.franco@durham.ac.uk,daniele.galloni@durham.ac.uk, alberto.mariotti@durham.ac.uk,trnka@caltech.edu}

\abstract{We initiate a comprehensive investigation of the geometry of the amplituhedron, a recently found geometric object whose volume calculates the integrand of scattering amplitudes in planar $\mathcal{N}=4$ SYM theory. We do so by introducing and studying its stratification, focusing on four-point amplitudes. The new stratification exhibits interesting combinatorial properties and positivity is neatly captured by permutations. As explicit examples, we find all boundaries for the two and three loop amplitudes and related geometries. We recover the stratifications of some of these geometries from the singularities of the corresponding integrands, providing a non-trivial test of the amplituhedron/scattering amplitude correspondence. We finally introduce a deformation of the stratification with remarkably simple topological properties. 
}

\preprint{
\begin{flushright}IPPP/14/75\end{flushright} \vspace{-0.9cm}
\begin{flushright}DCPT/14/150\end{flushright} \vspace{-0.9cm}
\begin{flushright}CALT-TH-2014-152\end{flushright}
}

\begin{document}

\maketitle

\section{Introduction}

Formidable progress in our understanding of scattering amplitudes in gauge theory has been achieved in the last two decades (see e.g.\cite{Bern:1994zx,Bern:1994cg,Bern:2005iz,Cachazo:2004kj,Britto:2004nc,Britto:2004ap,Britto:2005fq} and reviews \cite{Dixon:1996wi,Beisert:2010jr,Drummond:2011ic,Elvang:2013cua}). The progress is especially impressive for amplitudes in planar $\mathcal{N}=4$ super Yang-Mills theory where explicit results have been obtained up to high loop order \cite{Bern:2006ew,Bern:2007ct,Bourjaily:2011hi,ArkaniHamed:2010kv,
ArkaniHamed:2010gh,Dixon:2011nj,Dixon:2013eka,Dixon:2014xca}, and many interesting connections and dualities have been found including twistor strings \cite{Witten:2003nn}, the amplitude/Wilson loop correspondence \cite{CaronHuot:2010ek,Mason:2010yk,Alday:2010zy} and many others. Amazingly, this theory enjoys an infinite-dimensional Yangian symmetry \cite{Drummond:2009fd}, which results from the combination of superconformal and dual superconformal invariance \cite{Alday:2007hr,Drummond:2006rz} making an interesting connection to the integrability of the theory \cite{Beisert:2003yb,Beisert:2006ez}. This infinite symmetry is obscured in the standard Feynman diagram approach while it is completely manifest in the dual formulation of amplitudes in this theory using the positive Grassmannian \cite{ArkaniHamed:2012nw} (see also \cite{ArkaniHamed:2009dn,ArkaniHamed:2009vw,ArkaniHamed:2010kv,Kaplan:2009mh,Mason:2009qx} and recent work on a deformed version of the story \cite{Ferro:2012xw,Ferro:2013dga,Beisert:2014qba,Ferro:2014gca,Bargheer:2014mxa}) and the {\it amplituhedron} \cite{Arkani-Hamed:2013jha,Arkani-Hamed:2013kca}. This is a new algebraic geometric object which generalizes the positive Grassmannian and encodes scattering amplitudes in a maximally geometric way: they are simply given by its volume. The amplituhedron is the missing link explaining how to combine Yangian invariant building blocks to give rise to the amplitude. Different representations of the same amplitude are beautifully translated into different triangulations of the amplituhedron. In this approach the standard pillars of quantum field theory like locality or unitarity are derived properties from the geometry of the amplituhedron. The existence of such a structure in planar ${\cal N}=4$ SYM suggests that there might be a very different formulation of the field theory which does not use the standard Lagrangian description of physics.

The correspondence between scattering amplitudes and the amplituhedron has passed numerous tests \cite{Arkani-Hamed:2013jha,Arkani-Hamed:2013kca}, although it still remains conjectural and its study is at its infancy. In this article, we introduce new tools analyzing the amplituhedron and initiate the most comprehensive investigation of its geometry to date. A clear goal is to achieve a systematic understanding similar to the one available for cells in the positive Grassmannian \cite{2006math09764P}. Among other things, we expect our ideas to be instrumental for triangulating the amplituhedron, and hence contribute to its practical use in constructing scattering amplitudes. A beautiful interplay between experimental exploration of examples, discovery of new structures and theoretical new ideas has been a constant driving force for progress in the understanding of scattering amplitudes. It is reasonable to expect that the examples we study in this paper, and the ones which will be studied in the future with the help of the tools we introduce, will nicely fit into this trend.

This paper is organized as follows. \sref{section_intro_amplituhedron} provides a quick review of the basics of the amplituhedron. In \sref{section_first_stratification}, we introduce a stratification for it, which captures all detailed structures of the corresponding differential form and allows us to explore its geometry in depth. We also introduce a reduced version of the stratification, which we call mini stratification, which captures broader features of the geometry and is amenable to a combinatorial implementation. \sref{section_examples_first_approach} contains a first encounter with the stratification through simple examples. \sref{section_combinatorial_stratification} introduces a powerful combinatorial implementation of the mini stratification in terms of graphs and a new class of objects we denote hyper perfect matchings. The combinatorics of extended positivity is the subject of \sref{section_combinatorics_positivity}. Interestingly, we find that positivity can be neatly discussed in terms of permutations. \sref{section_examples} puts our techniques at work and investigates various geometries at 2 and 3-loops. 
In \sref{section_integrand_stratification} we study an alternative approach to stratification, based on the singularities of the integrand. For four particles, we find exact agreement with the geometric stratification of the amplitude and its log, providing new and significant evidence for the amplituhedron conjecture. In \sref{section_deformed_amplituhedron} we introduce and investigate the deformed amplituhedron, which seems to exhibit an outstandingly simple geometry. We conclude and present a vision for future work in \sref{section_conclusions}. We also include two appendices with supporting material.

\bigskip

\section{The Amplituhedron}

In this section we provide a brief introduction to the amplituhedron. We refer the reader to \cite{Arkani-Hamed:2013jha,Arkani-Hamed:2013kca} for further details.

\bigskip

\label{section_intro_amplituhedron}

\subsection{Tree-Level Amplituhedron}

The amplituhedron is a generalization of the positive Grassmannian conjectured to give all scattering amplitudes in planar $\mathcal{N}=4$ SYM theory when integrated over with an appropriate volume form. The amplituhedron can be regarded as a generalization of the interior of a set of $n$ vertices $Z^I$ of dimension $(k+4)$, where $(k+2)$ is the number of negative-helicity gluons, $I=1,2,\ldots,k+4$, and $n$ is the total number of external gluons. In this notation, $k=0$ corresponds to MHV amplitudes. These vertices can be combined into matrix $Z^I_a$, where $a=1,2,\ldots, n$. In order to have a notion of interior we need vertices to be ordered in a specific way. In the familiar 2-dimensional case of polygons, vertices must be cyclically ordered to avoid the crossing of external edges connecting consecutive vertices. The generalization of this cyclicity constraint takes the form of a positivity condition on the matrix $Z^I_a$: all maximal minors of $Z^I_a$ must be positive, i.e.\  $Z^I_a \in M_{+} (4+k,n)$ where $M_{+} (4+k,n)$ is the space of positive $(4+k)\times n$ matrices.

External vertices form a polytope. For $k=1$ we consider a point in the interior of this polytope, which corresponds to a linear combination of the external vertices, where the coefficients must be positive. Each of these points will be considered projectively, and can thus be seen as 1-planes (or lines) in $k+4$ dimensions. For general $k$, we consider a $k$-plane and impose positivity conditions on the matrix of coefficients of its expansion in terms of external points. Explicitly, a $k$-plane $Y$ in the interior of the tree-level amplituhedron is given by
\begin{equation}
Y= C \cdot Z \, ,
\end{equation}
where $Z$ is the $(k+4)\times n$ matrix of external vertices, $C$ is a $k\times n$ matrix in $G_{+}(k,n)$, and $Y$ is the tree-level amplituhedron interior, given by a $k \times (k+4)$ matrix.\footnote{A warning to the reader: whenever we refer to the positive Grassmannian $G_{+}(k,n)$, we mean the totally non-negative Grassmannian. The boundaries of this space arise when the positive degrees of freedom become zero. Similarly, we will use positive as a synonym of non-negative and emphasize when a given quantity is not zero. This slight abuse of terminology will persist throughout; we hope it will not cause any confusion.} We are not imposing positivity on each of the $k$ rows of the matrix $C$, but a condition on how the rows of $C$ interact with each other such that minors are positive. As a result, the amplituhedron is \textit{not} simply given by $k$ copies of ``the interior of the vertices'', but it is a more complicated geometric object. We can also think of the amplituhedron as a map:
\beq
G_+(k,n) \xrightarrow{Z} G(k,k+4) \, .
\eeq
The $GL(k)$ degree of freedom of the Grassmannian, which acts on $C$, must also apply to $Y$, thus implying the matrix $Y \in G(k,k+4)$. 

\bigskip

\subsection{Loop Geometry}

Each point of the tree-level amplituhedron spans a $k$-plane in $(k+4)$ dimensions; the full amplituhedron spans all possible $k$-planes in $(k+4)$ dimensions. For each point, the transverse space is 4-dimensional and this is where the loop-level part of the amplituhedron lives. The degrees of freedom of each loop span a 2-plane in this transverse space. Let us start our discussion with the $k=0$ case, which at tree-level is given by the empty projective space $\mathbb{P}^3$, since $Y$ is 0-dimensional. At loop level, it corresponds to what we call the pure {\it loop geometry}. In this case, every loop $\mathcal{L}_{(i)}$ is a different linear combination of the external vertices, which lies in $\mathbb{P}^3$:
\begin{equation}
\mathcal{L}_{(i)} = D_{(i)} \cdot Z \, ,
\end{equation}
where the $Z$'s are 4-dimensional vectors, $D_{(i)} \in G_{+}(2,n)$ maps the vertices in $Z$ to the transverse space, and so $\mathcal{L}_{(i)} \in G(2,4)$. Multiple loops are implemented by increasing the number of matrices $D_{(i)}$:
\begin{equation}
\begin{pmatrix}
\mathcal{L}_{(1)} \\
\mathcal{L}_{(2)} \\
\vdots \\
\mathcal{L}_{(L)}
\end{pmatrix} =
\begin{pmatrix}
D_{(1)} \\
D_{(2)} \\
\vdots \\
D_{(L)}
\end{pmatrix} \cdot Z \, .
\end{equation}

The matrices $D_{(i)}$ satisfy {\it extended positivity} conditions, i.e. for any subset of them we define
\begin{equation}
D_{(ij)} = \left(
\begin{array}{c}
D_{(i)} \\
D_{(j)} \\
\end{array}
\right) ,\qquad D_{(ijk)} = \left(
\begin{array}{c}
D_{(i)} \\
D_{(j)} \\
D_{(k)} \\
\end{array}
\right),\,\,\text{etc.}
\end{equation}
and demand all maximal minors of each of these extended matrices to be positive, namely $D_{(ij)}\in M_+(4,n)$, $D_{(ijk)}\in M_+(6,n)$, etc. In general, $D_{(a_1\ldots a_m)}\in M_+(2m,n)$. These conditions apply only for $m\leq n/2$. In the special case of $n=4$ and arbitrary $L$, the only surviving conditions are mutual positivities: $D_{(ij)}\in M_+(4,n)$ for all pairs of $i$ and $j$.

\bigskip

\subsection{The Full Amplituhedron}

To obtain the full amplituhedron for any $n,k,L$, we combine the tree-level space and the loop space into a larger matrix

\begin{equation}
\begin{pmatrix}
\mathcal{L}_{(1)} \\
\mathcal{L}_{(2)} \\
\vdots \\
\mathcal{L}_{(L)} \\
\hline
Y
\end{pmatrix} =  \begin{pmatrix}
D_{(1)} \\
D_{(2)} \\
\vdots \\
D_{(L)} \\
\hline
C
\end{pmatrix} \cdot Z
\end{equation}
or more neatly
\begin{equation}
\mathcal{Y} = \mathcal{C} \cdot Z \, ,
\label{eq_Y_CZ}
\end{equation}
where $\mathcal{C}$ is the $(k+2L)\times n$ matrix specifying the set of $(k+2L)$ different linear combinations of external vertices, and $\mathcal{Y}$ is the full amplituhedron interior. Here the positivity condition for $\mathcal{C}$ is not the same as the one for $C$: $\mathcal{C} \not\in G_{+}(k+2L,n)$ (in fact, $k+2L$ maybe be much larger than $n$). As for the pure loop geometry, the positivity condition is now an extended positivity. The requirements are that the combination of $C$ with any subset of the $D_{(i)}$ matrices is positive, i.e. all their maximal minors are positive, as long as the matrix has at least as many columns as rows, i.e.\ that

\begin{equation}
\begin{pmatrix}
C
\end{pmatrix} ,
\begin{pmatrix}
D_{(1)}\\
\hline
C
\end{pmatrix} , \cdots,
\begin{pmatrix}
D_{(L)}\\
\hline
C
\end{pmatrix} ,
\begin{pmatrix}
D_{(1)}\\
D_{(2)}\\
\hline
C
\end{pmatrix} , \cdots
\end{equation}
are all positive, where we stop stacking $D_{(i)}$'s onto $C$ when the resulting matrix has more rows than columns.\footnote{It is possible to stack more matrices but the maximal minors would be insensitive to this.} Note that there is no condition that only relates the various $D_{(i)}$'s to each other, except in the absence of $C$, i.e. for $k=0$. This novel space inhabited by $\mathcal{C}$, characterized by the extended positivity, is denoted $G_{+}(k,n;L)$.

\bigskip

\subsection{The Scattering Amplitude}

The scattering amplitude is obtained by integrating over all of the degrees of freedom of the amplituhedron, with a specific form constrained to have {\it logarithmic singularities} on the boundaries of the space. This form is the amplitude integrand, and can in principle be constructed using methods such as Feynman diagrams, unitary cuts or BCFW recursion relations. For arbitrary numbers of particles and loops such methods become very laborious, and it would be desirable to construct the integrand directly from the definition of the amplituhedron. There are several strategies for doing this: the first one is to try to triangulate the amplituhedron in terms of smaller elementary spaces which have trivial dlog forms. Recursion relations via on-shell diagrams provide examples of such triangulations, where the rules for triangulating are dictated by the physics rather than the amplituhedron geometry.\footnote{See \cite{Bai:2014cna} for alternative diagrammatic tools for addressing this problem and \cite{Enciso:2014cta} for interesting new ideas on the computation of volumes of polytopes associated to scattering amplitudes.} Another strategy is to nail down the integrand directly, by requiring that all spurious singularities (which do not correspond to amplituhedron boundaries) cancel. In either approach, an understanding of the boundary structure of the space will be crucial for systematically constructing the integrand form.

\bigskip

\section{Stratification of the Amplituhedron: Loop Geometry}

\label{section_first_stratification}

In this section we develop tools for {\it stratifying} the amplituhedron, by which we mean finding its boundary structure. 

In this paper, we focus our attention on the $k=0$ case, i.e. on the pure {\it loop geometry}, and also restrict to $n=4$. For $k=0$, the matrix $C$ disappears, and we are only left with the $D_{(i)}$ matrices:

\begin{equation}
\mathcal{C} = \begin{pmatrix}
D_{(1)} \\
D_{(1)} \\
\vdots \\
D_{(L)} 
\end{pmatrix} \, .
\end{equation}
The structure at loop level is rather non-trivial due to the extended positivity condition imposed on matrices. Note that $\mathcal{C}$ is \textit{not} an element of the positive Grassmannian, except for $L=1$. 

For $n=k+4$, the positivity of external data, encoded in the matrix $Z$, is trivial and the stratification of the amplituhedron corresponds to the stratification of $\mathcal{C}$.\footnote{This follows directly from the fact that when $Z$ is a square matrix we may choose a basis for which $Z$ equals the unit matrix. Then from \eref{eq_Y_CZ} we see that $\mathcal{Y} = \mathcal{C} \cdot Z = \mathcal{C}$.} Even in this simplified situation, the geometry of the amplituhedron will exhibit extraordinary richness. For general $n$, the process we will discuss can be regarded as the stratification of $G_+(0,n;L)$ rather than the stratification of the amplituhedron.  Independently of its relation to the amplituhedron, the stratification of $G_+(0,n;L)$ is an interesting geometric question in its own right. 

\bigskip

\subsection{The Degrees of Freedom of $\mathcal{C}$} \label{sec:dofC}

Each $D_{(i)} \in G_{+}(2,n)$ has $2(n-2)$ degrees of freedom, best parametrized by its $2\times 2$ minors, known as \pl coordinates. There are $\binom{n}{2}$ different \pl coordinates $\Delta^{(i)}_I$, with $I=\{a,b\}$ specifying which two columns $a$ and $b$ are involved in the minor. The $\Delta^{(i)}_I$'s are not all independent but are subject to relations, known as \pl relations. $\mathcal{C}$ gets a contribution from each $D_{(i)}$, giving a total of $2L(n-2)$ degrees of freedom.

Note that extended positivity, despite imposing a condition on the degrees of freedom of different $D_{(i)}$, does not decrease the dimension, for the simple reason that it is just an inequality and cannot determine any \pl coordinate in terms of the others. This is akin to the fact that the restriction to the positive Grassmannian, i.e. that $\Delta^{(i)}_I > 0$, does not create new relations between the coordinates $\Delta^{(i)}_I$, but simply constrains them to be positive. 

However, extended positivity can restrict the allowed domain of the $\Delta^{(i)}_I$ further than the simple $\Delta^{(i)}_I > 0$ condition. This additional restriction can in certain cases be quite non-trivial, and may even split the domain into disjoint {\it regions}. Later in this section, we will introduce a \textit{mini stratification} of $\mathcal{C}$ which is insensitive to this subtlety, and a \textit{full stratification} which refines the mini stratification and fully accounts for it. The full stratification in effect counts all domain regions of the amplituhedron.

Regardless of which stratification we are interested in, for the purposes of counting dimensions we only count the number of independent equalities between various $\Delta^{(i)}_I$'s. For example, when $\mathcal{C}$ is top-dimensional the only relations come from the \pl relations which are  independently present in each $D_{(i)}$, e.g.\ for $i=1$ there is a \pl relation between various $\Delta^{(1)}_I$'s, for $i=2$ there is a separate \pl relation between the $\Delta^{(2)}_I$'s, but we cannot write any $\Delta^{(1)}_I$ in terms of $\Delta^{(2)}_J$'s. 

\bigskip

\subsection{Extended Positivity and Boundaries}

\label{section_extended_positivity}

For $k=0$, extended positivity enforces the condition that all $D_{(i)}$ are positive, as well as all subsets of them when stacked onto each other (as long as the number of rows does not exceed the number of columns; these larger matrices produce no additional conditions), i.e.\ that

\begin{equation} \label{eq:deltaIstacks}
\begin{pmatrix} D_{(i)} \end{pmatrix} , \begin{pmatrix} D_{(i)} \\ D_{(j)}\end{pmatrix} , \cdots
\end{equation}
are all positive. This translates into various conditions on the \pl coordinates. To unify the conditions it is convenient to define $2m\times 2m$ minors $\Delta^{(i_1,\ldots,i_m)}_I$, $m=1,\ldots,L$, which are all the maximal minors when stacking the matrices $D_{i_1},\ldots D_{i_m}$.\footnote{This notation includes the $2\times 2$ \pl coordinates. In order to maintain an economic notation, we use a single subindex $I$ to indicate the set of columns in the larger minors.} First, all $\Delta^{(i)}_I$ must be positive. Extended positivity also requires the $\Delta^{(i_1,\ldots,\i_m)}_I$'s, which are polynomials of order $m$ in the $\Delta^{(i)}_I$'s, to be positive. In order to emphasize the contrast with \pl coordinates $\Delta_I^{(i)}$, we will often refer to the $m>1$ minors as {\it non-minimal minors}.

For a given number of loops $L$, there are $\binom{L}{m}$ ways of choosing $m$ matrices $D_{(i)}$ to form a $\Delta^{(i_1,\ldots,i_m)}_J$. For each of these choices, there are $\binom{n}{2m}$ ways of choosing the set $J$ of $2m$ columns out of all the $n$ external nodes. Hence, the number of non-minimal minors becomes

\begin{equation}
\label{eq:NumOfExtendedPosConditions}
\sum_{m=2}^{m\leq n/2} \binom{L}{m} \binom{n}{2m} \, .
\end{equation}
These larger minors are not all independent, there are Pl\"ucker-like relations among them.

Boundaries of $\mathcal{C}$ are reached by killing degrees of freedom in it by setting minors to zero. In other words, $\Delta^{(i_1,\ldots,i_m)}_I\geq 0$ has its boundary when $\Delta^{(i_1,\ldots,i_m)}_I= 0$. The more complicated inequalities arising from minors with $m>1$ give rise to relations between $\Delta^{(i)}_I$'s. Each independent relation of this form reduces the degrees of freedom by 1. A more precision characterization of boundaries is given below, when we discuss the stratification.

\medskip

\paragraph{Labels.}

To every boundary we can associate the corresponding list of vanishing $\Delta^{(i_1,\ldots,i_m)}_I$. In each list, all $\Delta^{(i_1,\ldots,i_m)}_I$, i.e. for both $m=1$ and $m>1$, are treated {\it democratically}. We will refer to such lists of minors as {\it labels}. The minors which are not in the label are not vanishing. Labels are very useful for characterizing boundaries and other configurations of minors, although they do not fully specify them.

These labels will form the basis of the mini stratification described in \sref{sec:ministrat}, which will only distinguish elements in the stratification by them. However, motivated by the physical problem of using the amplituhedron to identify all possible singularities of the integrand, we will refine this counting in \sref{sec:fullstrat} by noticing that there are several independent domain regions for each label, or equivalently by identifying {\it independent} solutions consistent with a given label.\footnote{As it will become clearer in \sref{sec:fullstrat}, and exemplified in \sref{section_full_stratification_L=2}, the definition automatically accounts for the information about the sequence or path in which minors are turned off to reach a given boundary.} It is thus important to emphasize that, generically, {\it labels do not fully specify boundaries}. 

However, labels are still subject to interesting restrictions, since not every arbitrary set of minors can be set to zero. There are two sources of hindrance:

\begin{itemize}
\item \pl relations relate different $\Delta^{(i)}_I$'s and hence it is sometimes impossible to kill a given \pl coordinate without some other coordinate also becoming zero. The same is in fact true for all $\Delta^{(i_1,\ldots,i_m)}_I$'s: they are not all independent, since there are Pl\"ucker-like relations between them. As a result, it is not possible to {\it exclusively} set any arbitrary combination of $\Delta^{(i_1,\ldots,i_m)}_I$'s to zero. 

\item Relations belonging to different levels of minors may be incompatible, i.e.\ the full extended positivity can become impossible to satisfy, despite only being given in terms of inequalities. This is because the relations arising from non-minimal minors typically contain positive and negative terms, and the sum must be non-negative. When all the \pl coordinates are turned on, extended positivity is easily satisfied. On the contrary if, for example, we kill a subset such that only the negative terms survive, we can no longer satisfy positivity. Similarly, setting a $\Delta^{(i_1,\ldots,i_m)}_I$ to zero becomes impossible if only positive terms in it are turned on. We shall later see explicit examples of both of these occurrences.
\end{itemize}

From the above discussion we conclude that while \pl relations and their generalizations for $m>1$ may invalidate boundaries in an automatic way, extended positivity does so more aggressively: it imposes by hand an ulterior check to determine whether a given boundary exists or not. This is analogous to what happens when imposing positivity on the Grassmannian: $G(k,n) \to G_{+}(k,n)$ kills ``by hand'' a subset of boundaries. In our case, we go from $G(k,n;L) \to G_{+}(k,n;L)$. For the tree-level case $G_{+}(k,n;0)\equiv G_{+}(k,n)$, it is a beautiful result that certain potential boundaries\footnote{By this we mean configurations in which some minors vanish.} are removed in such a way so as to generate an Eulerian poset \cite{2005arXiv09129W}.

\bigskip

\subsection{Mini Stratification} \label{sec:ministrat}

As mentioned above, the full stratification of the amplituhedron counts all independent solutions for a given positivity-preserving label. At this point in our discussion, it is natural to define an unrefined counting, which we call {\it mini stratification}, and serves as a close proxy of the full stratification introduced in next section. The mini stratification corresponds to only considering the labels of the boundaries. This counting can be used to generate a ``poor man's'' label stratification, in which multiple solutions for a given label are collapsed into a single point, which is assigned the highest dimension of all these solutions. In other words, the mini stratification combines boundaries into equivalence classes determined by the labels. For brevity, we will simply refer to these equivalence classes as the boundaries of the mini stratification.

While the mini stratification does not capture the full singularity structure of the amplitude, it is valuable for various reasons. First, it provides a rather complete geometric characterization of the amplituhedron. More importantly, as we discuss in \sref{section_combinatorial_stratification} and \sref{section_combinatorics_positivity}, its value follows from the fact that it admits a very efficient combinatorial implementation. We will present examples of the mini stratification in \sref{section_examples} and \sref{section_example_n4_L3}.

\bigskip

\subsection{Full Stratification} \label{sec:fullstrat}

As already discussed above, labels only include information on which minors are vanishing and which are non-vanishing. Their level of refinement is identical to that of the matroid strata for $G_{+}(k,n)$. It is often possible, however, that there are disjoint regions of domain for the minimal minors $\Delta^{(i)}_I$ which satisfy the equalities of a given label, i.e.\ that there are multiple solutions to the set of equalities described by the label. 

We are thus naturally led to the definition of a \textit{region}, which is a set of equalities and inequalities for the $\Delta^{(i_1,\ldots,i_m)}_I$, $m=1,...,L$, which has a unique solution. In general, the equalities and inequalities needed to describe a region are more than those specifying a label: given the label, we must also specify which of the solutions the region refers to. In the future, when we refer to a boundary of $G_{+}(k,n;L)$ we will mean a region as defined here. The {\it full stratification} is defined as the stratification which distinguishes all such regions. This suggests a natural extension of the labels introduced in the last section, to which we refer as {\it extended labels}. Extended labels correspond to specifying not only the vanishing ${\Delta}^{(i_1,\dots,i_m)}_I$'s but also all other relations between minors. Such an extended label then fully specifies a given boundary. While the mini stratification is based on labels, the full stratification uses extended labels.

For concreteness, let us focus on $n=4$, for which all non-minimal minors are $4\times 4$. Consider one such minor which, without loss of generality, we can assume to be $\Delta^{(1,2)}_{1234}$.\footnote{The simplest situation in which such a minor arises is for 2-loops, i.e. $G_+(0,4;2)$. In this case, this is the only non-minimal minor.} When all $\Delta^{(i)}_I$ are turned on, $\Delta^{(1,2)}_{1234}$ can be expressed in terms of \pl coordinates as follows: 
\beq
\Delta^{(1,2)}_{1234}=\Delta^{(1)}_{12}\Delta^{(2)}_{34} + \Delta^{(1)}_{23}\Delta^{(2)}_{14} + \Delta^{(1)}_{34}\Delta^{(2)}_{12} + \Delta^{(1)}_{14}\Delta^{(2)}_{23} - \Delta^{(1)}_{13}\Delta^{(2)}_{24} -\Delta^{(1)}_{24}\Delta^{(2)}_{13} \, .
\label{4x4_from_pl}
\eeq
After using the \pl relations $\Delta^{(i)}_{12}\Delta^{(i)}_{34}+\Delta^{(i)}_{23}\Delta^{(i)}_{14}=\Delta^{(i)}_{13}\Delta^{(i)}_{24}$ for $i=1,2$, this can be turned into the convenient form
\begin{eqnarray} \label{eq:4x4factorized}
\Delta^{(1,2)}_{1234}  & = & \frac{\left(\Delta^{(1)}_{12} \Delta^{(2)}_{13} - \Delta^{(1)}_{13} \Delta^{(2)}_{12} \right) \Big(\Delta^{(1)}_{13} \Delta^{(2)}_{34} - \Delta^{(1)}_{34} \Delta^{(2)}_{13} \Big)}{\Delta^{(1)}_{13} \Delta^{(2)}_{13}} \nonumber \\
& + & \frac{\Big(\Delta^{(1)}_{23} \Delta^{(2)}_{13} - \Delta^{(1)}_{13} \Delta^{(2)}_{23} \Big) \Big(\Delta^{(1)}_{13} \Delta^{(2)}_{14} - \Delta^{(1)}_{14} \Delta^{(2)}_{13} \Big)}{\Delta^{(1)}_{13} \Delta^{(2)}_{13}} \, \, .
\end{eqnarray}
If we now turn off $\Delta^{(1)}_{23} = \Delta^{(1)}_{14} =0$, we obtain 
\begin{equation}
\Delta^{(1,2)}_{1234} = \frac{\left(\Delta^{(1)}_{12} \Delta^{(2)}_{13} - \Delta^{(1)}_{13} \Delta^{(2)}_{12} \right) \Big(\Delta^{(1)}_{13} \Delta^{(2)}_{34} - \Delta^{(1)}_{34} \Delta^{(2)}_{13} \Big)}{\Delta^{(1)}_{13} \Delta^{(2)}_{13}} - \frac{\Delta^{(1)}_{13} \Delta^{(2)}_{23}  \Delta^{(2)}_{14}}{\Delta^{(2)}_{13}} 
\end{equation}

The mini stratification label for this is simply $\{ \Delta^{(1)}_{14} , \Delta^{(1)}_{23} \}$, which is the full set of vanishing minors. All other $\Delta^{(i)}_{I}$'s are strictly positive. However, we notice that there are two regions in which we may satisfy $\Delta^{(1,2)}_{1234} > 0$:
\begin{itemize}
\item {\bf Region 1:} $\left(\Delta^{(1)}_{12} \Delta^{(2)}_{13} - \Delta^{(1)}_{13} \Delta^{(2)}_{12} \right) >0$ and $\left(\Delta^{(1)}_{13} \Delta^{(2)}_{34} - \Delta^{(1)}_{34} \Delta^{(2)}_{13} \right) >0$
\item {\bf Region 2:} $\left(\Delta^{(1)}_{12} \Delta^{(2)}_{13} - \Delta^{(1)}_{13} \Delta^{(2)}_{12} \right) <0$ and $\left(\Delta^{(1)}_{13} \Delta^{(2)}_{34} - \Delta^{(1)}_{34} \Delta^{(2)}_{13} \right) <0$
\end{itemize}
These two regions are very easy to understand: denoting $x \equiv \big( \Delta^{(1)}_{12} \Delta^{(2)}_{13} - \Delta^{(1)}_{13} \Delta^{(2)}_{12} \big) $, $y \equiv \big( \Delta^{(1)}_{13} \Delta^{(2)}_{34} - \Delta^{(1)}_{34} \Delta^{(2)}_{13} \big)$ and $k \equiv \frac{\Delta^{(1)}_{13} \Delta^{(2)}_{23}  \Delta^{(2)}_{14}}{\Delta^{(2)}_{13}} $, we have the simple condition that 
\begin{equation}
\Delta^{(1,2)}_{1234} \geq 0 \ \ \Leftrightarrow \ \ xy \geq k \quad (k>0)
\end{equation}
which on the $x-y$ plane simply corresponds to two regions whose boundary is the hyperbolic curve $xy = k$. Here we see that to specify the regions within this label, all we need to do is additionally specify the sign of $x$ and $y$. The relations specifying regions 1 and 2 are explicit examples of the type of relations included in extended labels.

In this example, if we go to a different label where we have also shut off $\Delta^{(1,2)}_{1234}$, i.e.\ $\{ \Delta^{(1)}_{14} , \Delta^{(1)}_{23} , \Delta^{(1,2)}_{1234}\}$, we again have two regions: $xy = k$ with $x,y>0$, and $xy = k$ with $x,y<0$.

The full stratification contains all possible poles of the integrand. In fact, it is even more refined than the integrand: while there are several different integrand poles that correspond to the same label in the mini stratification, here it sometimes happens that there are several regions contained within the same integrand pole. The example above is an instance where this happens: as will be clear in subsequent sections, the pole of the integrand when we set $\Delta^{(1)}_{23} = \Delta^{(1)}_{14} =0 $ is 
\begin{equation}
\frac{\langle AB34 \rangle \langle CD12 \rangle+\langle AB12 \rangle \langle CD34 \rangle}{\langle ABCD \rangle \langle AB12 \rangle \langle AB34 \rangle \langle CD12 \rangle \langle CD14 \rangle \langle CD23 \rangle \langle CD34 \rangle} \, .
\end{equation}
We have just shown that this object is composed of two disjoint regions. Provided the amplituhedron proposal holds, 
identifying those regions in the full stratification which correspond to the same integrand pole exactly reproduces the pole structure of the integrand.

\bigskip

\subsection{Summary of the Method and Structure of the Stratification} \label{sec:methodsummary}

\label{section_summary_method}

In this section we summarize the general procedure for stratifying $\mathcal{C} \in G_{+}(0,n;L)$. As stated earlier, in this article we will almost exclusively focus on the case of $k=0$, $n=4$ and arbitrary $L$. This case is particularly simple owing the fact that for $n=4$ the $Z_I$ matrix can be chosen to be diagonal, and hence trivial, thus positivity of external data becomes unimportant and the stratification of $G_{+}(0,4;L)$ actually coincides with the one for the loop amplituhedron.\footnote{The case of $k>0$ is further complicated by the fact that the minors of the $D_{(i)}$ matrices do not have a definite sign, and tuning these to zero does not constitute a boundary of the amplituhedron. Boundaries are only obtained by shutting off degrees of freedom that have a definite sign.} 

As previously mentioned, every boundary of $G_{+}(0,n;L)$ has an associated label, i.e. a list of vanishing minors. For any given label, there is one boundary (or region) for each independent solution giving rise to it, in general specified by some additional inequalities.

All minors should be treated democratically. When implementing the stratification, however, it is natural to give the \pl coordinates $\Delta_I^{(i)}$ a special treatment. The reasons for this choice include the facts that every minor $\Delta_I^{(i_1, \ldots, i_m)}$ is an order $m$ polynomial in $\Delta_I^{(i)}$'s and, as we will discuss in \sref{section_combinatorial_stratification}, the $\Delta_I^{(i)}$'s are related to certain collections of edges, denoted perfect matchings, of simply connected graphs. Moreover, the \pl coordinates for each $D_{(i)}$ scale with a common factor under the $GL(2)$ acting on $D_{(i)}$. The dimension of each boundary is given by the number of degrees of freedom in the $\Delta_I^{(i)}$'s:
\begin{equation}
\label{eqs:dimension}
d = N_{\Delta_I} - N_{\text{rel}} -L \, ,
\end{equation}
where $N_{\Delta_I}$ is the number of non-vanishing $\Delta_I^{(i)}$ on the boundary and $N_{\text{rel}}$ is the number of independent equations relating the $\Delta_I^{(i)}$.\footnote{The subtraction of $L$ degrees of freedom follows from the fact that \pl coordinates are projectively defined.} These equations may be \pl relations or follow from non-minimal minors that have been independently set to zero on a given boundary. In the mini stratification, each label is assigned the dimension of the top-dimensional region associated to it.

In this way we split the positivity constraint on the matrix $\mathcal{C}$ in two:
\begin{itemize}
\item $\Delta_I^{(i)} \geq 0$.
\item Larger minors $\Delta_I^{(i_1, \ldots, i_m)}$, expressed as sums of products of $\Delta_I^{(i)}$, also satisfy\\ 
$\Delta_I^{(i_1,\ldots, i_m)} \geq 0$.
\end{itemize}

The aforementioned distinction between \pl coordinates and non-minimal minors reflects into a natural separation of the stratification of $G_{+}(0,n;L)$ into two stages. First, we obtain all possible sets of vanishing \pl coordinates $\Delta_I^{(i)}$, subject to extended positivity conditions. At this step larger minors are not set to zero, unless they trivially vanish as a result of the vanishing \pl coordinates. If we are considering the full stratification, some of these configurations can be further divided in different regions, specified by inequalities among the non-vanishing \pl coordinates.
Next, we introduce for each of these elements a further structure corresponding to the vanishing of non-minimal minors. This second stage reduces the dimension of boundaries by imposing constraints on the non-vanishing $\Delta_I^{(i)}$'s. Depending on whether we are interested in the mini or the full stratification, it is implemented slightly differently.

The first stage in the stratification thus corresponds to the following two steps:
\begin{enumerate}
\item Classify potential boundaries according only to the vanishing \pl coordinates. This corresponds to independently performing the positroid stratification of each $D_{(i)}$, i.e. of each $G_+(2,n)$.
\item Some of these collections violate the extended positivity of the larger minors $\Delta_I^{(i_1,\ldots, i_m)} \geq 0$ and are thus removed. The surviving collections of $\Delta_I^{(i)}$ represent all the labels of $G_{+}(0,n;L)$ for which non-minimal minors can be non-negative.
\end{enumerate}

Step $1$ produces the $L^{th}$ power of the positroid stratification of $G_+(2,n)$ and is independent of what type of stratification we are considering. We will denote the numbers of potential boundaries with dimension $d$ obtained at this first step as $\mathbb{N}^{(d)}$, where $d$ is determined using \eref{eqs:dimension}. Step $2$ represents a further refinement of this decomposition, removing some of the potential boundaries obtained at step $1$ by demanding extended positivity. We refer to the number of remaining boundaries as $\mathcal{N}^{(d)}$. These boundaries can be organized in a poset that we denote $\Gamma_0$, where at the top element corresponds to all minors non-vanishing. Every element in $\Gamma_0$ is associated to a set of vanishing $\Delta_I^{(i)}$'s. In the case of the full stratification, this information might not uniquely fix the element of $\Gamma_0$, due to the multiplicity of regions. A combinatorial approach for constructing $\Gamma_0$ in the mini stratification will be introduced in \sref{section_combinatorial_stratification}.

Independently of whether we are constructing the mini or the full stratification, for each element in $\Gamma_0$ there are, generally, multiple boundaries, which arise from setting  to zero non-minimal minors which are not automatically vanishing due to vanishing \pl coordinates. The procedure for systematically constructing these boundaries is:

\begin{enumerate}
\item[3.] For each element of $\Gamma_0$  and its collections of surviving $\Delta_I^{(i)}$, we first classify non-minimal minors $\Delta_I^{(i_1,\ldots, i_m)} \geq 0$, $m>1$, into three categories:
\begin{enumerate}
\item[(i)] Those that are trivially zero given the list of vanishing $\Delta_I^{(i)}$.
\item[(ii)] Those that are manifestly positive, because only positive terms are turned on by the given collection of non-zero $\Delta_I^{(i)}$.
\item[(iii)] Those that have both positive and negative terms turned on.
\end{enumerate}
\item[4.] Given the previous classification, for each element of $\Gamma_0$ the additional boundary structure is obtained by turning off combinations of type (iii) $\Delta_I^{(i_1,\ldots, i_m)}$. Additionally, for the full stratification we may sometimes obtain additional boundaries from type (i) non-minimal minors. The mini and the full stratifications differ in the structure arising from this step.
\end{enumerate}

\begin{figure}[h]
\begin{center}
\includegraphics[width=15cm]{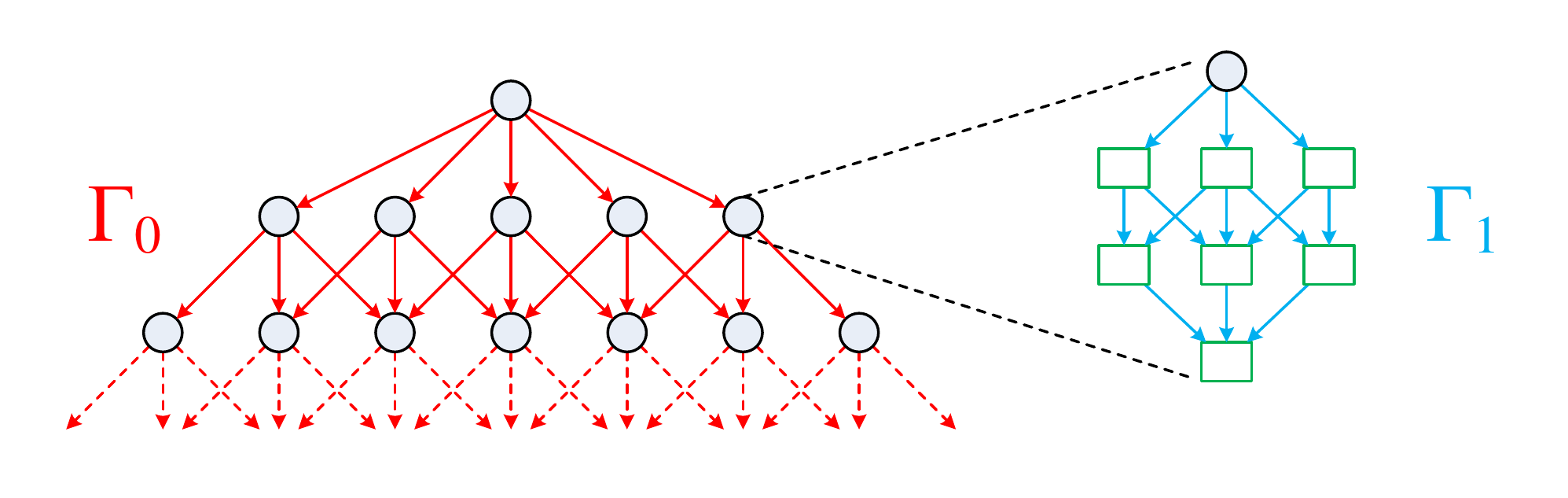}
\caption{A natural decomposition of the poset associated to the stratification. $\Gamma_0$ corresponds to $2\times 2$ minors and $\Gamma_1$ corresponds to non-minimal ones.}
\label{cartoon_gammas}
\end{center}
\end{figure}

This new set of boundaries can be nicely captured by additional posets $\Gamma_1$ emanating from every point in $\Gamma_0$. It is important to emphasize that, in general, each point in $\Gamma_0$ can have a different $\Gamma_1$. In addition, the explicit form of $\Gamma_0$ and the $\Gamma_1$'s generically depends on whether we are considering the mini or full stratification. The top element of each $\Gamma_1$ is characterized by having all non-minimal minors of types (ii) and (iii) non-vanishing. \fref{cartoon_gammas} shows a cartoon of the structure of the full stratification poset.

Note that the construction of the $\Gamma_1$'s requires caution. First, not all type (iii) minors can always be set to zero. Non-minimal minors are in general not independent and it is necessary to explicitly check whether it is possible to shut them off while preserving the positivity of the type (ii) and type (iii) larger minors and of the \pl coordinates $\Delta_I^{(i)}$. This becomes particularly important when trying to turn off combinations of them. Moreover, if considering the full stratification, for every label we should consider all separate regions. Finally, the computation of the dimension of the boundaries via equation (\ref{eqs:dimension}) can be subtle. The vanishing of the larger minors should be taken into account as extra relations among \pl coordinates, and hence contribute to $N_{\text{rel}}$ in (\ref{eqs:dimension}), only if they are independent from the other conditions, i.e. \pl relations plus the possible vanishing of other larger minors. Explicit examples of all these issues are given in \sref{section_examples}.

\bigskip

\section{Simple Examples: Basic Properties}

\label{section_examples_first_approach}

This section further illustrates some of the basic properties of positivity in terms of simple examples.

\bigskip

\subsection{Stratification of $G_{+}(0,n;1)=G_{+}(2,n)$}

\label{section_stratification_G2n}

Let us first consider the 1-loop geometry. A top-dimensional cell of $G_+(0,n,1)\equiv G_{+}(2,n)$ has all $\binom{n}{2}=\frac{1}{2}n(n-1)$ \pl coordinates turned on. There are $(\frac{n^2}{2}-\frac{n}{2}-2n+3)$ independent \pl relations; together with the $GL(2)$ invariance which removes one extra degree of freedom by rescaling the coordinates, we get

\begin{equation}
\frac{1}{2}n(n-1)-(\frac{n^2}{2}-\frac{n}{2}-2n+3)-1=2(n-2)
\end{equation}
degrees of freedom. Boundaries are obtained by setting some $\Delta_I$'s to zero in a way that is compatible with the \pl relations and $\Delta_J > 0$. Since in this case there are no non-minimal minors, there is no distinction between mini and full stratification. From each boundary it is then possible to further set more $\Delta_I$ to zero in a way compatible with the \pl relations and $\Delta_J > 0$ to obtain all of the sub-boundaries. Iterating this procedure until reaching the zero-dimensional boundaries produces the stratification of $G_{+}(2,n)$. There are efficient combinatorial techniques that can be employed for doing this in a quick and systematic way \cite{Franco:2013nwa}, which will be briefly reviewed in \sref{sec:pmGivedecomposition}.

\begin{figure}[h]
\begin{center}
\includegraphics[scale=1.1]{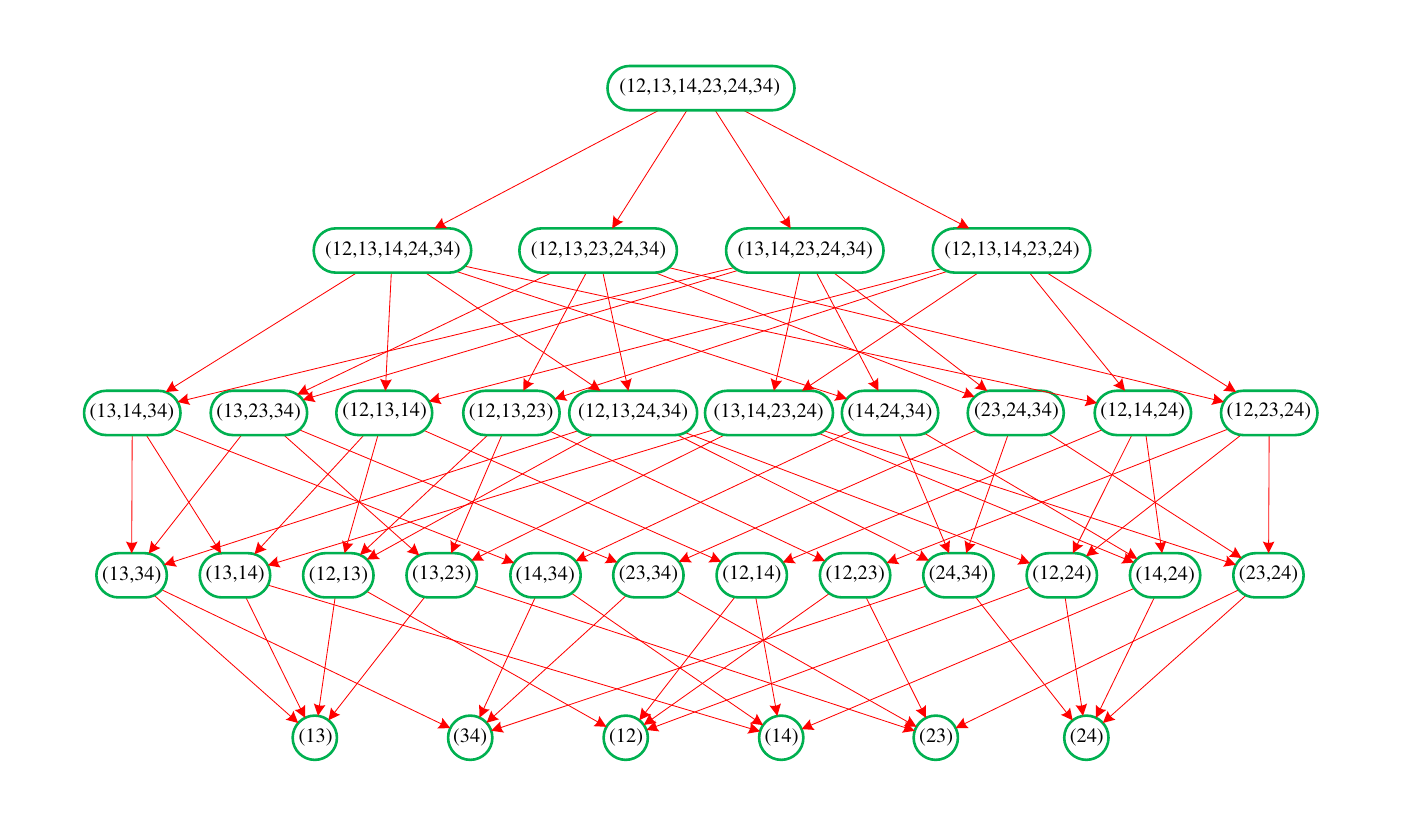}
\vspace{-1cm}\caption{Boundaries of $G_{+}(2,4)$. The parentheses indicate which \pl coordinates are turned on. The top level has all 6 coordinates turned on and has dimension 4, the bottom level has only one coordinate turned on and has dimension 0.}
\label{fig:G24decomp}
\end{center}
\end{figure}

The boundaries can be conveniently organized into levels according to their dimensions. Connecting with arrows each boundary to its sub-boundaries creates a poset. An example is provided in \fref{fig:G24decomp}, where we illustrate the stratification of $G_{+}(2,4)$.\footnote{This poset has already appeared in the literature, see e.g.\ \cite{ArkaniHamed:2012nw,Franco:2013nwa}.} In this example there are 6 \pl coordinates: $\Delta_{12}$, $\Delta_{13}$, $\Delta_{14}$, $\Delta_{23}$, $\Delta_{24}$, $\Delta_{34}$ and one \pl relation: 
\begin{equation}
\label{eq:G24pluckerRel}
\Delta_{12}\Delta_{34}+\Delta_{23}\Delta_{14}=\Delta_{13}\Delta_{24} \, .
\end{equation}
%
%
Some remarks are already in order:
\begin{itemize}
\item At the first step, going to the 3-dimensional boundaries, we only turn off one \pl coordinate. Since there are six \pl coordinates that can be turned off, we would naively expect six different 3-dimensional boundaries. Instead, as shown in \fref{fig:G24decomp}, there are only four of them. This is because once we restrict the $\Delta_I$'s to be positive, two of these would-be boundaries are inconsistent with the \pl relations. For example, killing $\Delta_{13}$ gives
\begin{equation}
\Delta_{12}\Delta_{34}+\Delta_{23}\Delta_{14}=0 \, ,
\end{equation}
which can only be satisfied if we do not restrict ourselves to the strictly positive domain. This is the first example of positivity killing boundaries ``by hand''. This phenomenon was already studied in \cite{Franco:2013nwa} and emerged naturally from the methods therein. We note that this is not imposing extended positivity yet, which imposes compatibility of relations from different loops; this is positivity at a single loop level.
\item For several 2-dimensional boundaries some extra $\Delta_I$ had to be set to zero in order to satisfy the \pl relation. For example, starting from the boundary with non-vanishing $(12,13,14,24,34)$, i.e.\ where we have turned off $\Delta_{23}$, it is not possible to only kill $\Delta_{12}$, because the \pl relation would then become
\begin{equation}
\Delta_{13}\Delta_{24}=0 \, ,
\end{equation}
which is not possible on \textit{any} non-zero domain. Note here that positivity is not the issue, it is the violation of the \pl relation.
\item As mentioned, the boundaries constructed in this way form a poset. Moreover, this poset is Eulerian, i.e.\
\begin{equation}
\sum_{d=0}^4 (-1)^d \mathbb{N}^{(d)} = 1 \, ,
\end{equation}
where $\mathbb{N}^{(d)}$ is the number of boundaries of dimension $d$. We note that for this simple example there is no distinction between mini and full stratification.

\item The full extent of extended positivity never comes into play in this example. Having only one matrix, we never need to consider whether minors of different matrices are compatible. This will however not be the case for the example of $G_{+}(0,n;L=2)$.
\end{itemize}

\bigskip

\subsection{Non-Minimal Minors} \label{sec:G24twoloops}

Before developing a practical implementation for it in the coming section, it is illuminating to consider a few explicit examples of the classification of non-minimal minors introduced in \sref{sec:methodsummary}.

Let us consider the simple case of $G_{+}(0,4;2)$, which has 12 \pl coordinates. From \fref{fig:G24decomp}, we see that $G_{+}(0,4;1)$ has 33 boundaries. The square of this positroid stratification then has $33^2=1\, 089$ configurations, the top-dimensional one being that with all 12 $\Delta^{(i)}_I$'s turned on, giving dimension 8. All these configurations automatically satisfy the two \pl relations, both of the form \eref{eq:G24pluckerRel}, as well as the non-negativity of all \pl coordinates.

Some of these configurations, however, do not satisfy the extended positivity $\Delta^{(1,2)}_{1234} \geq 0$, with $\Delta^{(1,2)}_{1234}$ given in terms of \pl coordinates in \eref{4x4_from_pl}. One such configurations corresponds to the set of vanishing \pl coordinates, i.e. label, $\{ \Delta^{(2)}_{12},\Delta^{(2)}_{23},\Delta^{(2)}_{14},\Delta^{(2)}_{34},$ $\Delta^{(2)}_{24}\}$. In this case, we have
\beq
\Delta^{(1,2)}_{1234}=0+0+0+0+0- \Delta^{(1)}_{24} \Delta^{(2)}_{13} \, ,
\eeq
which is explicitly negative. We hence conclude that this label does not correspond to a boundary.

Let us now present examples of the three different types of behaviors identified in \sref{sec:methodsummary}.

\begin{itemize}
\item {\bf Type (i)}: for the label $\{\Delta^{(1)}_{12},\Delta^{(2)}_{12},\Delta^{(1)}_{14},\Delta^{(2)}_{14},\Delta^{(1)}_{13},\Delta^{(2)}_{13} \}$, we automatically have
\beq
\Delta^{(1,2)}_{1234}=0 \, .
\eeq

\item {\bf Type (ii)}: for the label $\{\Delta^{(2)}_{12},\Delta^{(2)}_{23},\Delta^{(2)}_{14},\Delta^{(2)}_{13},\Delta^{(2)}_{24}\}$, we have
\beq
\Delta^{(1,2)}_{1234}=\Delta^{(1)}_{12}\Delta^{(2)}_{34} + 0 + 0 + 0 - 0 - 0 \,  ,
\eeq
which is strictly positive. We then cannot reach new boundaries by only turning off $\Delta^{(1,2)}_{1234}$.

\item {\bf Type (iii)}: for the label $\{\Delta^{(1)}_{12},\Delta^{(1)}_{34}\}$, we obtain 
\beq
\Delta^{(1,2)}_{1234}=0+0+\Delta^{(1)}_{23} \Delta^{(2)}_{14}+\Delta^{(1)}_{14} \Delta^{(2)}_{23}-\Delta^{(1)}_{13} \Delta^{(2)}_{24}-\Delta^{(1)}_{24} \Delta^{(2)}_{13} \, ,
\eeq
which has both positive and negative contributions. This type of non-minimal minor can in principle be turned off without turning off \pl coordinates. This is possible whenever there are no obstructions coming from relations with other non-minimal minors, which in this particular case do not exist.

\end{itemize}

In the combinatorial approach we will introduce in the coming sections, the building blocks naturally correspond to entire terms in the non-minimal minors rather than only factors within them.

\bigskip

\section{Combinatorial Stratification}

\label{section_combinatorial_stratification} 
 
There is a natural, combinatorial implementation of the mini stratification of the loop geometry, to which we will refer to as {\it combinatorial stratification}, which generalizes the graphical stratification first introduced by Postnikov for $G_{+}(k,n)$ \cite{2006math09764P}. This extension includes the more general cases that appear in $G_{+}(0,n;L)$, for which extended positivity can be systematically incorporated as explained in \sref{section_combinatorics_positivity}. The language of this stratification is not matroids, positroids, \pl coordinates, and permutations, but is simply that of perfect matchings and perfect orientations. The combinatorial structures discussed in this section only depend on labels and hence correspond to the mini stratification.

\bigskip

\subsection{Perfect Matchings and the Stratification of $G_{+}(k,n)$} \label{sec:pmGivedecomposition}

The stratification illustrated in \fref{fig:G24decomp} can be achieved through a variety of methods, extensively discussed in \cite{Franco:2013nwa}. Here we provide a brief summary of its graphical implementation.

\begin{figure}[h]
\begin{center}
\includegraphics[width=15cm]{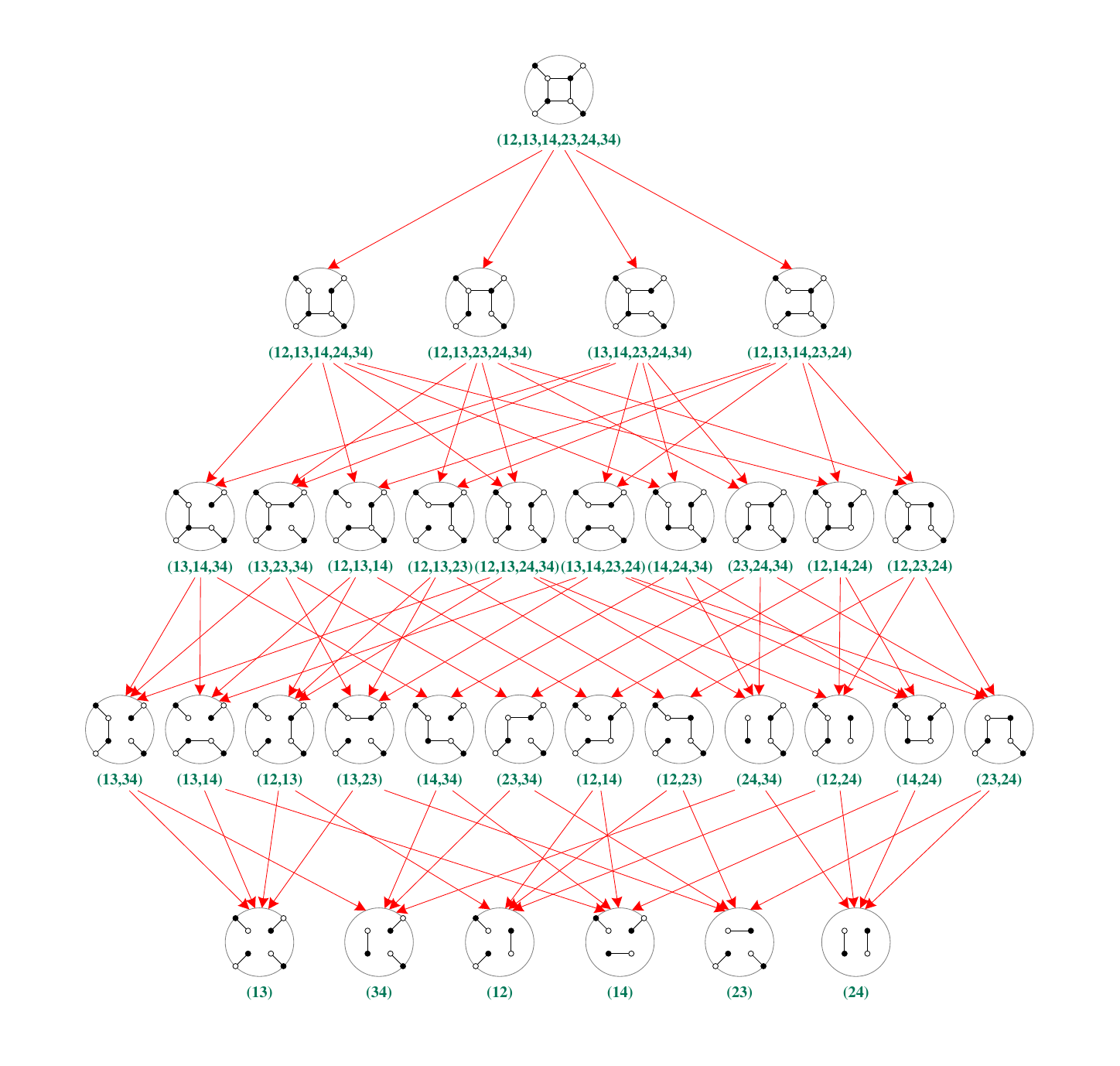}
\vspace{-1cm}\caption{Boundary structure of $G_{+}(2,4)$ and the graphs associated to each boundary. For each graph we indicate the set of 
non-vanishing \pl coordinates.}
\label{fig:G24decompgraphmatroid}
\end{center}
\end{figure}

Following \cite{2006math09764P}, every cell of the positive Grassmannian $G_+(k,n)$ can be associated to a planar bicolored graph,\footnote{To be precise, it is associated to an equivalence class of graphs, which differ by certain moves and reductions.} which in turn determines a specific set of totally positive \pl coordinates. Furthermore, it is also possible, as we do in this paper, to restrict to graphs which are not only bicolored but that are bipartite. \fref{fig:G24decompgraphmatroid} shows the graphical representation of the top-dimensional cell of $G_+(2,4)$ and its lower dimensional boundaries.

\begin{figure}[h]
\begin{center}
\includegraphics[scale=0.7]{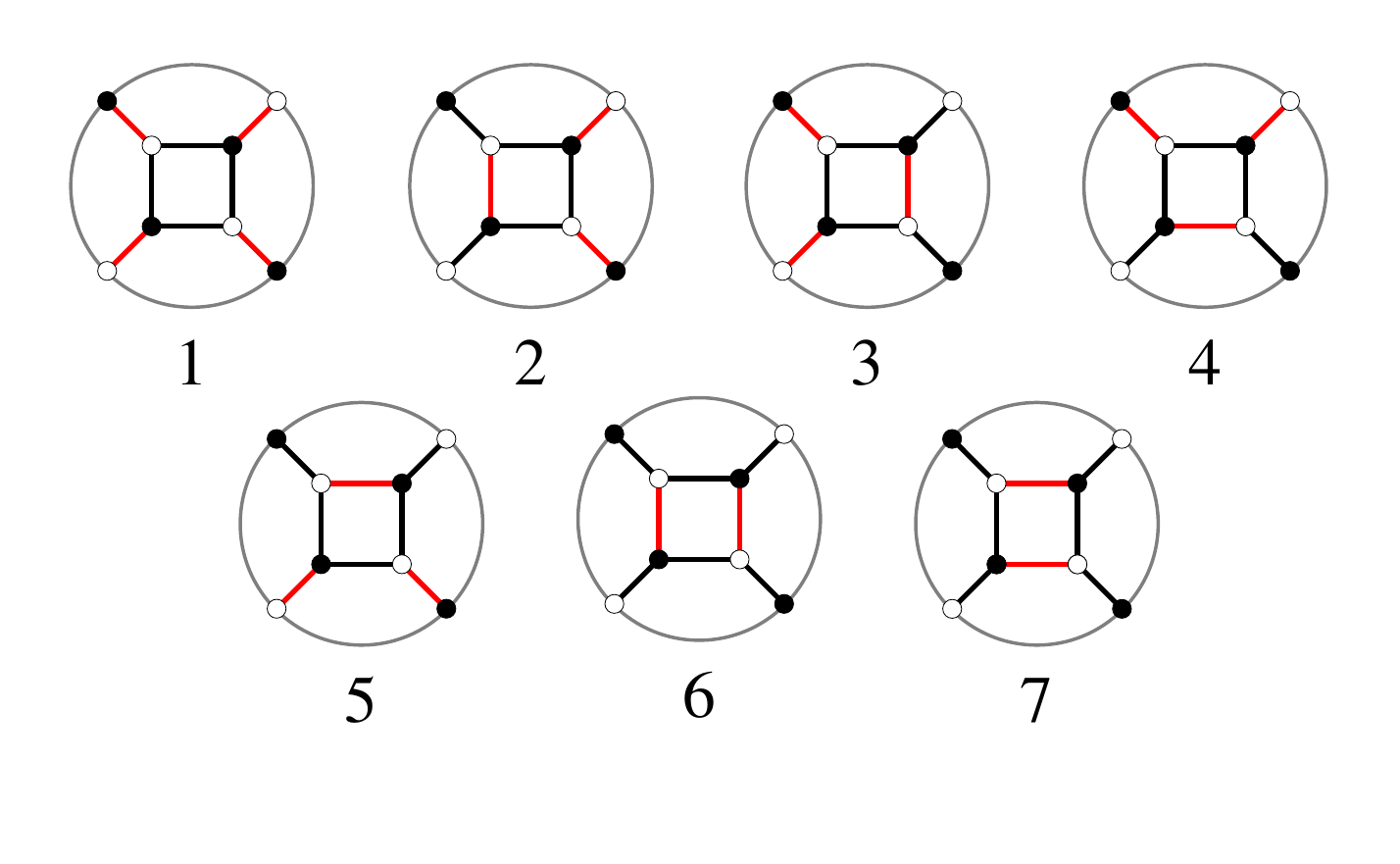}
\vspace{-1cm}\caption{The seven perfect matchings for the bipartite graph associated to the top-dimensional cell of $G_+(2,4)$. Edges in the perfect matchings are shown in red. The graph is embedded into a disk, whose boundary is shown in gray.}
\label{fig:sqbpms}
\end{center}
\end{figure}

%
\textit{Perfect matchings} are fundamental objects in the study of bipartite graphs. A perfect matching is a sub-collection of edges such that every internal node is the endpoint of only one edge, while external nodes may or may not be contained in the perfect matching.\footnote{External nodes are those that lie on the boundary. The objects we have just defined are, more precisely, denoted {\it almost perfect matchings} in the literature. For brevity, we will simply refer to them as perfect matchings. Similarly, we refer to edges as external or internal depending on whether they terminate on external nodes or not.} As an example, the top-dimensional cell of $G_{+}(2,4)$ has 7 perfect matchings, which we present in \fref{fig:sqbpms}.\footnote{There are powerful methods for obtaining the perfect matchings of a graph, see e.g.\ \cite{Franco:2012mm}.}

\begin{figure}[h]
\begin{center}
\includegraphics[scale=0.7]{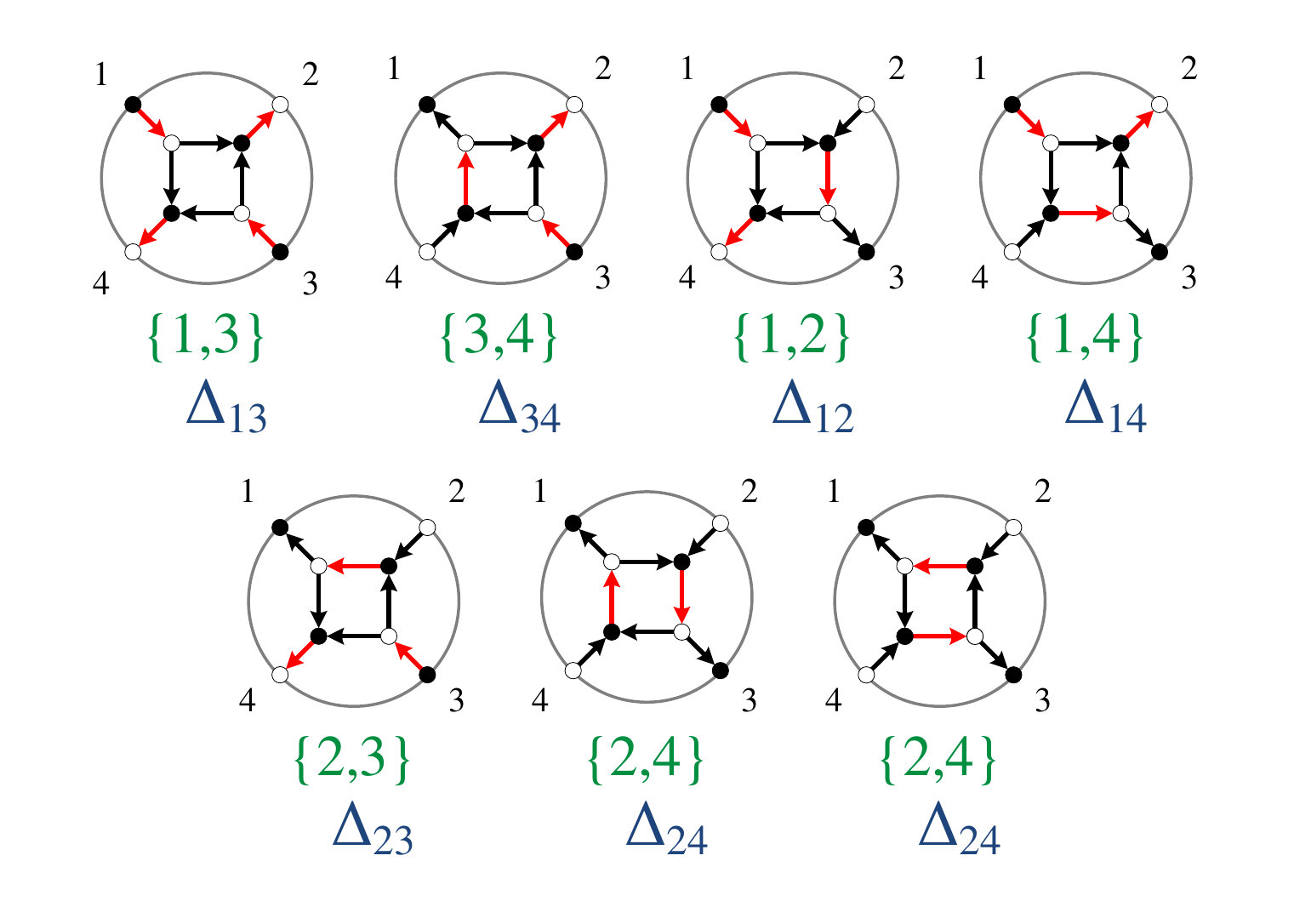}
\caption{Perfect orientations corresponding to the perfect matchings shown in \fref{fig:sqbpms}. The edges of the perfect matchings are shown in red, the source set is labeled underneath each graph in green and the \pl coordinate associated to each perfect flow is in blue. The last two perfect orientations have the same sources and hence contribute to the same \pl coordinate.}
\label{fig:sqbPerforient}
\end{center}
\end{figure}

There exists a precise map between perfect matchings and \pl coordinates. The map is based on perfect orientations, which are flows over the edges of the graph constructed according to the following rules:
\begin{itemize}
\item White nodes must have one incoming arrow and the rest outgoing.
\item Black nodes must have one outgoing arrow and the rest incoming.
\end{itemize}
Going from a perfect matching to a perfect orientation is a simple matter of drawing an arrow pointing from black node to white node over those edges that the perfect matching occupies, i.e.\ the red edges in \fref{fig:sqbpms}, and the rest of the arrows according to the above rules. 
Given a perfect orientation, {\it its source set} is the set of external nodes whose edges point into the graph. 
The label $I$ of the source set of a perfect orientation corresponds to the index of the associated \pl coordinate $\Delta_I$. Multiple perfect matchings can share the same source set, which indicates that they represent contributions to the same \pl coordinate. Such perfect matchings correspond to the same point in the matroid polytope. The perfect orientations and source sets associated to \fref{fig:sqbpms} are shown in \fref{fig:sqbPerforient}.

\begin{figure}[h]
\hspace{-1.15cm}\includegraphics[width=18cm]{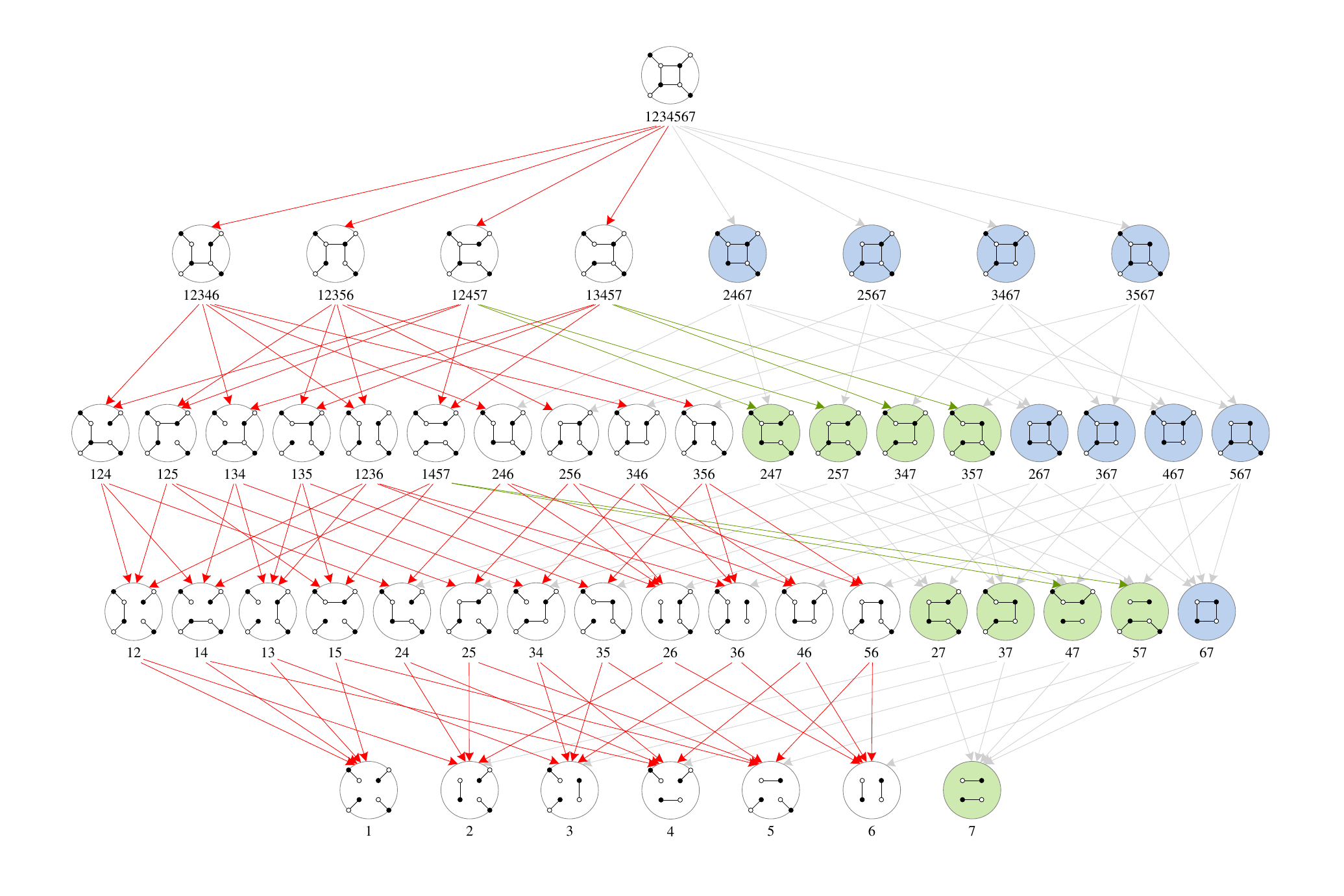}
\begin{center}
\caption{Stratification of the graph associated with the top-dimensional cell of $G_{+}(2,4)$. Below each graph we indicate the surviving perfect matchings. When 6 and 7 are identified, green and blue nodes in the poset are subject to horizontal and vertical identifications, respectively.}
\label{fig:sqbfacelattice}
\end{center}
\end{figure}

It is possible to obtain the stratification by using the graph as a starting point. The way to proceed is to successively remove edges, following the prescription in \cite{2007arXiv0706.2501P,Franco:2013nwa}. This kills the perfect matchings that occupied those edges. Doing this for the example under consideration we obtain the lattice shown in \fref{fig:sqbfacelattice}.

\begin{figure}[h]
\begin{center}
\includegraphics[width=15cm]{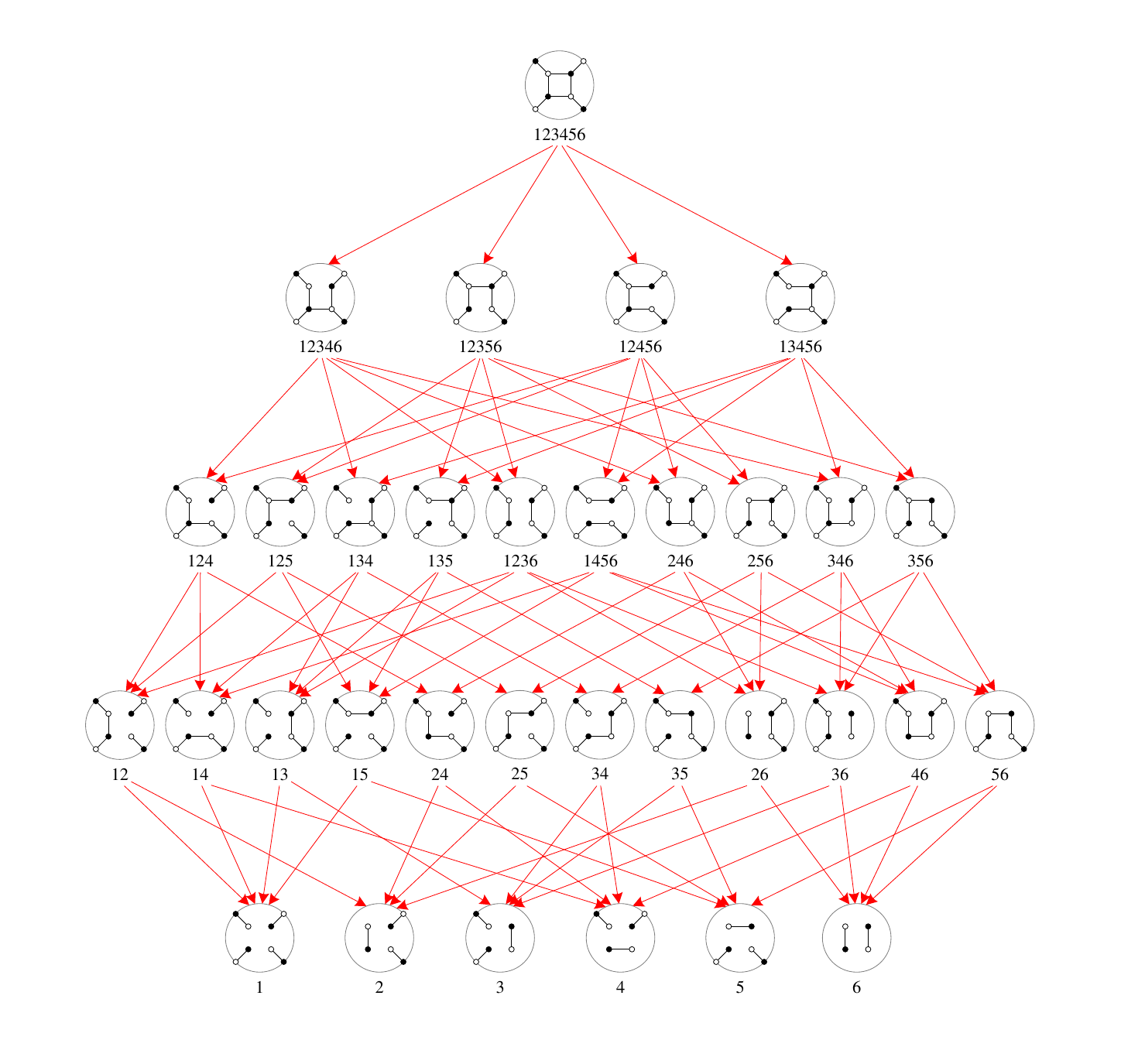}
\caption{Boundary structure of $G_{+}(2,4)$, obtained through identification of perfect matchings $6 \leftrightarrow 7$. Below each graph we indicate the surviving perfect matchings.}
\label{fig:G24decompgraphpms}
\end{center}
\end{figure}

%
The stratification of $G_{+}(k,n)$ is then achieved by \textit{identifying} those perfect matchings that only differ by internal edges, equivalently those perfect matchings which contribute to the same \pl coordinate.
To obtain the stratification of the example at hand, $G_{+}(2,4)$, we identify the perfect matchings 6 and 7. This in turn causes the boundaries colored in green to be identified with other boundaries of the same dimension, and the boundaries colored in blue with other boundaries of lower dimension. Following \cite{Franco:2013nwa}, we refer to these processes as horizontal and vertical identifications, respectively. The result of this identification is illustrated in \fref{fig:G24decompgraphpms}, which perfectly coincides with Figures \ref{fig:G24decomp} and \ref{fig:G24decompgraphmatroid}.

\bigskip

\subsection{Multi-Loop Geometry and Hyper Perfect Matchings}

\label{section_hyper_pms}

Based on our previous discussion, the natural approach for treating the $k=0$, $L$-loop geometry $G_+(0,n;L)$ is to introduce one bipartite graph associated to the top dimensional cell of $G_+(2,n)$ per loop, and to regard the union of these $L$ identical disjoint graphs as a unified object in its own right.

\begin{figure}[h]
\begin{center}
\includegraphics[scale=0.5]{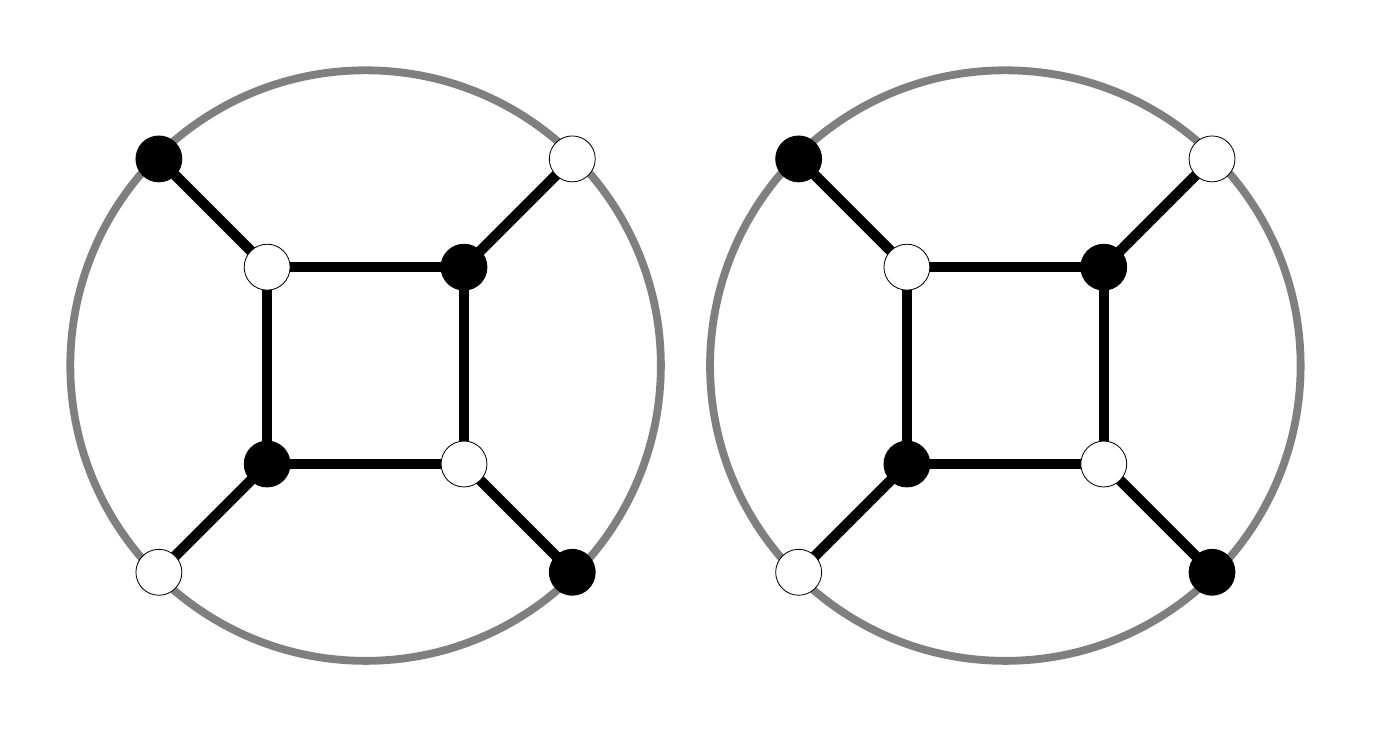}
\caption{The starting graph for the stratification of two loops is simply two separate identical planar graphs for the top-dimensional cell of $G_+(2,n)$ (here $n=4$), each representing one loop.}
\label{fig:twoG24graph}
\end{center}
\end{figure}

As for $G_+(k,n)$, perfect matchings of the multi-component bipartite graph play a central role. In order to emphasize the disjoint nature of the underlying graphs we will refer to them as {\it hyper perfect matchings}, reserving the term perfect matching for those on each component. Denoting $p_i$ the perfect matchings on the first component, $q_j$ the ones on the second component, etc, an hyper perfect matching takes the form

\beq
P_{i,j,k,\ldots} = p_i q_j r_k \ldots .
\eeq

The first step, before incorporating the effect of extended positivity, is to produce the $L^{th}$ power of the 1-loop stratification, as done in \sref{sec:G24twoloops}. This can be done in two ways:

\smallskip

\begin{itemize}
\item Performing the combinatorial stratification introduced in \cite{2007arXiv0706.2501P,Franco:2013nwa} of the $L$-component graph, considered as a unified object. This involves constructing the face lattice of the matching polytope and identifying hyper perfect matchings that correspond to the same point in the matroid polytope or equivalently, in more practical terms, those differing only at internal edges. Here matching and matroid polytopes indicate their obvious generalizations to disjoint graphs. In practice, the matroid polytope identification corresponds to identifying hyper perfect matchings which only differ on internal edges. This method is straightforward to implement.

\item Taking $L$ copies of the 1-loop stratification in which perfect matchings from different loops are given a distinct name and multiplying them together. Effectively, this is equivalent to directly taking the $L^{th}$ power of the 1-loop result, whilst keeping track of which graph component perfect matchings belong to.
\end{itemize} 

\medskip

\noindent The second method is computationally much easier to implement and faster to execute, and will therefore be adopted from here on. However, it is often conceptually useful to think in terms of the first one.

Like the positroid stratification of the positive Grassmannian, its $L^{th}$ power automatically gives rise to a poset with Euler number $\mathcal{E}=1$. This can be understood in different ways. First, as we mentioned above, this is in fact the positroid stratification of a graph made out of $L$ disjoint components. Alternatively, one can understand this by thinking that there are $L$ nested Eulerian posets. Our explicit results in \sref{section_examples}, \sref{section_example_n4_L3} and \sref{section_deformed_amplituhedron} confirm this general result.

Let us see how these ideas work for $G_+(0,4;2)$. In this case, we need to consider two graphs for the top-dimensional cell of $G_{+}(2,4)$ as shown in  \fref{fig:twoG24graph}. Each of them has 7 perfect matchings, which we call $p_i$ and $q_j$, $i,j=1,\ldots, 7$. The combined graph thus has $7^2=49$ hyper perfect matchings $P_{i,j}=p_i q_j$. The matroid identification of perfect matchings on each loop, $p_6 \leftrightarrow p_7$ and $q_6 \leftrightarrow q_7$, implies the identification of hyper perfect matchings $P_{6,j} \leftrightarrow P_{7,j}$ and $P_{i,6} \leftrightarrow P_{i,7}$. The identifications arising from $p_6 \leftrightarrow p_7$ and $q_6 \leftrightarrow q_7$ are automatically implemented if we only use the labels in \fref{fig:G24decompgraphpms}: hyper perfect matchings $P_{7,j}$ and $P_{i,7}$ simply do not appear.

\bigskip

\section{The Combinatorics of Extended Positivity}

\label{section_combinatorics_positivity}

The procedure explained in the previous section automatically implements the \pl relations and the positivity of the $\Delta^{(i)}_I$'s, but not yet the full extended positivity. The next step of the process is to shrink the poset we have just generated by eliminating those points which violate extended positivity. The purpose of this section is to introduce efficient combinatorial methods to deal with positivity based on the properties of hyper perfect matchings.

\bigskip

\subsection{Further Thoughts on Extended positivity}

Before introducing a combinatorial approach, it is useful to revisit our discussion of extended positivity from \sref{section_extended_positivity} and the observations made for explicit examples in \sref{section_examples_first_approach}. 

Boundaries can be associated to labels, i.e. to lists of vanishing minors, generally of different dimensions, $\Delta^{(i_1,\ldots,i_m)}_J$, $m=1,\ldots,L$. Extended positivity demands the non-vanishing ones to be strictly positive. The $\Delta^{(i_1,\ldots,\i_m)}_J$'s, are polynomials in which every term is an order $m$ product of $\Delta^{(i)}_I$'s coming from different loops. For illustration purposes, consider the single $4\times 4$ minor that exists for $G_+(0,4;2)$, which was presented in \eref{4x4_from_pl}. It is given by
\begin{equation}
\Delta^{(1,2)}_{1234}=\Delta^{(1)}_{12}\Delta^{(2)}_{34} + \Delta^{(1)}_{23}\Delta^{(2)}_{14} + \Delta^{(1)}_{34}\Delta^{(2)}_{12} + \Delta^{(1)}_{14}\Delta^{(2)}_{23} - \Delta^{(1)}_{13}\Delta^{(2)}_{24} -\Delta^{(1)}_{24}\Delta^{(2)}_{13} .
\label{example_4_x_4_minor}
\end{equation}
This example illustrates the  behavior of general minors. {\it From the point of view of a given $2m\times 2m$ minor}, there is a rather obvious distinction among those terms which: appear with a positive sign, appear with a negative sign or do not appear. In the coming section we will translate the different types of terms into a classification of hyper perfect matchings.

\bigskip

\subsection{Hyper Perfect Matchings: Good, Bad and Neutral}  \label{sec:goodandbadSourcesets}

The different types of contributions to a given minor can be translated into a classification of hyper perfect matchings.

Following the discussion in \sref{sec:pmGivedecomposition}, for every loop there is a correspondence between \pl coordinates $\Delta^{(i)}_{\ell_a \ell_b}$ in $G_+(2,n)$ and perfect matchings.\footnote{As explained in \sref{section_combinatorial_stratification}, the map becomes a bijection after the identification of perfect matchings following the matroid polytope has been implemented.}  The \pl coordinate associated to a given perfect matching is determined by the source set of the corresponding perfect orientation.

Since every term in a $2m\times 2m$ minor is a product of $m$ \pl coordinates coming from different loops, the previous map implies that every such term can be identified with a hyper perfect matching. Extending what we did for perfect matchings, here we also discuss hyper perfect matchings after identifications following from the matroid polytope or, equivalently, distinguishing them only by their external edge content. For $m>1$, however, the sign of terms vary, as e.g. in \eref{example_4_x_4_minor}.

For every non-minimal minor, we will thus define the following classification of hyper perfect matchings:

\begin{itemize}
\item {\bf Good:} it corresponds to a positive term in the minor.
\item {\bf Bad:}  it corresponds to a negative term in the minor.
\item {\bf Neutral:} it does not appear in the minor.
\end{itemize}

Let us investigate in more detail how these concepts work for the example in \eref{example_4_x_4_minor}. The corresponding graph is shown in \fref{fig:twoG24graph} and the map between perfect matchings for each loop and \pl coordinates is given in \fref{fig:sqbPerforient}. In terms of perfect matchings and hyper perfect matchings, we have

\beq
\begin{array}{rcccccccccccr}
\Delta^{(1,2)}_{1234} = & \Delta^{(1)}_{12}\Delta^{(2)}_{34} & +  & \Delta^{(1)}_{23}\Delta^{(2)}_{14} & + & \Delta^{(1)}_{34}\Delta^{(2)}_{12} & + & \Delta^{(1)}_{14}\Delta^{(2)}_{23} & - & \Delta^{(1)}_{13}\Delta^{(2)}_{24} & - & \Delta^{(1)}_{24}\Delta^{(2)}_{13} & . \\
& p_3 \, q_2 & & p_5 \, q_4 & & p_2 \, q_3 & & p_4 \, q_5 & & p_1 \, q_6 & &  p_6 \, q_1 & \\
& P_{3,2} & & P_{5,4} & & P_{2,3} & & P_{4,5} & & P_{1,6} & & P_{6,1} & 
\end{array}
\label{eq:2Lnequal4extposWITHpmterms}
\eeq
For this minor, we thus have:
\begin{itemize}
\item {\bf Good:} $P_{3,2}$, $P_{5,4}$, $P_{2,3}$, $P_{4,5}$ 
\item {\bf Bad:} $P_{1,6}$, $P_{6,1}$ 
\end{itemize}
while all other hyper perfect matchings are neutral.

Specifying the label completely determines which hyper perfect matchings are present. The converse is, however, not true.

We now have a powerful technology for incorporating extended positivity into our stratification. For a given minor to be positive, some of its good hyper perfect matchings must survive. Conversely, a minor violates positivity if only bad hyper perfect matchings are present. We can also see how to, in the language of \sref{sec:methodsummary}, go from $\Gamma_0$ to $\Gamma_1$ by turning off $m>1$ minors. Such minors can vanish {\it without sending to zero additional \pl coordinates} only if both good and bad hyper perfect matchings are simultaneously present. Note that this condition is necessary but not sufficient.

\bigskip

\paragraph*{Practical Implementation.} In cases with multiple $m>1$ minors, a good approach for implementing extended positivity is as follows:

\begin{itemize}
\item For every minor, determine whether a given hyper perfect matching $P_i$ is good, bad or neutral. For each hyper perfect matching, this information is easily stored in a vector whose length is the number of non-minimal minors. If $P_i$ is bad for a given minor, the corresponding entry is set to be the complex number $i$; if $P_i$ is good, the entry is set to $1$; if $P_i$ is neutral, the entry is $0$.

\item We then generate a single vector for each boundary, by adding the vectors associated to all hyper perfect matchings in it.

\item If in the final vector the argument of the complex number in any entry is $\pi / 2$, the boundary has at least one relation with only negative terms turned on, so it violates extended positivity and should be removed. If the argument is $0$, the corresponding minor has only positive terms turned on or none at all, and hence cannot be further turned off to go to a lower dimensional boundary.
\end{itemize}

\noindent It is straightforward to implement this method with any algebraic manipulation software. We stress that sticking to this method is however not strictly necessary to obtain the stratification. For it, only knowledge of vanishing minors is necessary and, as we have just seen, hyper perfect matchings provide a highly efficient language for dealing with them.

\bigskip

\subsection{Extended Positivity and the Return of Permutations}

\label{extended_positivity_permutations}

Permutations play a central role in the classification of cells in the positive Grassmannian. Remarkably, as we explain in this section, extended positivity in $G_+(0,n;L)$ is also beautifully linked to permutations.

Consider a hyper perfect matching $P_{i,j,k,\ldots}=p_i q_j r_k \ldots$. Let us call  $\{s_j,t_j\}$, $\{s_k,t_k\}$, $\{s_l,t_l\}, \ldots$ the pairs of sources for each of the constituent perfect matchings. The columns identifying the minor that the hyper perfect matching contributes to are given by the union of these source sets. The classification of the hyper perfect matching is determined by the parity of the number of crossings in the source set. Let us denote $a_1, a_2$ the ordered source set for the first loop under consideration, $b_1,b_2$ the ordered source set for the second loop, etc. Then, define $\epsilon^{a_1 a_2 b_1 b_2 \cdots}$ to be the ordinary antisymmetric tensor, with the slight modification that the ordered indices are not necessarily consecutive, but do need to be monotonically increasing. For example, $\epsilon^{1256}=\epsilon^{1234}=1$ and $\epsilon^{5739}=1$ but $\epsilon^{2648}=-1$ and $\epsilon^{4849}=0$. The classification of hyper perfect matchings then reduces to:

\begin{equation}
\epsilon^{a_1 a_2 b_1 b_2 \cdots} = \Bigg\{ \begin{array}{rl} 1 \Rightarrow & \text{ good} \\ -1 \Rightarrow & \text{ bad} \\ 0 \Rightarrow & \text{ neutral} \end{array}
\end{equation}

Let us discuss in further detail the graphical implementation of extended positivity. For doing so, we draw a line connecting the pairs of sources for each perfect matching in a given hyper perfect matching and superimpose them on a single graph. 

\medskip

\begin{figure}[h]
\begin{center}
\includegraphics[scale=0.5]{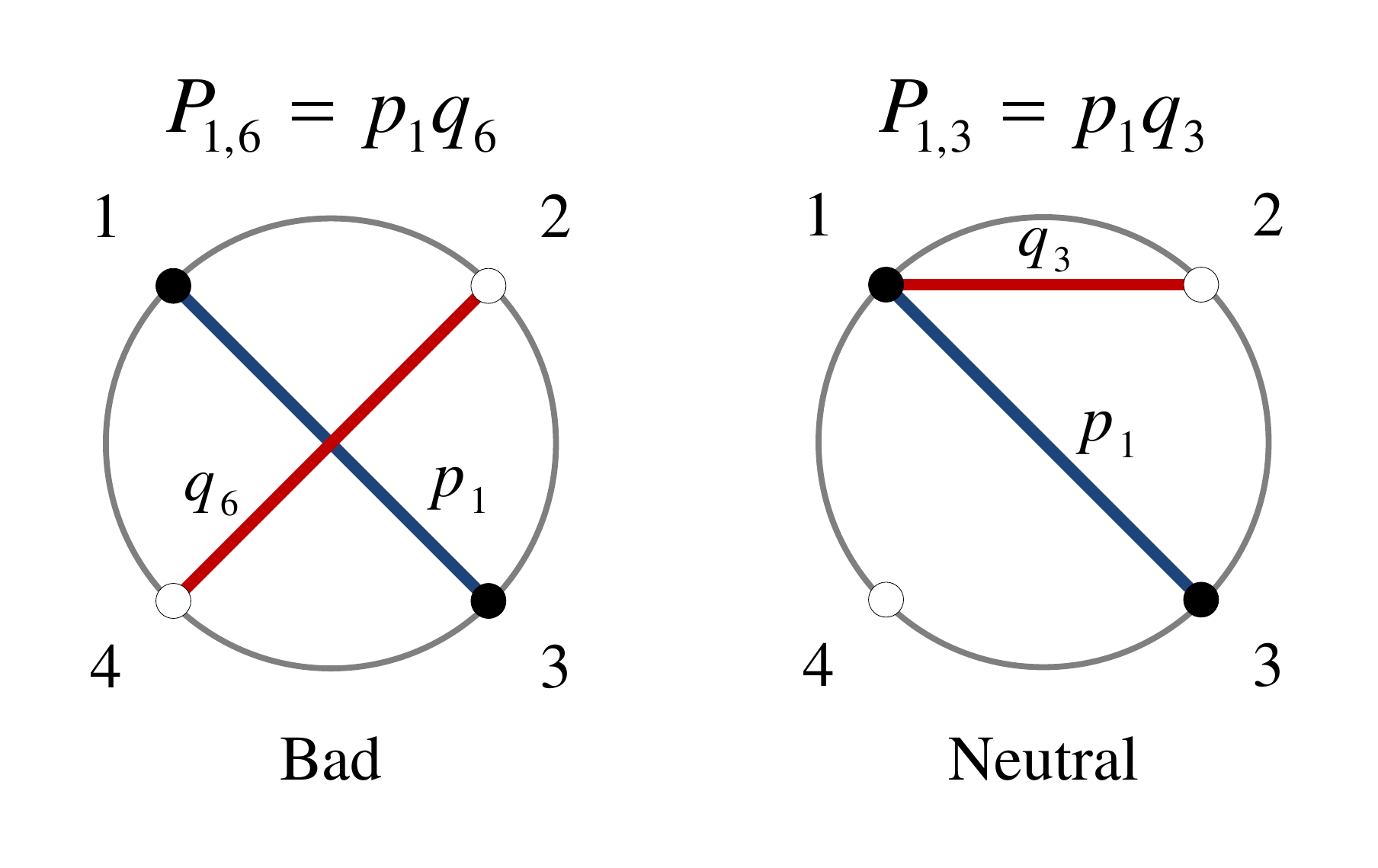}
\caption{$P_{1,6}$ is a bad perfect matching. $P_{1,3}$  is instead neutral, since the crossing does not occur in the interior of the graph. In fact 
$P_{1,3}$ does not occupy all four external nodes, equivalently all columns in the minor.}
\label{fig:badperfmatching}
\end{center}
\end{figure}

\paragraph*{Bad hyper perfect matchings.} Bad hyper perfect matchings are those for which the lines between sources intersect an odd number of times in the interior of the graph. Edges touching at external nodes do not count towards the intersections. \fref{fig:badperfmatching} shows an example of a bad perfect matching for the $n=4$, 2-loop case, $P_{1,6}=p_1 q_6$.\footnote{Notice that $P_{1,7}=p_1 q_7$ is also a bad perfect matching, but it coincides with $P_{1,6}$ after the matroid polytope identification.} The sources for $p_1$ are $\{ 1,3 \}$ and the ones for $q_6$ are $\{ 2,4 \}$. Their union occupies all 4 external nodes and hence all the columns in the minor. The lines between sources cross once.

\begin{figure}[h]
\begin{center}
\includegraphics[scale=0.5]{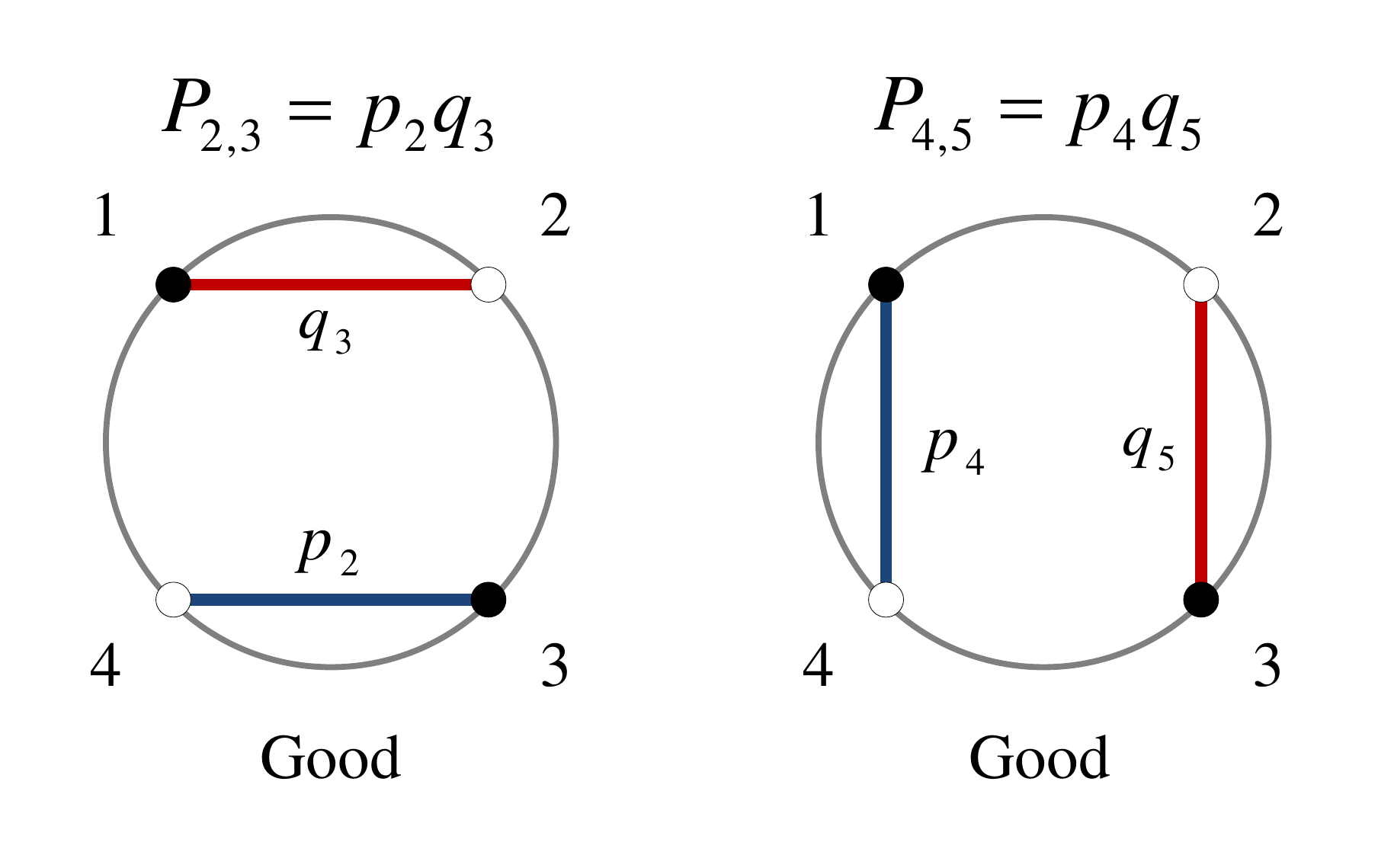}
\caption{$P_{2,3}$ and $P_{4,5}$ are two examples of good perfect matchings.}
\label{fig:goodperfmatching}
\end{center}
\end{figure}

\paragraph*{Good hyper perfect matchings.} They are those whose lines intersect an even number of times in the interior of the graph. Two examples are presented in \fref{fig:goodperfmatching}.

\bigskip

\paragraph*{Neutral hyper perfect matchings.} When the lines joining sources touch on external points, the configuration does not occupy all columns in the minor and hence it does not contribute to it. An example is shown in \fref{fig:badperfmatching}.

\begin{figure}[h]
\begin{center}
\includegraphics[scale=0.5]{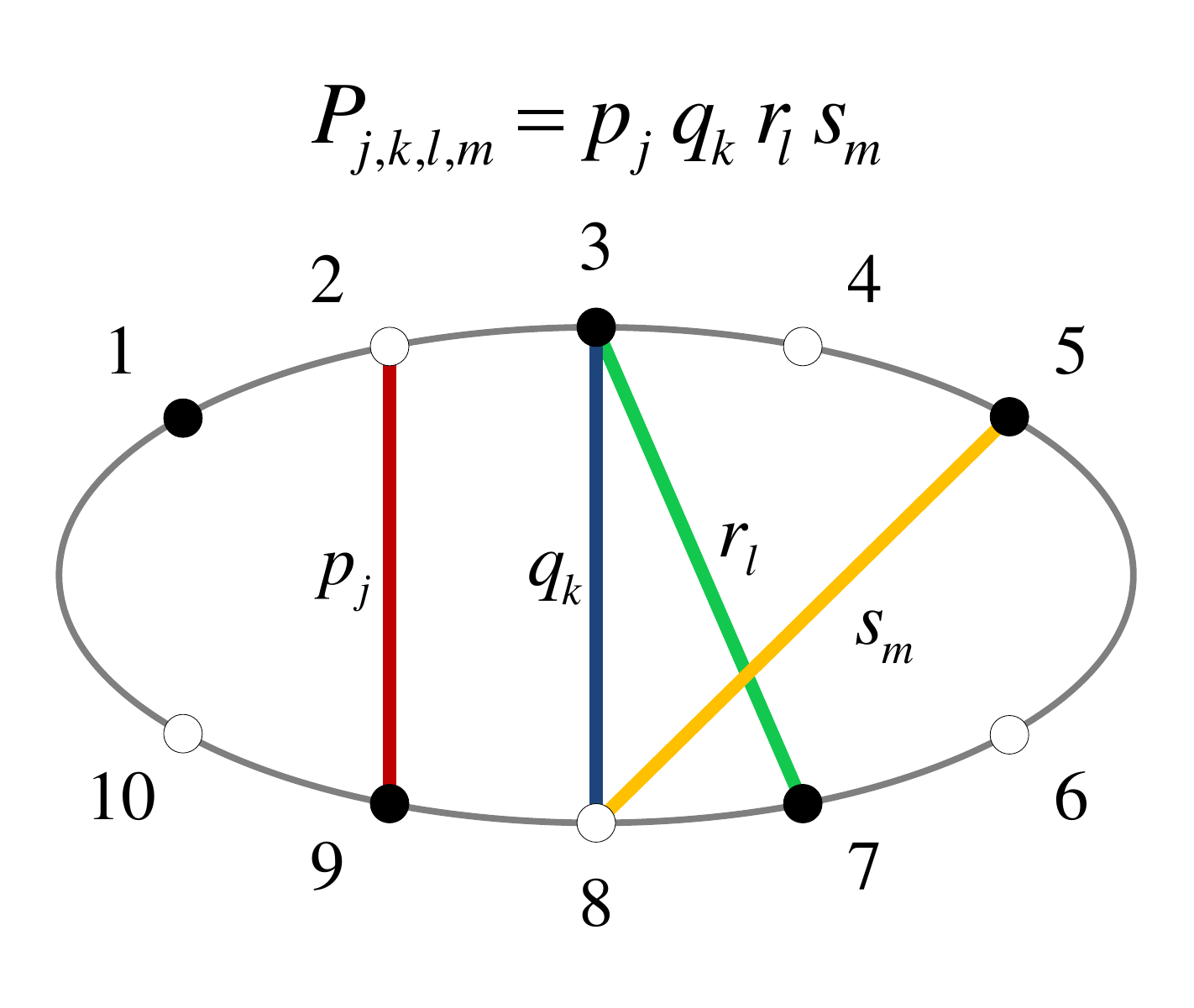}
\caption{This hyper perfect matching is good with regards to the $4 \times 4$ minor involving loops $p$ and $q$ and matrix columns $2,3,8,9$ and is bad with regards to loops $r$ and $s$ and columns $3,5,7,8$.}
\label{fig:goodandbadperfmatching}
\end{center}
\end{figure}

\bigskip
We would like to emphasize that, generally, a hyper perfect matching can be good with regards to a non-minimal minor but bad with regards to another one. An example of this situation is provided in \fref{fig:goodandbadperfmatching}.

\bigskip

\section{Two Loops}

\label{section_examples}

To illustrate the techniques presented above, we stratify the amplituhedron and the log of the amplitude in the case of $k=0$ for 4 particles at 2-loops. We first present the mini stratification introduced in \sref{section_label_counting}. As a crosscheck, the results have been derived both in terms of hyper perfect matchings and directly using \pl coordinates and the relations between them. The full stratification, accounting for all solutions arising from factorization, is presented in \sref{section_full_stratification_L=2}.

\bigskip

\subsection{Mini Stratification}

Let us begin our analysis by classifying boundaries according to their labels.

\label{section_label_counting}

\subsubsection{The Amplituhedron} \label{sec:4particle2loopEX}

The starting point is the graph in \fref{fig:twoG24graph}, which has $7^2=49$ hyper perfect matchings. The 1-loop stratification is shown in \fref{fig:G24decompgraphpms}. To square it, we produce an equivalent set of boundaries for the second graph; the boundaries of both are summarized in \tref{tab:2oneloops}.
%
\begin{table}
\begin{center}
\bigskip 
{\small $\begin{array}{|c|c|}
\hline
\textbf{Dim} & \text{\bf Boundaries of graph 1} \\ \hline \hline
\multirow{1}{*} {\bf 4} & \{p_1, p_2, p_3, p_4, p_5, p_6 \} \\ \hline
\multirow{2}{*} {\bf 3}  & \{p_1, p_2, p_3, p_4, p_6 \} , \{p_1, p_2, p_3, p_5, p_6 \} ,  \\ 
 & \{p_1, p_2, p_4, p_5, p_6 \} , \{p_1, p_3, p_4, p_5, p_6 \} \\ \hline
\multirow{4}{*} {\bf 2} & \{p_1, p_2, p_4 \} , \{p_1, p_2, p_5 \} , \{p_1, p_3, p_4 \} , \\
 & \{p_1, p_3, p_5 \} , \{p_1, p_2, p_3, p_6 \} , \\
 & \{p_1, p_4, p_5, p_6 \} , \{p_2, p_4, p_6 \} , \\
 & \{p_2, p_5, p_6 \} , \{p_3, p_4, p_6 \} , \{p_3, p_5, p_6 \}    \\ \hline
\multirow{3}{*} {\bf 1} & \{p_1, p_2 \} , \{p_1, p_4 \} , \{p_1, p_3 \} , \{p_1, p_5 \} , \\
 & \{p_2, p_4 \} , \{p_2, p_5 \} , \{p_3, p_4 \} , \{p_3, p_5 \} , \\
 & \{p_2, p_6 \} , \{p_3, p_6 \} , \{p_4, p_6 \} , \{p_5, p_6 \}  \\ \hline
\multirow{1}{*} {\bf 0} & \{p_1 \} , \{p_2 \} , \{p_3 \} , \{p_4 \} , \{p_5 \} , \{p_6 \}  \\ \hline
\end{array}$}
\quad \quad
{\small $\begin{array}{|c|c|}
\hline
\textbf{Dim} & \text{\bf Boundaries of graph 2} \\ \hline \hline
\multirow{1}{*} {\bf 4} & \{q_1, q_2, q_3, q_4, q_5, q_6 \} \\ \hline
\multirow{2}{*} {\bf 3}  & \{q_1, q_2, q_3, q_4, q_6 \} , \{q_1, q_2, q_3, q_5, q_6 \} ,  \\ 
 & \{q_1, q_2, q_4, q_5, q_6 \} , \{q_1, q_3, q_4, q_5, q_6 \} \\ \hline
\multirow{4}{*} {\bf 2} & \{q_1, q_2, q_4 \} , \{q_1, q_2, q_5 \} , \{q_1, q_3, q_4 \} , \\
 & \{q_1, q_3, q_5 \} , \{q_1, q_2, q_3, q_6 \} , \\
 & \{q_1, q_4, q_5, q_6 \} , \{q_2, q_4, q_6 \} , \\
 & \{q_2, q_5, q_6 \} , \{q_3, q_4, q_6 \} , \{q_3, q_5, q_6 \}    \\ \hline
\multirow{3}{*} {\bf 1} & \{q_1, q_2 \} , \{q_1, q_4 \} , \{q_1, q_3 \} , \{q_1, q_5 \} , \\
 & \{q_2, q_4 \} , \{q_2, q_5 \} , \{q_3, q_4 \} , \{q_3, q_5 \} , \\
 & \{q_2, q_6 \} , \{q_3, q_6 \} , \{q_4, q_6 \} , \{q_5, q_6 \}  \\ \hline
\multirow{1}{*} {\bf 0} & \{q_1 \} , \{q_2 \} , \{q_3 \} , \{q_4 \} , \{q_5 \} , \{q_6 \}  \\ \hline
\end{array}$}
\bigskip
\caption{List of boundaries, in terms of perfect matchings, for each component of the graph in \fref{fig:twoG24graph}.
\label{tab:2oneloops}}
\end{center}
\end{table}
%
Every boundary in the left table must be multiplied by all boundaries in the right table. This automatically accounts for the \pl relations and the positivity of all \pl coordinates $\Delta^{(i)}_I > 0$. For amusement, we pictorially show all $33^2 = 1\, 089$ boundaries in \fref{fig:4particle2loop}. Organizing these boundaries according to their dimension we obtain the results summarized in the first column of \tref{tab:4particle2loop}, where we show the number of boundaries $\mathbb{N}$ of each dimension. This corresponds to performing step (1) in \sref{sec:methodsummary}.

\begin{table}
\begin{center}
\bigskip
{\small
\begin{tabular}{|c|c|c|c|}
\hline
\textbf{Dim} & $\boldsymbol{\mathbb{N}}$ &  $\boldsymbol{\mathcal{N}_M}$ &  $\boldsymbol{\mathfrak{N}_M}$ \\
\hline
\textbf{8} & 1 & 1 & 1 \\
\hline
\textbf{7} & 8 & 8 & 9 \\
\hline
\textbf{6} & 36 & 36 & 44 \\
\hline
\textbf{5} & 104 & 104 & 140 \\
\hline
\textbf{4} & 208 & 178 & 274 \\
\hline
\textbf{3} & 288 & 224 & 330 \\
\hline
\textbf{2} & 264 & 216 & 264 \\
\hline
\textbf{1} & 144 & 128 & 136 \\
\hline
\textbf{0} & 36 & 34 & 34 \\
\hline
\end{tabular}
}
\bigskip
\caption{Number of boundaries $\mathfrak{N}_{M}$ of the $n=4$, 2-loop amplituhedron, of various dimensions. $\mathbb{N}$ is the number of boundaries before the positivity of $\Delta^{(1,2)}_{1234}$ is implemented. $\mathcal{N}_M$ is the surviving number of boundaries after this condition is enforced, but before non-trivial vanishing of $\Delta^{(1,2)}_{1234}$ is considered. We use a subindex $M$ to emphasize quantities which are computed in the mini stratification. \label{tab:4particle2loop}}
\end{center}
\end{table}

In agreement with our general statement in \sref{section_hyper_pms}, the poset for the square of the positroid stratification of $G_+(2,4)$ is Eulerian:
\begin{equation}
\sum_{i=0}^8 (-1)^i \, \mathbb{N}^{(i)} = 36-144+264-\ldots -8+1=1 \quad .
\end{equation}

Extended positivity only imposes one additional condition: that the $4 \times 4$ minor $\Delta^{(1,2)}_{1234} \geq 0$. The bad perfect matchings here are quickly found to be the one shown in \fref{fig:badperfmatching} and the one where $p$ and $q$ are swapped, i.e.\ $P_{1,6}$ and $P_{6,1}$; the good perfect matchings are those shown in \fref{fig:goodperfmatching} and their $p \leftrightarrow q$ counterparts, i.e.\ $P_{2,3}, P_{4,5}, P_{3,2}$ and $P_{5,4}$, c.f.\ \eref{eq:2Lnequal4extposWITHpmterms}.

Next, we remove all boundaries containing $P_{1,6}$ or $P_{6,1}$, unless they also contain any of $P_{2,3}, P_{4,5}, P_{3,2}$ or $P_{5,4}$. This procedure corresponds to performing step (2) in \sref{sec:methodsummary} and yields the middle column in \tref{tab:4particle2loop}. It is very interesting to see that this column also forms an Eulerian poset:
\begin{equation}
\sum_{i=0}^8 (-1)^i \mathcal{N}_M^{(i)} = 34-128+216-\ldots -8+1=1 \quad .
\end{equation}
This is in general not true at higher loops. However, we will later observe in \sref{section_deformed_4-loops} that this is also the case at 4-loops.

Finally, we construct new boundaries by further imposing the vanishing of the $4\times 4$ minor $\Delta^{(1,2)}_{1234}$ on those boundaries on which it is possible and not automatic due to the vanishing of \pl coordinates. Its expression in terms of \pl coordinates is given in \eref{example_4_x_4_minor}. This corresponds to steps (3) and (4) in \sref{sec:methodsummary}. For every boundary in the $\mathcal{N}_M$ column of \tref{tab:4particle2loop} for which it is possible to impose the equality in \eref{eq:2Lnequal4extposWITHpmterms}, we get an additional boundary of one dimension less. The final answer for the total number of boundaries of the amplituhedron is displayed in the right-hand column in \tref{tab:4particle2loop}. The poset is no longer Eulerian:
\begin{equation}
\sum_{i=0}^8 (-1)^i \mathfrak{N}_M^{(i)} = 34-136+264-\ldots -9+1=2 \quad .
\label{Euler_L=2}
\end{equation}
Remarkably, in \sref{section_integrand_stratification} we will reproduce the right column of \tref{tab:4particle2loop} by studying the singularities of the integrand.

\bigskip

\subsubsection{The Log of the Amplitude}

\label{section_log_combinatorial}

Let us now investigate the geometric properties of another object related to the amplitude. While the fundamental object of interest in field theory is the amplitude, in order to make a connection with the S-matrix we are really interested in its \textit{log}, \mbox{$S \sim \log(A)$}. Writing the loop expansion of $A$ as
\begin{equation}
A = 1 + g A_1 + g^2 A_2 + \ldots \, ,
\end{equation}
where $A_L$ is the $L$-loop contribution, and expanding $\log (A) $ we find the second-order correction to the log of the amplitude to be $g^2 (A_2 -\frac{A_1^2}{2})$. 

Physically, the log of the amplitude is a very interesting object. All amplitudes are IR divergent, with the divergence going as $\frac{1}{\epsilon^{2L}}$ for the $L$-loop contribution, in dimensional regularization. However, the divergence of the log of the amplitude has a fixed order, always going as $\frac{1}{\epsilon^2}$. In the 2-loop case this manifests itself in an exact cancellation of higher order divergences between the $A_2$ and $\frac{A_1^2}{2}$ terms.

Let us continue focusing on $k=0$, $n=4$ and $L=2$. The amplitude $A_2$ can be viewed as two $D_{(i)} \in G_{+}(2,4)$ with the additional condition that the $4 \times 4$ minor $\Delta_{1234}^{(1,2)} \geq 0$. On the other hand, $A_1^2$ is simply the square of the 1-loop amplitude, and corresponds to two $D_{(i)} \in G_{+}(2,4)$ with no extra condition imposed (the factor of $\frac{1}{2}$ corresponds to the symmetrization of loop variables and is of no geometric importance). Then, the \textit{difference} between these two objects is clearly given by two $D_{(i)}$ with $\Delta_{1234}^{(1,2)} \leq 0$. We thus conclude that, from a geometric standpoint, the log of the amplitude at 2-loops can be seen as a complement of the amplitude. At higher loops the story is more complicated, so we shall here only focus on understanding the geometric significance of the complement of the 2-loop amplituhedron. 

It is straightforward to modify our combinatorial methods to incorporate the change from $\Delta_{1234}^{(1,2)} \geq 0$ to $\Delta_{1234}^{(1,2)} \leq 0$. The results of the stratification of the log of the amplitude are summarized in \tref{tab:4particle2loopLOG}. Very interestingly, $\mathcal{E}$ is once again
\begin{equation}
\sum_{i=0}^8 (-1)^i \mathfrak{N}_{M, \text{Log}}^{(i)} = 32-120+220-\ldots -9+1=2 \quad .
\end{equation}

\begin{table}
\begin{center}
\bigskip 
{\small
\begin{tabular}{|c|c|}
\hline
\textbf{Dim} &  $\boldsymbol{\mathfrak{N}_{M, \text{Log}}}$ \\
\hline
\textbf{8} & 1 \\
\hline
\textbf{7} & 9 \\
\hline
\textbf{6} & 44 \\
\hline
\textbf{5} & 132 \\
\hline
\textbf{4} & 240 \\
\hline
\textbf{3} & 274 \\
\hline
\textbf{2} & 220 \\
\hline
\textbf{1} & 120 \\
\hline
\textbf{0} & 32 \\
\hline
\end{tabular}
}
\caption{Number of boundaries $\boldsymbol{\mathfrak{N}_{M, \text{Log}}}$ of various dimensions of the log of the $k=0$, $n=4$, 2-loop amplituhedron.\label{tab:4particle2loopLOG}}
\end{center}
\end{table}

\bigskip

\subsubsection{Gluing the Amplitude to its Log}

\label{section_gluing_amplitude_log}

The amplitude and its log are characterized by having $\Delta^{(1,2)}_{1234} \geq 0$ and $\Delta^{(1,2)}_{1234} \leq 0$, respectively. Their gluing corresponds to the square of the positroid stratification of $G_{+}(2,4)$, since it is obtained by not imposing any restriction on $\Delta^{(1,2)}_{1234}$. Here we discuss in detail the emergence of this simple geometric object from its components.

The 7-dimensional gluing subspace is characterized by $\Delta^{(1,2)}_{1234} =0$. We can study its structure by demanding $\Delta^{(1,2)}_{1234} =0$ and proceeding with our standard stratification. The numbers of boundaries at different dimensions $\mathfrak{N}_{M,\Delta^{(1,2)}=0}$ are given in \tref{tab:4particle2loopGLUING}. These boundaries can be divided into two disjoint categories: 
\begin{itemize}
\item Boundaries on which the condition $\Delta^{(1,2)}_{1234} =0$ imposes a constraint on $2 \times 2$ minors.
\item Boundaries on which the condition $\Delta^{(1,2)}_{1234} =0$ is trivially satisfied because at least six of the $2 \times 2$ minors vanish, c.f. \eref{example_4_x_4_minor}. 
\end{itemize}
The first category corresponds to boundaries of both the amplitude and its log, but which are not present in the square of the positroid stratification of $G_{+}(2,4)$. It is given by the first column on the left of \tref{tab:4particle2loopGLUING}. The second category consists of boundaries of the amplitude, its log, and the square of the positroid stratification of $G_{+}(2,4)$. The corresponding number of boundaries is simply the difference of the two columns in this table. Note that the first category also represents the difference between the last two columns of Table \ref{tab:4particle2loop}, and for this reason we have denoted it $\mathfrak{N}_M-\mathcal{N}_M$.

\begin{table}
\begin{center}
\bigskip
{\small
\begin{tabular}{|c|c|c|}
\hline
\textbf{Dim} & $\boldsymbol{\mathfrak{N}_M}-\boldsymbol{\mathcal{N}_M}$ &    $\boldsymbol{\mathfrak{N}_{M,\Delta^{(1,2)} =0}}$ \\
\hline
\textbf{7} & 1  & 1 \\
\hline
\textbf{6} & 8  & 8 \\
\hline
\textbf{5} & 36  & 36 \\
\hline
\textbf{4} & 96 &  104 \\
\hline
\textbf{3} & 106  & 162 \\
\hline
\textbf{2} & 48  & 164 \\
\hline
\textbf{1} & 8  & 104 \\
\hline
\textbf{0} & 0  & 30 \\
\hline
\end{tabular}
}
\qquad \qquad 
{\small
\begin{tabular}{|c|c|}
\hline
\textbf{Dim} & $(\boldsymbol{\mathfrak{N}_M}-\boldsymbol{\mathcal{N}_M})^{\boldsymbol{(+1)}}$   \\
\hline
\textbf{8} & 1   \\
\hline
\textbf{7} & 8   \\
\hline
\textbf{6} & 36   \\
\hline
\textbf{5} & 96  \\
\hline
\textbf{4} & 106   \\
\hline
\textbf{3} & 48   \\
\hline
\textbf{2} & 8   \\
\hline
\textbf{1} & 0   \\
\hline
\textbf{0} & 0   \\
\hline
\end{tabular}
}

\caption{
On the left: number of boundaries $\boldsymbol{\mathfrak{N}_{M,\Delta^{(1,2)} =0}}$ for the space with $\Delta^{(1,2)}_{1234} =0$ in the $n=4$, 2-loop case. The first column $\boldsymbol{\mathfrak{N}_M}-\boldsymbol{\mathcal{N}_M}$ lists those boundaries where the condition $\Delta^{(1,2)}_{1234} =0$  imposes a non-trivial constraint among the $2 \times 2$ minors. On the right: the list of boundaries $\boldsymbol{\mathfrak{N}_M}-\boldsymbol{\mathcal{N}_M}$ considered as of one dimension larger, following the explanation in the text.\label{tab:4particle2loopGLUING}
}
\end{center}
\end{table}

Let us investigate the interplay among the boundaries of the two components and the gluing region. One should be particularly careful in not double counting boundaries which are present in both the amplitude and its log. Moreover, there are boundaries of the gluing subspace which are not boundaries of the square of the positroid stratification of $G_{+}(2,4)$. \tref{tab:4x4cases} presents a useful classification of the boundaries of all the objects under consideration based on the properties of the $4\times 4$ minor.

The last row in \tref{tab:4x4cases} corresponds to the $(\mathfrak{N}_M-\mathcal{N}_M)$ boundaries of \tref{tab:4particle2loopGLUING}. The first row in the table specifies those boundaries for which $\Delta^{(1,2)}_{1234}$ contains both positive and negative terms but it is not set to zero. Starting from such configurations, $\Delta^{(1,2)}_{1234}$ can be turned off non-trivially, reducing the dimension by one and producing the boundaries in the last row of \tref{tab:4x4cases}. We thus conclude that the list of the boundaries in the first row is also equal to $(\mathfrak{N}_M-\mathcal{N}_M)$, but where the dimensions of the boundaries is increased by 1. We denote this set $(\mathfrak{N}_M-\mathcal{N}_M)^{(+1)}$ and show it on the right of \tref{tab:4particle2loopGLUING}.

\begin{table}
\begin{center}
\bigskip
{\small
\begin{tabular}{|c|c|c|c|c|}
\hline
 \bf{$\Delta^{(1,2)}_{1234}$ property }& \bf{Square of $G_{+}(2,4)$} & \bf{Amplitude}  & \bf{Log} & \bf{Gluing space} \\
&$ \boldsymbol{\mathbb{N}}$ &$\boldsymbol{\mathfrak{N}_M}$ &$\boldsymbol{\mathfrak{N}_{M,\text{Log}}}$& 
$\boldsymbol{\mathfrak{N}_{M,\Delta^{(1,2)} =0}}$ \\
\hline
$\neq 0$, $(+)$ and $(-)$ terms & $\times$ & $\times$  & $\times$  &\\
\hline
$> 0$, only $(+)$ terms & $\times$ & $\times$  &   &\\
\hline
$< 0$, only $(-)$ terms & $\times$ &   & $\times$ &\\
\hline
$=0$ trivially & $\times$ & $\times$  & $\times$  & $\times$ \\
\hline
$=0$ non-trivially &  & $\times$  & $\times$  & $\times$ \\
\hline
\end{tabular}
}
\bigskip
\caption{
Boundaries of the different geometries, classified in terms of the properties of $\Delta^{(1,2)}_{1234}$: whether it is vanishing (trivially or not once vanishing \pl coordinates have been fixed), and if it contains positive negative or both types of \pl coordinates, c.f. \eref{example_4_x_4_minor}.
\label{tab:4x4cases}
}
\end{center}
\end{table}

Given the structure shown in \tref{tab:4x4cases}, the relation between the number of boundaries {\it at each dimension} is
\be
\label{eq:boundary_relation}
\mathbb{N}=\mathfrak{N}_M+\mathfrak{N}_{M,\text{Log}}-\mathfrak{N}_{M,\Delta^{(1,2)} =0}-(\mathfrak{N}_M-\mathcal{N}_M)
-(\mathfrak{N}_M-\mathcal{N}_M)^{(+1)} \, .
\ee
The validity of this equation can be explicitly checked using Tables \ref{tab:4particle2loop}, \ref{tab:4particle2loopLOG} and \ref{tab:4particle2loopGLUING}. For instance, at dimension $4$ we have $274+240-104-96-106= 208$. The relation extends to the Euler numbers of the different objects. $\mathcal{E}=2$ for $\mathfrak{N}_{M}$ and $\mathfrak{N}_{M,\text{Log}}$, the Euler numbers of $(\mathfrak{N}_M-\mathcal{N}_M)$ and $(\mathfrak{N}_M-\mathcal{N}_M)^{(+1)}$ are opposite by construction and cancel in \eref{eq:boundary_relation}, while $\mathcal{E}=3$ for $\mathfrak{N}_{M,\Delta^{(1,2)}=0}$. The combination of all these pieces beautifully produces the $\mathcal{E}=1$ for the square of the positroid stratification of $G_+(2,4)$.

\bigskip

\subsection{Full Stratification}

\label{section_full_stratification_L=2}

Let us now consider the full stratification of $G_+(0,4;2)$. As explained in \sref{sec:fullstrat}, the full stratification refines the mini stratification by distinguishing the different regions satisfying each positivity condition. In the $G_+(0,4;2)$ case, the positivity condition being satisfied in different regions is the extended positivity of the $4 \times 4$ minor $\Delta^{(1,2)}_{1234}$, and the domains are characterized by additional inequalities imposed on (combinations of) $2\times2$ \pl coordinates. In this way, each boundary is specified by a list of minors, and by a set of inequalities for the $2\times 2$ minors.

The refinement to obtain the full stratification changes the mini stratification in two ways:
\begin{itemize}
\item The boundaries in $\Gamma_0$ are now distinguished by the set of vanishing \pl coordinates and the region. For every set of vanishing \pl coordinates, the minor $\Delta^{(1,2)}_{1234}$ may or may not be trivially zero; if it is not, the separate regions are generated by the condition $\Delta^{(1,2)}_{1234}>0$ which can be satisfied on disjoint regions of the $\Delta^{(i)}_I$ parameter space. If instead $\Delta^{(1,2)}_{1234}=0$ trivially, there may still be multiple regions: they descend from higher-dimensional configurations where the $4 \times 4$ minor is different from zero and splits into separate regions. Of course, it is also possible that $\Delta^{(1,2)}_{1234}=0$ trivially and we only have a single region. We will illustrate explicit examples of each of these phenomena in the examples below.
\item The structure of the $\Gamma_1$, which is obtained by setting $\Delta^{(1,2)}_{1234}=0$ non-trivially when it is possible to do so, changes in general. The new $\Gamma_1$ takes into account the explicit form of the regions in $\Gamma_0$. 
\end{itemize}
For convenience we again reproduce the expression for the only $4 \times 4$ minor present at 2-loops, expressed in terms of the $2 \times 2$ \pl coordinates:
\be \label{remind4x4}
\Delta^{(1,2)}_{1234}=\Delta^{(1)}_{12} \Delta^{(2)}_{34}+\Delta^{(1)}_{34}\Delta^{(2)}_{12}+\Delta^{(1)}_{23}\Delta^{(2)}_{14}+\Delta^{(1)}_{14}\Delta^{(2)}_{23}-\Delta^{(1)}_{13}\Delta^{(2)}_{24}-\Delta^{(1)}_{24}\Delta^{(2)}_{13} \, .
\ee
By using the \pl relations this may be turned into the convenient form
\bea
\label{factors}
\Delta^{(1,2)}_{1234}&=&\frac{1}{\Delta^{(1)}_{24} \Delta^{(2)}_{24}} \Big[ (\Delta_{23}^{(1)} \Delta^{(2)}_{24}-\Delta_{24}^{(1)} \Delta^{(2)}_{23}) (\Delta^{(2)}_{14} \Delta_{24}^{(1)}-\Delta^{(2)}_{24}\Delta_{14}^{(1)})  +\nonumber \\
&& \qquad \qquad \quad
(\Delta_{12}^{(1)} \Delta^{(2)}_{24}-\Delta_{24}^{(1)}\Delta^{(2)}_{12})(\Delta^{(2)}_{34} \Delta_{24}^{(1)}-\Delta^{(2)}_{24}\Delta_{34}^{(1)}) \Big] \, .
\eea
An equivalent expression exists where all $\{ 24 \}$ indices are replaced by $\{ 13 \}$ indices; this simply amounts to solving for the \pl relations in terms of $\Delta^{(i)}_{13}$ instead of $\Delta^{(i)}_{24}$. To avoid ambiguity, when the \pl relations are non-trivial we shall always explicitly solve for them, and plug the answer into $\Delta^{(1,2)}_{1234}$, in a form similar to \eref{factors}.

The inequalities that characterize the full stratification only involve the factors in the expression for $\Delta^{(1,2)}_{1234}$ shown in \eref{factors}. Explicitly, the inequalities specifying the regions can only be one or more of the following:
\bea
\label{inequalities}
&&
(\Delta_{23}^{(1)} \Delta^{(2)}_{24}-\Delta_{24}^{(1)} \Delta^{(2)}_{23}) \gtrless 0 \qquad , \qquad
(\Delta^{(2)}_{14} \Delta_{24}^{(1)}-\Delta^{(2)}_{24}\Delta_{14}^{(1)}) \gtrless 0  \nonumber \\
&&
(\Delta_{12}^{(1)} \Delta^{(2)}_{24}-\Delta_{24}^{(1)}\Delta^{(2)}_{12}) \gtrless 0 \qquad , \qquad
(\Delta^{(2)}_{34} \Delta_{24}^{(1)}-\Delta^{(2)}_{24}\Delta_{34}^{(1)})  \gtrless 0
\eea
or their equivalent counterparts where $\Delta^{(i)}_{24}$ is replaced by $\Delta^{(i)}_{13}$. The choice of whether we need to consider the expressions with $\Delta^{(i)}_{13}$ or $\Delta^{(i)}_{24}$ is determined by which ones are equal to zero: if any $\Delta^{(i)}_{13}=0$ we need to use the expression with $\Delta^{(i)}_{24}$'s, and vice-versa. If both $\Delta^{(i)}_{13}=\Delta^{(j)}_{24}=0$ are zero (where $i=1,2$ and $j=1,2$), there are no non-trivial inequalities which we may consider. When there are no non-trivial inequalities, we only have a single region for the label in question.

Given a set of vanishing \pl coordinates, the full list of cases for which there cannot be any non-trivial inequalities is the following:
\begin{itemize}
\item Configurations where the expression \eref{remind4x4} for $\Delta^{(1,2)}_{1234}$ only has positive terms.
\item Configurations where $\Delta^{(i)}_{13}=\Delta^{(j)}_{24}=0$, where $i$ and $j$ are individually free to be 1 or 2.
\item Configurations where the following combination of $2 \times 2$ minors is vanishing: $\Delta^{(i)}_{12}=\Delta^{(j)}_{14}=\Delta^{(k)}_{23}=\Delta^{(l)}_{34}=0$, where $i,j,k,l$ are individually free to be 1 or 2. These configurations ruin all 4 inequalities in \eref{inequalities}.
\end{itemize}
For these cases, the construction of $\Gamma_1$ is identical to that of the mini stratification.

For the remaining cases we now identify eight prototypical configurations, which exhaust all possibilities which may arise at 2-loops. In each separate case, we specify the $\Gamma_1$ structure, and in this way construct the full stratification. We indicate with $(\ldots)$ the factors in the $4 \times 4$ determinant which are ``non-trivial'', e.g. $(\Delta_{23}^{(1)} \Delta^{(2)}_{24}-\Delta_{24}^{(1)} \Delta^{(2)}_{23})$, and which may thus define a region through the inequalities \eref{inequalities}.  We indicate with $k_i$ a positive quantity made up of a product of $2 \times 2$ \pl coordinates, e.g. $k=(\Delta_{24}^{(1)} \Delta^{(2)}_{24})$. 

The eight possible configurations are the following:
\begin{enumerate}
\item Cases where the $4 \times 4$ is different from zero and has the expression\footnote{For notational convenience we suppress the subindex of the $4 \times 4$ minor, since for four particles it can only be $\{1234\}$.}
$$
\Delta^{(1,2)}= \frac{1}{k} \Big[ \left( \ldots \right)\left( \ldots \right)  +\left( \ldots \right) \left( \ldots \right) \Big] \, . 
$$
At 2-loops there is in fact only one such case in $\Gamma_0$, which is the 8-dimensional element. Here $\Delta^{(1,2)}>0$ specifies a single region, with a single boundary at $\Delta^{(1,2)}=0$. Thus, $\Gamma_1$ only gives rise to one additional boundary of dimension 7, precisely as in the mini stratification.
\item Cases where the $4 \times 4$ is different from zero and has the expression
$$
\Delta^{(1,2)}= \frac{1}{k_1} \Big[  \left( \ldots \right)\left( \ldots \right)  +k_2 \left( \ldots \right) \Big] \, .
$$
All 7-dimensional elements in $\Gamma_0$ are of this type, e.g.\ the configuration with $\Delta_{23}^{(1)}=0$. $\Delta^{(1,2)}>0$ specifies a single region, with a single 6-dimensional boundary at $\Delta^{(1,2)}=0$, similarly to the case above.
\item Cases where the $4 \times 4$ is different from zero and has the expression
$$
\Delta^{(1,2)}= \frac{1}{k_1} \Big[  \left( \ldots \right)\left( \ldots \right)  -k_2 \big] \, .
$$
Here $\Delta^{(1,2)}>0$ is divided into two regions, each bounded by a hyperbolic curve, as explained in \sref{sec:fullstrat}. The regions are specified by the parentheses being both positive or both negative. The condition $\Delta^{(1,2)}=0$ gives rise to two boundaries of one dimension less, because we can solve $\Delta^{(1,2)}=0$ on these two different regions, each region being one of the two hyperbolic curves. An example of this type is $\Delta_{23}^{(1)}=\Delta_{14}^{(1)}=0$.
\item Cases where the $4 \times 4$ is different from zero and has the expression
$$
\Delta^{(1,2)}= \frac{1}{k_1} \Big[  \left( \ldots \right)\left( \ldots \right)  + k_2 \big] \, .
$$
This is a single connected region, bounded by two hyperbolic curves. Hence, the condition $\Delta^{(1,2)}=0$ gives rise to two extra boundaries of one dimension less. As an example for this category, consider the case
$$
\Delta_{12}^{(1)}=\Delta_{34}^{(2)}=0 \, .
$$
Using \eref{factors}, the region with $\Delta^{(1,2)}>0$ is defined by the inequality
$$
(\Delta_{23}^{(1)} \Delta^{(2)}_{24}-\Delta_{24}^{(1)} \Delta^{(2)}_{23}) 
(\Delta^{(2)}_{14} \Delta_{24}^{(1)}-\Delta^{(2)}_{24}\Delta_{14}^{(1)}) >
-(\Delta_{24}^{(1)}\Delta^{(2)}_{12})(\Delta^{(2)}_{24}\Delta_{34}^{(1)})  \, .
$$
Parameterizing $x=(\Delta_{23}^{(1)} \Delta^{(2)}_{24}-\Delta_{24}^{(1)} \Delta^{(2)}_{23}) $, $y=(\Delta^{(2)}_{14} \Delta_{24}^{(1)}-\Delta^{(2)}_{24}\Delta_{14}^{(1)})$ and $k=(\Delta_{24}^{(1)}\Delta^{(2)}_{12})(\Delta^{(2)}_{24}\Delta_{34}^{(1)}) $, this is the connected region in the $xy$ plane inside the hyperbola $xy=-k$. The two extra boundaries of one dimension less are the two branches of the hyperbola.
\item Cases where the $4 \times 4$ is different from zero and factorizes as
$$
\Delta^{(1,2)}=\frac{1}{k} \Big[  \left( \ldots \right)\left( \ldots \right)  \Big] \, .
$$
This type of configuration is a bit more subtle, as it is the limit of the hyperbolic cases above where the two branches of the hyperbola meet at the origin. Parametrizing the first $\left( \ldots \right)$ as $x$ and the second one as $y$, the $\Delta^{(1,2)}>0$ condition is satisfied in the first and third quadrant of the $xy$ plane, thus giving rise to two regions. Here there are four boundaries of one dimension less, where $\Delta^{(1,2)}=0$, corresponding to the positive and negative $x$ and $y$ axes. The origin corresponds to a single boundary of two dimensions less. These boundaries may be seen as setting $x=0$ while remembering that $y \neq 0$ was composed of two separate regions, or setting $y=0$ and $x \neq 0$, and finally setting $x=y=0$. An example for this category is
$$
\Delta_{12}^{(1)}=\Delta_{12}^{(2)}=0 \, .
$$
\item Cases where the $4 \times 4$ is different from zero and does not contain parentheses $\left( \ldots \right)$ that are multiplied together, i.e.\
$$
\Delta^{(1,2)}= \frac{1}{k_1}\Big[ \left( \ldots \right)k_2 +  \left( \ldots \right) k_3 \Big] ~ \text{or} ~  \Delta^{(1,2)}= \frac{1}{k_1}\Big[ \left( \ldots \right)k_2 \pm k_3 \Big] ~ \text{or} ~ \Delta^{(1,2)}= \left( \ldots \right) k \, .
$$
Each of these cases consist of a single region and the condition $\Delta^{(1,2)}=0$ gives rise to a single boundary of one dimension less. This can most clearly be seen by studying the $xy$ plane as done above. An example of this category is 
$$
\Delta^{(1)}_{12}=\Delta^{(2)}_{23}=0 \, .
$$
\item Cases where the $4 \times 4$ trivially vanishes but two of the four inequalities in \eref{inequalities} (or their $\{13\} \leftrightarrow \{24\}$ counterparts) remain untouched. This is most transparently written as
$$
\Delta^{(1,2)}= \frac{1}{k} \Big[ 0 \times \left( \ldots \right)+ 0 \times \left( \ldots \right) \Big] \, .
$$
These cases are the most subtle of all. Although the $4\times 4$ minor vanishes, we still have four separate regions, specified by the two possible inequalities which are still present in each $\left( \ldots \right)$. To see why this is the case, we need to know how these configurations arose from higher dimensional ones: here the path taken to reach this configuration will specify the region.

To this end, let us denote the first bracket as $x$ and the second one as $y$. A detailed investigation shows that all these cases arise from Type 5 cases described above, where additionally one of the brackets is trivially shut off by turning off some $\Delta^{(i)}_I$'s. Here, the remaining bracket is still split into two regions, while the brackets that do not appear in Type 5 are completely free. 

Thus, the only possibilities are as follows: either $x$ is split into two regions while $y$ is free, or $y$ is split into two regions while $x$ is free. In total we then have four regions. 

From these four regions descend two extra boundaries of one dimension less: either $x=0$ and $y$ is free, or $y=0$ and $x$ is free. From here there are no further boundaries, as we may not set a free variable to zero.

An example for this category is given by the following set of vanishing \pl coordinates
$$
\Delta_{12}^{(1)}=\Delta_{23}^{(1)}=\Delta_{13}^{(1)}=\Delta_{12}^{(2)}=\Delta_{23}^{(2)}=\Delta_{13}^{(2)}=0 \, .
$$
Here the four 4-dimensional regions are
\bea
&&
\text{Regions 1 and 2:} \qquad (\Delta^{(2)}_{14} \Delta_{24}^{(1)}-\Delta^{(2)}_{24}\Delta_{14}^{(1)}) \gtrless 0 \nonumber \\
&&
\text{Regions 3 and 4:} \qquad (\Delta^{(2)}_{34} \Delta_{24}^{(1)}-\Delta^{(2)}_{24}\Delta_{34}^{(1)})  \gtrless 0 \nonumber
\eea
while the two extra lower dimensional boundaries of dimension 3 are characterized by the conditions
\bea
&&
\text{Region A:} \qquad (\Delta^{(2)}_{14} \Delta_{24}^{(1)}-\Delta^{(2)}_{24}\Delta_{14}^{(1)})=0 \nonumber \\
&&
\text{Region B:}  \qquad (\Delta^{(2)}_{34} \Delta_{24}^{(1)}-\Delta^{(2)}_{24}\Delta_{34}^{(1)}) =0\nonumber
\eea
\item 
Cases where the $4 \times 4$ trivially vanishes but one of the four inequalities in \eref{inequalities} can be imposed. These are most transparently written as
$$
\Delta^{(1,2)}=\frac{1}{k_1} \Big[ 0 \times \left( \ldots \right)+ k_2 \times 0 \Big] \qquad \text{or}  \qquad \Delta^{(1,2)}= \frac{1}{k_1} \Big[ 0 \times \left( \ldots \right) \Big]
$$
and can be obtained from the Type 7, above. These cases consist of two regions, determined by the sign of the non-vanishing parenthesis. They give rise to one extra boundary of one dimension less, when we saturate the inequality.
\end{enumerate}

The results of the full stratification are summarized in \tref{tab:4n2Lsolutions}. To give an example of how these numbers are obtained, let us discuss in detail the 6-dimensional boundaries of $\mathfrak{N}_{F}$. At dimension 6, there are four possible sets of vanishing \pl coordinates which are cases of Type 3, four cases of Type 4, four cases of Type 5, and 24 cases of Type 6. On top of that, there are other 8 boundaries descending from eight 7-dimensional configurations of Type 2, where we have imposed $\Delta^{(1,2)}=0$. In total this gives the entry at dimension 6 in \tref{tab:4n2Lsolutions}, i.e. $4 \times 2+4+ 4 \times 2+24+8=52$.

We can then adopt the same strategy to obtain the full stratification of the log of the amplitude; the only difference is that we have to impose $\Delta^{(1,2)} \leq 0$ to identify the different regions. This modification takes a very simple form on the classification described here: we only need to interchange Types 3 and 4. \tref{tab:4n2Lsolutions} also shows the results for the log of the amplitude, as well as the gluing region defined by $\Delta^{(1,2)}=0$, which is obtained in a very similar way.

We note that for the full stratification, the relation \eref{eq:boundary_relation} which connects the amplitude, the log and the gluing region is no longer valid. 

\begin{table}
\begin{center}
\bigskip
{\small
\begin{tabular}{|c|c|c|c|}
\hline
\textbf{Dim} &  $\boldsymbol{\mathfrak{N}_{F}}$ & $\boldsymbol{\mathfrak{N}_{F,\textbf{Log}}}$  &  $\boldsymbol{\mathfrak{N}_{F,\Delta^{(1,2)}=0}}$ \\
\hline
\textbf{8} & 1 & 1 & 0 \\
\hline
\textbf{7} & 9 & 9 & 1 \\
\hline
\textbf{6} & 52 & 52 & 8 \\
\hline
\textbf{5} & 168 & 160 & 56 \\
\hline
\textbf{4} & 328 & 294 & 156 \\
\hline
\textbf{3} & 392 & 336 & 224 \\
\hline
\textbf{2} & 306 & 262 & 206 \\
\hline
\textbf{1} & 144 & 128 & 112 \\
\hline
\textbf{0} & 34 & 32 & 30 \\
\hline
\end{tabular}
}
\bigskip
\caption{Full stratification of the $n=4$, 2-loop amplituhedron. $\mathfrak{N}_{F}$ gives the number of boundaries for the amplitude.
$\mathfrak{N}_{F,\text{Log}}$ gives the number of boundaries for the log of the amplitude, and $\mathfrak{N}_{F,\Delta^{(1,2)}=0}$ describes the full stratification
of the gluing space.  \label{tab:4n2Lsolutions}}
\end{center}
\end{table}

The Euler numbers for the full stratification of  the different spaces can be easily computed to be:
\begin{itemize}
\item $\mathfrak{N}_{F}$: $\mathcal{E} = 8$,
\item $\mathfrak{N}_{F,\text{Log}}$: $\mathcal{E} = 8$,
\item $\mathfrak{N}_{F,\Delta^{(1,2)}=0}$: $\mathcal{E} = 7$
\end{itemize}
Interestingly, the Euler number of the amplitude and of the log of the amplitude coincide; the reason for this is that there is an equal number of cases of Types $3$ and $4$.

\bigskip

\section{Three loops}

\label{section_example_n4_L3}

In this section we initiate the investigation of $L=3$, for which we construct the mini stratification. Our results should be valuable for any future study of this geometry.

\bigskip

\subsection{Mini Stratification}

\label{section_mini_stratification_L=3}

The matrix $\mathcal{C}$ takes the form
\begin{equation}
\mathcal{C} = \begin{pmatrix} D_{(1)} \\ D_{(2)} \\ D_{(3)} \end{pmatrix}.
\end{equation} 
Its largest minors are $4\times 4$ and we have three of them. $\mathcal{C}$ has $3 \times 4 = 12$ degrees of freedom.

Taking three identical copies of the graph in \fref{fig:G24decompgraphpms} and doing the decomposition followed by identification as done in \sref{sec:pmGivedecomposition}, we obtain the left-hand column of \tref{tab:4particle3loop}. This is the same as taking the 3$^{rd}$ power of the 1-loop stratification, which could be pictorially illustrated by replacing each of the $1\,089$ sites in \fref{fig:4particle2loop} with the decomposition given in \fref{fig:G24decompgraphpms}, representing the fact that for each of the $1\,089$ sites there is a full decomposition of the third graph. In total we get $33^3= 35\, 937$ different boundaries. At this stage extended positivity has not yet been fully implemented; we have only performed step (1) in \sref{sec:methodsummary}. Again, we note in agreement with the general discussion in \sref{section_hyper_pms}, we obtain an Eulerian poset:
\begin{equation}
\sum_{i=0}^{12} (-1)^i \mathbb{N}^{(i)} = 216-1296+\ldots -12+1=1 \, .
\end{equation}

\begin{table}
\begin{center}
\bigskip 
{\small
\begin{tabular}{|c|c|c|c|c|}
\hline
\textbf{Dim} & $\boldsymbol{\mathbb{N}}$ &  $\boldsymbol{\mathcal{N}_M}$ & $\boldsymbol{\mathfrak{N}_{M}}$ \\
\hline
\textbf{12} & 1 & 1 & 1 \\
\hline
\textbf{11} & 12 & 12 & 15 \\
\hline
\textbf{10} & 78 & 78 & 117 \\
\hline
\textbf{9} & 340 & 340 & 611 \\
\hline
\textbf{8} & 1\,086 & 1\,002 & 2\,244 \\
\hline
\textbf{7} & 2\,640 & 2\,160 & 5\,908 \\
\hline
\textbf{6} & 4\,960 & 3\,490 & 10\,996 \\
\hline
\textbf{5} & 7\,200 & 4\,440 & 13\,956 \\
\hline
\textbf{4} & 7\,956 & 4\,656 & 12\,044 \\
\hline
\textbf{3} & 6\,480 & 3\,960 & 7\,488 \\
\hline
\textbf{2} & 3\,672 & 2\,520 & 3\,504 \\
\hline
\textbf{1} & 1\,296 & 1\,008 & 1\,128 \\
\hline
\textbf{0} & 216 & 186 & 186 \\
\hline
\end{tabular}
}
\bigskip
\caption{Number of boundaries $\boldsymbol{\mathfrak{N}_{M}}$ of $G_+(0,4;3)$, of various dimensions. $\mathbb{N}$ is the number of boundaries before the extended positivity conditions on the larger minors are implemented, and $\mathcal{N}_M$ is the surviving number of boundaries after these conditions are enforced, but before taking into account the boundaries arising from the $\Delta^{(i,j)}_I \geq 0$.\label{tab:4particle3loop}}
\end{center}
\end{table}

Next, we need to impose three additional conditions from extended positivity: $\Delta^{(1,2)}_I \geq 0$, $\Delta^{(1,3)}_I \geq 0$ and $\Delta^{(2,3)}_I \geq 0$, where $I=1234$ as in the rest of this section. This can be done either by checking them individually or employing the method expounded in \sref{section_combinatorial_stratification}. Deleting the boundaries that violate extended positivity gives the second column in \tref{tab:4particle3loop}. We note that this column does not correspond to an Eulerian poset:
\begin{equation}
\sum_{i=0}^{12} (-1)^i \mathcal{N}_M^{(i)} = 186-1008+\ldots -12+1=13 \, .
\end{equation}

Let us now perform a complete classification of the possible $\Gamma_1$ sub-posets in the mini stratification of $G_+(0,4;3)$, i.e. the new structure arising from turning off $4 \times 4$ minor. Points in $\Gamma_0$ can be discriminated according to the number of $\Delta^{(i,j)}_I$'s with both positive and negative terms, i.e. of type (iii) in the discussion of \sref{sec:methodsummary}. We denote the three possibilities as $N \Delta^{(i,j)}_I$, where $N=1,2,3$.

\begin{figure}[h]
\begin{center}
\includegraphics[width=8cm]{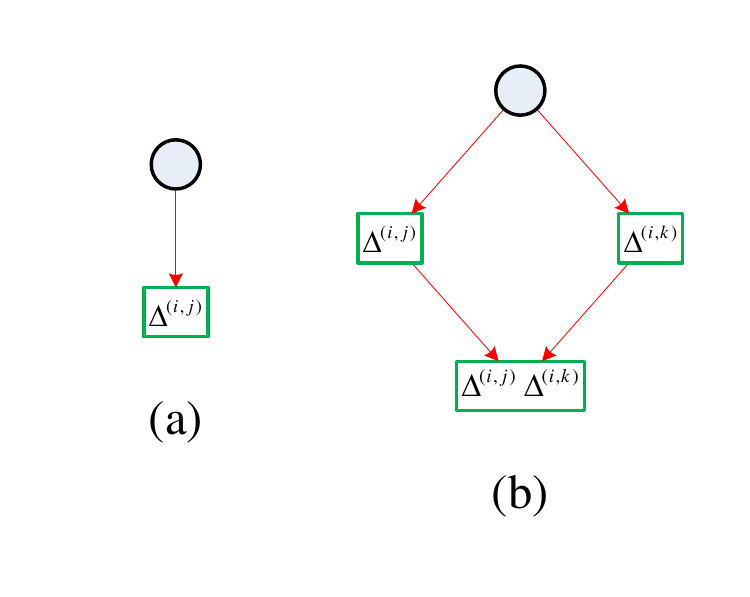}
\vspace{-.6cm}\caption{The general structure of $\Gamma_1$'s emanating from: a) $1\Delta^{(i,j)}_I$ and b) $2\Delta^{(i,j)}_I$ points}
\label{1Delta2Delta}
\end{center}
\end{figure}

\fref{1Delta2Delta} shows the possible $\Gamma_1$'s emanating from $1\Delta^{(i,j)}_I$ and $2\Delta^{(i,j)}_I$ points. This is a result of careful analysis which shows that in both cases, all type (iii) $\Delta^{(i,j)}_I$ can be independently turned off.

The possible structures become far richer for $3\Delta^{(i,j)}_I$ points. In general the determination of $\Gamma_1$'s is challenging, because it requires solving equations in which variables and certain combinations of them are restricted to the positive domain. To illustrate the subtleties involved, let us consider a $3\Delta^{(i,j)}_I$ example, i.e. one in which it naively seems possible that any of the three $4\times 4$ minors can be turned off, but this is not the case once relations are properly taken into account. For example, if we have a relation like

\begin{equation}
\Delta^{(1,3)}_I = a \, \Delta^{(1,2)}_I - b \, \Delta^{(2,3)}_I  \, , \quad a,b>0 \, ,
\end{equation}

\begin{figure}[h]
\begin{center}
\includegraphics[width=14cm]{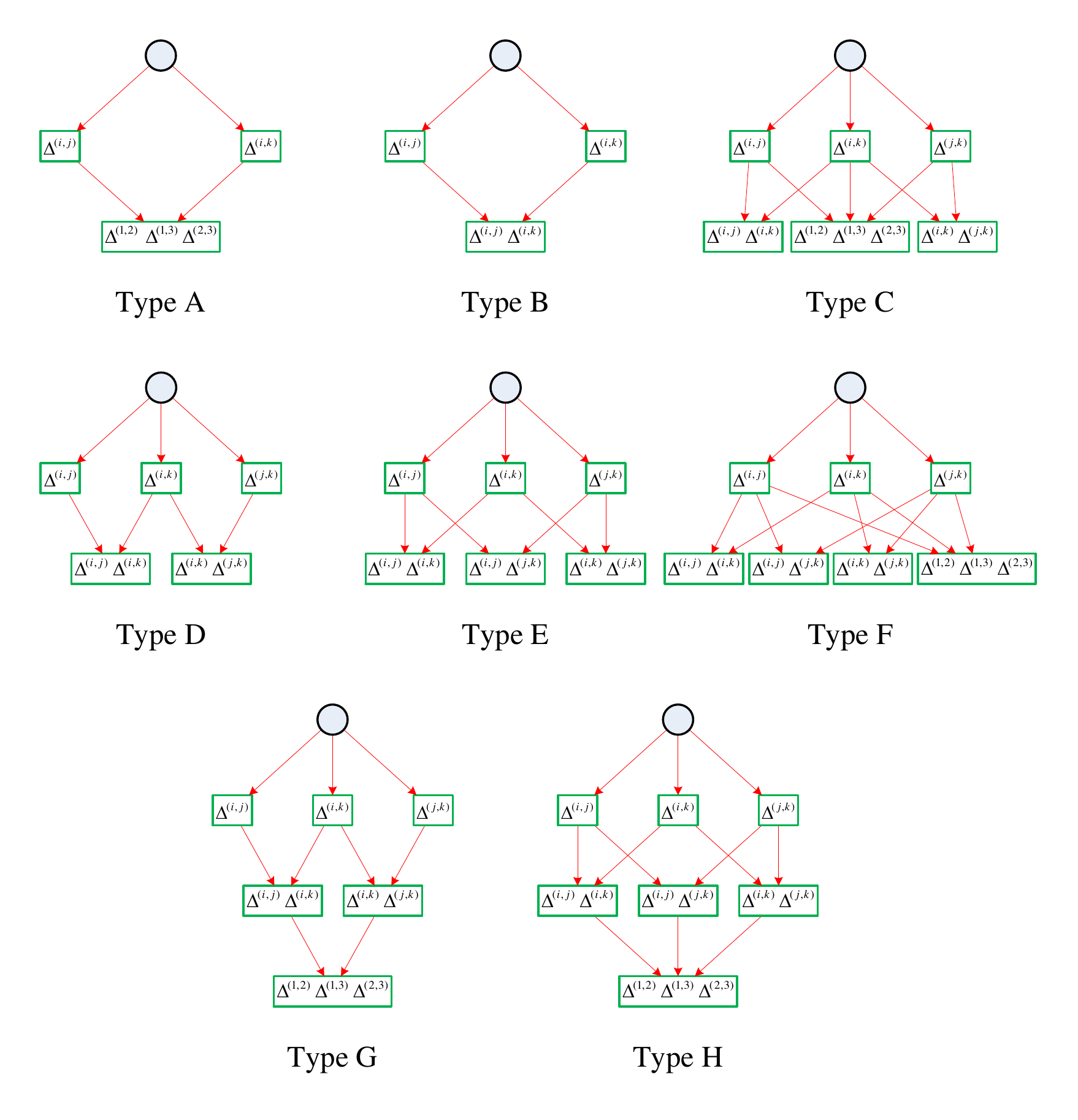}
\caption{Full classification of possible $\Gamma_1$'s emanating from $3\Delta^{(i,j)}_I$ points in $\Gamma_0$ in the mini stratification of $G_+(0,4;3)$. In each green box we indicate which $4 \times 4$ minors have been set to zero. Interestingly, for Type A it is not possible to turn off only two of them due to positivity. Furthermore, for types B, D and E it is also impossible to turn off the three $4 \times 4$ minors.}
\label{fig:3Ltrees}
\end{center}
\end{figure}

\medskip

\noindent we see that it is not possible to turn off $\Delta^{(1,2)}_I$ while keeping both $\Delta^{(1,3)}_I$ and $\Delta^{(2,3)}_I$ positive. In this expression, $a$ and $b$ are functions of non-vanishing \pl coordinates. We also see that it is not possible to turn off only two of the three $\Delta^{(i,j)}_I$. From any boundary that has a reduced set of \pl coordinates from the maximum possible, such that the larger minors $\Delta^{(i,j)}_I$ satisfy the relation above, we expect a $\Gamma_1$ as in \fref{fig:3Ltrees} Type A.  

\bigskip
\bigskip

Other structures in \fref{fig:3Ltrees} result from relations of the following general forms
\beq
{\renewcommand{\arraystretch}{1.3}
\begin{array}{llcrcl}
\text{Type B:} \quad & \Delta^{(i,j)}_I = a \, \Delta^{(j,k)}_I - b \, \Delta^{(i,k)}_I - c \; , & \quad & a,b,c & >& 0 \\ 
\text{Type C:} \quad & \Delta^{(i,k)}_I = k \left( a \, \Delta^{(i,j)}_I - b \, \Delta^{(j,k)}_I \right) \; , & \quad & a,b & > & 0 ,\quad k \text{ free}\\ 
\text{Type D:} \quad & \Delta^{(i,k)}_I = k \left( a \, \Delta^{(i,j)}_I - b \, \Delta^{(j,k)}_I \right) - c\; , & \quad \quad  & a,b,c & > & 0 ,\quad k \text{ free}
\end{array}}
\label{obstruction_types}
\eeq
and so on. Here $a$, $b$, $c$ and $k$ represent functions of non-vanishing \pl coordinates. For Type H structures, all the $\Delta^{(i,j)}_I$'s may be turned off completely independently. In \sref{section_deformed_amplituhedron} we consider an explicit example of these relations and discuss it in more detail.

\fref{fig:3Ltrees} provides a comprehensive treatment of $3\Delta^{(i,j)}_I$ boundaries. We stress that all the boundaries in a given $\Gamma_1$ have the same set of non-vanishing \pl coordinates; different sites only differ by $\Delta^{(i,j)}$'s that have been set to zero.

\tref{tab:3LtreeContributions} shows the number of boundaries of each dimension with the structures in \fref{fig:3Ltrees}, and the added contribution to the total number of boundaries. This contribution must be added to those boundaries in column $\mathcal{N}_M$ of \tref{tab:4particle3loop}, to yield the total $\mathfrak{N}_{M}$, also quoted in \tref{tab:4particle3loop}. This procedure implements step (4) in \sref{sec:methodsummary}.

\begin{table}
\begin{center}
\bigskip
{\small
\begin{tabular}{c|c|c|c|c|c|c|c|c|c|c|c|}
\cline{2-12}
 & \multicolumn{8}{c|}{\textbf{3} $\boldsymbol{\Delta_I^{(i,j)}}$} & \multirow{2}{*}{\textbf{2} $\boldsymbol{\Delta_I^{(i,j)}}$} & \multirow{2}{*}{\textbf{1} $\boldsymbol{\Delta_I^{(i,j)}}$} & \textbf{Total} \\
 \cline{1-9}
\multicolumn{1}{|c|}{\textbf{Dim}} & \textbf{A} & \textbf{B} & \textbf{C} & \textbf{D} & \textbf{E} & \textbf{F} & \textbf{G} & \textbf{H} &  & & \textbf{contribution} \\
\hline
\multicolumn{1}{|c|}{\textbf{12}} & 0 & 0 & 0 & 0 & 0 & 0 & 0 & 1 & 0 & 0 & $+0$ \\
\hline 
\multicolumn{1}{|c|}{\textbf{11}} & 0 & 0 & 0 & 0 & 0 & 0 & 0 & 12 & 0 & 0 & $+3$ \\
\hline 
\multicolumn{1}{|c|}{\textbf{10}} & 0 & 0 & 0 & 0 & 0 & 0 & 0 & 78 & 0 & 0 & $+39$ \\
\hline 
\multicolumn{1}{|c|}{\textbf{9}} & 0 & 0 & 0 & 0 & 0 & 4 & 0 & 324 & 0 & 12 & $+271$ \\
\hline 
\multicolumn{1}{|c|}{\textbf{8}} & 0 & 12 & 48 & 0 & 0 & 12 & 0 & 726 & 96 & 108 & $+1\,242$ \\
\hline 
\multicolumn{1}{|c|}{\textbf{7}} & 48 & 96 & 144 & 96 & 48 & 12 & 12 & 600 & 576 & 528 & $+3\,748$ \\
\hline 
\multicolumn{1}{|c|}{\textbf{6}} & 144 & 120 & 144 & 96 & 0 & 2 & 0 & 144 & 1\,080 & 1\,584 & $+7\,506$ \\
\hline 
\multicolumn{1}{|c|}{\textbf{5}} & 144 & 0 & 24 & 0 & 0 & 0 & 0 & 0 & 792 & 2\,424 & $+9\,516$ \\
\hline 
\multicolumn{1}{|c|}{\textbf{4}} & 24 & 0 & 0 & 0 & 0 & 0 & 0 & 0 & 240 & 1\,848 & $+7\,388$ \\
\hline 
\multicolumn{1}{|c|}{\textbf{3}} & 0 & 0 & 0 & 0 & 0 & 0 & 0 & 0 & 24 & 672 & $+3\,528$ \\
\hline 
\multicolumn{1}{|c|}{\textbf{2}} & 0 & 0 & 0 & 0 & 0 & 0 & 0 & 0 & 0 & 96 & $+984$ \\
\hline 
\multicolumn{1}{|c|}{\textbf{1}} & 0 & 0 & 0 & 0 & 0 & 0 & 0 & 0 & 0 & 0 & $+120$ \\
\hline 
\multicolumn{1}{|c|}{\textbf{0}} & 0 & 0 & 0 & 0 & 0 & 0 & 0 & 0 & 0 & 0 & $+0$ \\
\hline 
\end{tabular}
}
\bigskip
\caption{Number of boundaries with $N=1,2,3$ number of $4 \times 4$ minors which have both positive and negative terms, and may hence be set to zero non-trivially. The cases with 3 $\Delta_I^{(i,j)}$ are refined according to which type they are, c.f.\ \fref{fig:3Ltrees}. The final column contains the added contribution to the total number of boundaries.\label{tab:3LtreeContributions}}
\end{center}
\end{table}

We can use these results to compute an Euler number, which is
\begin{equation}
\mathcal{E}=\sum_{i=0}^{12} (-1)^i \mathfrak{N}_M^{(i)} = 186-1128+\ldots -15+1=-14 \, .
\end{equation}
This, however, should only be interpreted as a possible characterization of the space based on the mini stratification. It should not be assigned much geometric significance beyond this. In fact, as we have seen for $L=2$, the value of $\mathcal{E}$ associated to the full stratification will most likely be different.

\bigskip

\section{An Alternative Path to Stratification: Integrand Poles}

\label{section_integrand_stratification}

The amplituhedron was introduced as a geometric object whose properties replicate those of the amplitude integrand. In particular, boundaries of the amplituhedron directly correspond to singularities of the integrand. The same holds for the log of the amplitude. This implies that the corresponding integrands provide an alternative way of obtaining the stratification of these spaces. 

In this section we will focus on $n=4$ and $L=2$ and discuss how the stratification of the amplitude and its log can be derived from the corresponding integrands. In particular, we will manage to obtain the entire mini stratifications for the two objects. The full agreement with the ones attained via the amplituhedron constitutes substantial non-trivial evidence for the amplituhedron conjecture. It should be straightforward to extend our analysis to the full stratification. It may be possible that agreement at the level of the mini stratifications implies agreement of the full stratifications. While very interesting, investigating this claim is beyond the scope of this article. 

We stress that looking for poles of the integrand is a substantially different approach to the one adopted in previous sections involving minors and positivity, and it is very satisfactory to see that the two methods agree beautifully. From the integrand perspective, positivity is not an ingredient that is introduced by hand; the integrand accounts for positivity in an automatic way, and positivity emerges as a result.

\bigskip

\subsection{The Amplitude}

\label{section_integrand_stratification_amplitude}

For the amplitude, the integrand in question is
\begin{equation} \label{eq:topdimIntegrand}
\frac{\langle AB34 \rangle \langle CD12 \rangle+\langle AB23 \rangle \langle CD14 \rangle+\langle AB14 \rangle \langle CD23 \rangle+\langle AB12 \rangle \langle CD34 \rangle}{\langle ABCD \rangle \langle AB12 \rangle \langle AB14 \rangle \langle AB23 \rangle \langle AB34 \rangle \langle CD12 \rangle \langle CD14 \rangle \langle CD23 \rangle \langle CD34 \rangle} \, .
\end{equation}
The stratification results from looking for poles of this integrand.

We have seen in previous sections that positivity eliminates many of the potential boundaries which one might naively expect from just taking square of the positroid stratification of $G_+(2,4)$. The integrand achieves this through the presence of a nontrivial numerator, which for certain would-be boundaries cancels with factors in the denominator, to eliminate those poles which would violate positivity. Conversely, positivity eliminates configurations for which the integrand is non-singular.

It is useful to highlight that for $n=4$ at arbitrary $L$ there is a very simple map between brackets and minors, as shown in \cite{Arkani-Hamed:2013kca}. For $L=2$ it is
\begin{align} \label{eq:bracketpluckermap}
\langle AB12 \rangle = \Delta^{(1)}_{34} \quad \quad \quad \langle AB13 \rangle = \Delta^{(1)}_{24} \quad \quad \quad \langle CD12 \rangle = \Delta^{(2)}_{34} \quad \quad \quad \langle CD13 \rangle = \Delta^{(2)}_{24} \nonumber \\
\langle AB14 \rangle = \Delta^{(1)}_{23} \quad \quad \quad \langle AB23 \rangle = \Delta^{(1)}_{14} \quad \quad \quad \langle CD14 \rangle = \Delta^{(2)}_{23} \quad \quad \quad \langle CD23 \rangle = \Delta^{(2)}_{14} \nonumber \\
\langle AB24 \rangle = \Delta^{(1)}_{13} \quad \quad \quad \langle AB34 \rangle = \Delta^{(1)}_{12} \quad \quad \quad \langle CD24 \rangle = \Delta^{(2)}_{13} \quad \quad \quad \langle CD34 \rangle = \Delta^{(2)}_{12} \nonumber \\
\langle ABCD \rangle = \Delta^{(1,2)}_{1234} \quad \quad \quad \quad \quad \quad \quad \quad \quad \quad \quad \quad \quad \quad 
\end{align}
This map generalizes in the obvious way for higher loops. In this language, \eref{4x4_from_pl} translates into an expression for $\langle ABCD \rangle$ in terms of $\langle ABij \rangle$ and $\langle CDij \rangle$ brackets:
\begin{eqnarray}
\langle ABCD \rangle & = & \langle AB34 \rangle \langle CD12 \rangle-\langle AB24 \rangle \langle CD13 \rangle +\langle AB23 \rangle \langle CD14 \rangle  \nonumber \\ 
& + & \langle AB14 \rangle \langle CD23 \rangle-\langle AB13 \rangle \langle CD24 \rangle+\langle AB12 \rangle \langle CD34 \rangle \, .
\label{4_by_4_brackets}
\end{eqnarray} 
Similarly,\eref{eq:4x4factorized}, which was obtained by using \pl relations, becomes
{\small
\begin{eqnarray}
\langle ABCD \rangle & = & \frac{\Big(\langle AB24 \rangle \langle CD34 \rangle - \langle AB34 \rangle \langle CD24 \rangle \Big) \Big(\langle AB12 \rangle \langle CD24 \rangle - \langle AB24 \rangle \langle CD12 \rangle \Big)}{\langle AB24 \rangle \langle CD24 \rangle} \nonumber \\ 
& + & \frac{\Big(\langle AB24 \rangle \langle CD23 \rangle - \langle AB23 \rangle \langle CD24 \rangle \Big) \Big(\langle AB14 \rangle \langle CD24 \rangle - \langle AB24 \rangle \langle CD14 \rangle \Big)}{\langle AB24 \rangle \langle CD24 \rangle} \, . \nonumber \\
\end{eqnarray}}
It is possible to use the integrand to construct both the mini and the full stratifications. As usual, for the latter it is necessary to properly account for the possible factorization of $\langle ABCD \rangle$. This can be done exactly as explained in \sref{section_full_stratification_L=2}.

When going to poles by shutting off brackets, it is necessary to take into account the \pl relations associated to each of the 2-loops. In bracket language, they become
\beq \label{eq:pluckequ}
\begin{array}{ccccc}
\langle AB14 \rangle \langle AB23 \rangle & + & \langle AB12 \rangle \langle AB34 \rangle & = & \langle AB13 \rangle \langle AB24 \rangle \\ 
\langle CD14 \rangle \langle CD23 \rangle & + & \langle CD12 \rangle \langle CD34 \rangle & = & \langle CD13 \rangle \langle CD24 \rangle
\end{array}
\eeq
We do not substitute these relations explicitly, but account for them \textit{implicitly}, by only shutting off allowed combinations of brackets. For example, when shutting off $\langle AB12 \rangle = 0$ and $\langle AB14 \rangle = 0$ we see that we are forced to also shut off $\langle AB13 \rangle = 0$ and/or $\langle AB24 \rangle = 0$. 

The main result of this section is that we have implemented the procedure described above and, focusing on labels, reproduced the entire mini stratification of $G_+(0,4;2)$ given by the third column of \tref{tab:4particle2loop} starting from \eref{eq:topdimIntegrand}.  It is important to emphasize that we have not
only reproduced the counting of boundaries obtained with amplituhedron, but have managed to establish a {\it one-to-one} map between all boundaries constructed with both methods. In order to illustrate this, in Appendix \ref{appendix_comparison_integrand_geometric} we present representative subsets of of the boundaries at each dimension. The examples have been chosen to showcase the conceptually different scenarios that might arise. Each of them is presented in geometric and integrand language.

The procedures for deriving the mini stratification based on the integrand and the amplituhedron are path-independent: the order in which minors are turned off to arrive at a given boundary is irrelevant. However, in a few cases, it is logically simpler to arrive at a given boundary using one route rather than another. In particular, it is usually preferable to set $\langle ABCD \rangle \to 0$ as late as possible.

\bigskip

\subsection{The Log of the Amplitude}

\label{section_log_from_integrand}

Let us now investigate the log of the amplitude in terms of the integrand. Using the integrand for $A_2$ given in \eref{eq:topdimIntegrand} and the square of the 1-loop
{\small
\begin{equation}
\frac{1}{\langle AB12 \rangle \langle AB14 \rangle \langle AB23 \rangle \langle AB34 \rangle \langle CD12 \rangle \langle CD14 \rangle \langle CD23 \rangle \langle CD34 \rangle} \, ,
\end{equation}}
the integrand for the 2-loop log of the amplitude becomes  
{\small \begin{align} \label{eq:LOGtopdimIntegrand}
& \frac{\langle AB34 \rangle \langle CD12 \rangle+\langle AB23 \rangle \langle CD14 \rangle+\langle AB14 \rangle \langle CD23 \rangle+\langle AB12 \rangle \langle CD34 \rangle - \langle ABCD \rangle \langle 1234 \rangle}{\langle ABCD \rangle \langle AB12 \rangle \langle AB14 \rangle \langle AB23 \rangle \langle AB34 \rangle \langle CD12 \rangle \langle CD14 \rangle \langle CD23 \rangle \langle CD34 \rangle} \nonumber  \\
& = \frac{\langle AB13 \rangle \langle CD24 \rangle+\langle AB24 \rangle \langle CD13 \rangle}{\langle ABCD \rangle \langle AB12 \rangle \langle AB14 \rangle \langle AB23 \rangle \langle AB34 \rangle \langle CD12 \rangle \langle CD14 \rangle \langle CD23 \rangle \langle CD34 \rangle} \, .
\end{align}}
We still have the two \pl relations \eref{eq:pluckequ}. For convenience, we shall usually use the form in \eref{eq:LOGtopdimIntegrand}; this makes it explicit that once $\langle ABCD \rangle $ is zero, the singularities of the log integrand are the same as those of the ordinary integrand.

As in the previous section, we obtain the singularities by setting to zero brackets which explicitly appear in the denominator of the integrand. Due to \pl relations, this may force other brackets to turn off. Again, we stress that the order in which we turn off minors to arrive at a given singularity is irrelevant. But as previously done, it is often simpler to set $\langle ABCD \rangle \to 0$ as late as possible.

Using the singularities of \eref{eq:LOGtopdimIntegrand}, we have managed to derive the mini stratification of the log of the amplitude previously obtained by geometric methods and summarized in \tref{tab:4particle2loopLOG}. As for the amplitude, we stress that we have not only reproduced the counting of boundaries, but have managed to establish a one-to-one map between all boundaries constructed with both methods. This matching provides additional strong support for the amplituhedron conjecture.

\bigskip

\section{The Deformed $G_+(0,n;L)$}

\label{section_deformed_amplituhedron}

A remarkable property of cells in the positive Grassmannian is that they are topologically balls. In other words, it is possible to prove that the posets encoding the positroid stratification of the Grassmannian are Eulerian, i.e. have $\mathcal{E}=1$ \cite{2005arXiv09129W}. The same is true for the $L^{th}$ power of such positroid stratification, the initial step for the stratification $G_+(0,n;L)$.

Given the detailed information of the boundary structure of the amplituhedron (or more precisely of $G_+(0,n;L)$ when discussing general values of $n$) we have gathered it is natural to ask whether general statements regarding the topology of the amplituhedron can be made.

In this section we would like to report on some striking experimental evidence based on explicit examples suggesting that there is a simple generalization of $G_+(0,n;L)$ which might exhibit a remarkably simple topology. 

Let us introduce the {\it deformed} $G_+(k,n;L)$. It is convenient to define it through its stratification as we explain below. For our purposes, it is equivalent to think we are considering the original $G_+(k,n;L)$, but a modified or deformed stratification. All the discussion in this section will be in the context of the mini stratification.\footnote{It would be interesting to investigate how the full stratification is affected by the deformation. In order to do this, however, a more detailed definition of the deformation is necessary.}

Recalling the general discussion in \sref{section_summary_method}, given a point in $\Gamma_0$, which is defined by a list of vanishing \pl coordinates, we can identify non-minimal minors of type (iii). These are minors that, at least initially, can be turned off. In fact, in general, sometimes some of these minors cannot be switched off due to relations. For example, turning off one of them might impose a relation that forces another one to be strictly non-zero, or might be forbidden because it would force another minor to violate positivity. We have already encountered this kind of restrictions in \sref{section_mini_stratification_L=3}, when constructing the mini stratification of $G_+(0,4;3)$. The deformed $G_+(0,n;L)$ corresponds to assuming that {\it all} such minors can be independently switched off at will in the $\Gamma_1$ that emanates from that point in $\Gamma_0$. Of course we know that this is not true for $G_+(0,n;L)$: as we turn off non-minimal minors, relations between them generically become important and determine the actual structure of $\Gamma_1$.

\medskip

\paragraph{An Example.} Let us demonstrate the difference between the deformed and standard stratifications with an explicit example from $G_+(0,4;3)$, for which a general discussion of all possible relations which can arise between non-minimal minors was presented in \sref{section_mini_stratification_L=3}. Consider the point in $\Gamma_0$ corresponding to the vanishing of
\beq
\begin{array}{c}
\Delta^{(1)}_{14},\Delta^{(1)}_{23},\Delta^{(1)}_{24},\Delta^{(1)} _{34} \\
\Delta^{(2)} _{12},\Delta^{(2)}_{13},\Delta^{(2)}_{14},\Delta^{(2)}_{23} \\
\Delta^{(3)}_{14},\Delta^{(3)}_{23}
\label{vanishing_pl_example}
\end{array}
\eeq
with all other \pl coordinates being non-zero. In this case, only the \pl relation associated to the third loop remains non-trivial and reduces to
\beq
\Delta^{(3)}_{12} \Delta^{(3)}_{34} = \Delta^{(3)}_{13}\Delta^{(3)}_{24}  \,.
\label{pl_relation_3rd_loop}
\eeq
The $4\times 4$ minors become
\begin{eqnarray}
\Delta^{(1,2)}_{1234} & = & \Delta^{(1)}_{12}\Delta^{(2)}_{34} - \Delta^{(1)}_{13}\Delta^{(2)}_{24} \nonumber \\
\Delta^{(1,3)}_{1234} & = & \Delta^{(1)}_{12}\Delta^{(3)}_{34} - \Delta^{(1)}_{13}\Delta^{(3)}_{24} \nonumber \\
\Delta^{(2,3)}_{1234} & = & \Delta^{(2)}_{34}\Delta^{(3)}_{12} - \Delta^{(2)}_{24}\Delta^{(3)}_{13}
\end{eqnarray}
The three of them are of type (iii) in the classification of \sref{sec:methodsummary}, i.e. they contain both positive and negative contributions and it naively appears that any of them can be independently set to zero while preserving extending positivity. However, this is not the case. Imagine we set to zero only $\Delta^{(1,2)}_{1234}$. In this case, the remaining $4\times 4$ minors take the form
\begin{eqnarray}
\Delta^{(1,3)}_{1234} & = & {\Delta^{(1)}_{13} \over \Delta^{(2)}_{34}} \left( \Delta^{(2)}_{24} \Delta^{(3)}_{34} - \Delta^{(2)}_{34}\Delta^{(3)}_{24} \right) \nonumber \\
\Delta^{(2,3)}_{1234} & = & {\Delta^{(3)}_{12} \over \Delta^{(3)}_{24}} \left( \Delta^{(2)}_{34}\Delta^{(3)}_{24}- \Delta^{(2)}_{24} \Delta^{(3)}_{34} \right)
\end{eqnarray}
We have rewritten the first one using $\Delta^{(1,2)}_{1234}=0$ and the second one using \eref{pl_relation_3rd_loop}. Since the prefactors are positive, we conclude it is impossible for $\Delta^{(1,3)}_{1234}$ and $\Delta^{(2,3)}_{1234}$ to be simultaneously positive. 

An alternative way of reaching the same conclusion is as follows. Using \eref{pl_relation_3rd_loop} to rewrite $\Delta^{(2,3)}_{1234}$ as before, it is possible to prove the following relation

\beq
\Delta^{(1,2)}_{1234} = {\Delta^{(2)}_{24} \over \Delta^{(3)}_{24}} \, \Delta^{(1,3)}_{1234} + {\Delta^{(1)}_{12} \over \Delta^{(2)}_{12}} \, \Delta^{(2,3)}_{1234} \,. 
\label{relation_4x4s}
\eeq
This is an explicit realization of the relations of Type C of \eref{obstruction_types}. Once again, we see we cannot turn off $\Delta^{(1,2)}_{1234}$ while preserving the positivity of the other two $4\times 4$ minors. We conclude that the $\Gamma_1$ emanating from this point in the underformed mini stratification does not contain a point in which only $\Delta^{(1,2)}_{1234}$ vanishes. In contrast, the deformed stratification is precisely defined such that all type (iii) minors can be independently turned off in $\Gamma_1$. 

This example illustrates why we refer to the object defined by the new stratification as a {\it deformation}. The relaxation of the constraint imposed by each relation between non-minimal minors can be regarded as the introduction of a new degree of freedom, i.e. a {\it deformation parameter}. Very schematically, each relation gets an independent deformation of the form\footnote{As in \eref{relation_4x4s}, these relations generally depend on smaller minors, too.}
\beq
R(\Delta^{(i,j)}_I)=0 \ \ \ \ \to \ \ \ \ R(\Delta^{(i,j)}_I)=\epsilon
\eeq
Similar deformations are possible in the presence of higher dimensional minors. In what follows, we assume all relations between non-minimal minors can be independently relaxed. Determining how many independent deformation parameters are necessary for achieving this for each geometry is certainly an interesting problem that we will not pursue here.

\begin{figure}[h]
\begin{center}
\includegraphics[width=12cm]{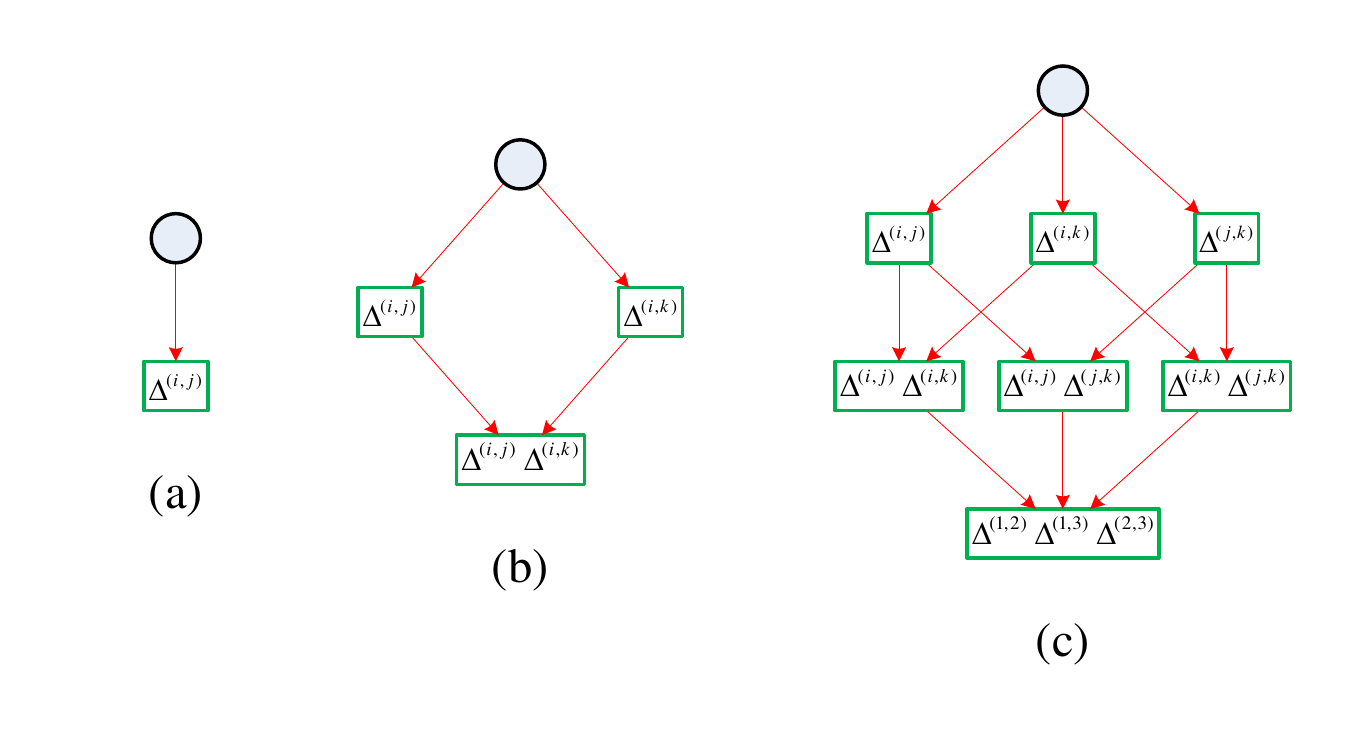}
\vspace{-.8cm}\caption{$\Gamma_1$'s for the deformed $G_+(0,n;L)$ in the cases of 1, 2 and 3 type (iii) $\Delta^{ij}_I$'s.}
\label{1Delta2Delta3Delta}
\end{center}
\end{figure}

As a result of the relaxation of relations in the deformed stratification, the structure of $\Gamma_1$'s is considerably simplified. \fref{1Delta2Delta3Delta} shows the $\Gamma_1$'s for the cases of 1, 2 and 3 type (iii) $\Delta^{(i,j)}_I$’s. They coincide with types (a) and (b) of \fref{1Delta2Delta} and type H of \fref{fig:3Ltrees}, from the mini stratification of the undeformed $G_+(0,4,3)$. We see the deformation substantially reduces the number of possible $\Gamma_1$'s.

\bigskip

\subsection{Examples}

We will now stratify the deformed $G_+(0,4;L)$ for $1 \leq L \leq 4$. Taking an experimental approach, we will observe that the resulting data gives rise to a natural conjecture about the topology.

\bigskip

\subsubsection{1-loop}

For $L=1$, there are no non-minimal minors and hence the deformed $G_+(0,4;1)$ is equal to the standard $G_+(0,4;1)\equiv G_+(2,4)$, which was discussed in detail in \sref{section_stratification_G2n} and \sref{sec:pmGivedecomposition}. The resulting poset has $\mathcal{E}=1$.

\bigskip

\subsubsection{2-loops}

$G_+(0,4;2)$ coincides with its deformation, since this example contains a single $4 \times 4$ minor $\Delta^{(1,2)}_{1234}$. Then, the right-hand column of \tref{tab:4particle2loop} also gives the boundaries of the deformed $G_+(0,4;2)$, which we reproduce in \tref{tab:4particle2loop_resolved} for easy reference. The total number of boundaries is $1232$. As noted in \eref{Euler_L=2}, the Euler number is equal to 2:
\begin{equation}
\mathcal{E}=\sum_{i=0}^8 (-1)^i \mathfrak{N}_M^{(i)} = 34-136+264-\ldots -9+1=2 \, .
\end{equation}

\begin{table}
\begin{center}
\bigskip
{\small
\begin{tabular}{|c|c|}
\hline
\textbf{Dim} & $\boldsymbol{\mathfrak{N}_{M, \text{deformed}}}$ \\
\hline
\textbf{8} & 1 \\
\hline
\textbf{7} & 9 \\
\hline
\textbf{6}  & 44 \\
\hline
\textbf{5} & 140 \\
\hline
\textbf{4} & 274 \\
\hline
\textbf{3} & 330 \\
\hline
\textbf{2} & 264 \\
\hline
\textbf{1} & 136 \\
\hline
\textbf{0} & 34 \\
\hline
\end{tabular}
}
\bigskip
\caption{Number of boundaries at each dimension for $G_+(0,4;2)$, which coincides with its deformation.\label{tab:4particle2loop_resolved}}
\end{center}
\end{table}

\bigskip

\subsubsection{3-loops}

It is straightforward to directly construct the stratification of the deformed $G_+(0,4;3)$. However, for illustration, here we take a shortcut and derive it from a detailed analysis of the undeformed mini stratification presented in \sref{section_mini_stratification_L=3}. In the deformation, we simply assume that the non-minimal minors $\Delta_I^{(i,j)}$ are completely independent. Thus, we just need to know how many $\Delta_I^{(i,j)}$ naively appear to be tunable to zero, i.e.\ the total number of $N \Delta_I^{(i,j)}$'s. We can determine this by just collapsing the various types of $3 \Delta_I^{(i,j)}$'s in \tref{tab:3LtreeContributions} into a single total number. The boundaries in this column are assigned the structure of Type H in \fref{fig:3Ltrees}. The remaining two columns do not change, and give rise to the same additional contributions as before.

The result of this modification is displayed in \tref{tab:4particle3loopResolved}. The final column adds up all of the contributions from the first three columns. Adding these contributions to the $\mathcal{N}_M$ column in \tref{tab:4particle3loop} will indeed give the number of boundaries $\mathfrak{N}_{M, \text{deformed}}$ of the deformed $G_+(0,4;3)$. The total number of boundaries is $61\, 354$ and, once again, the Euler number is
\begin{equation}
\sum_{i=0}^{12} (-1)^i \mathfrak{N}_{M, \text{deformed}}^{(i)} = 186-1152+3720-\ldots -15+1=2 \, .
\end{equation}

\begin{table}
\begin{center}
\bigskip
{\small
\begin{tabular}{|c|c|c|c|c|c|}
\hline
\textbf{Dim} & \textbf{3} $\boldsymbol{\Delta_I^{(i,j)}}$ &  \textbf{2} $\boldsymbol{\Delta_I^{(i,j)}}$ & \textbf{1} $\boldsymbol{\Delta_I^{(i,j)}}$ & \textbf{Total contribution} & $\boldsymbol{\mathfrak{N}_{M, \text{deformed}}}$ \\
\hline
\textbf{12} & 1 & 0 & 0 & $+0$ & 1 \\
\hline
\textbf{11} & 12 & 0 & 0 & $+3$ & 15 \\
\hline
\textbf{10} & 78 & 0 & 0 & $+39$ & 117 \\
\hline
\textbf{9} & 328 & 0 & 12 & $+271$ & 611 \\
\hline
\textbf{8} & 798 & 96 & 108 & $+1\,242$ & 2\,244 \\
\hline
\textbf{7} & 1\,056 & 576 & 528 & $+3\,756$ & 5\,916 \\
\hline
\textbf{6} & 650 & 1\,080 & 1\,584 & $+7\,666$ & 11\,156 \\
\hline
\textbf{5} & 168 & 792 & 2\,424 & $+10\,236$ & 14\,676 \\
\hline
\textbf{4} & 24 & 240 & 1\,848 & $+8\,598$ & 13\,254 \\
\hline
\textbf{3} & 0 & 24 & 672 & $+4\,346$ & 8\,306 \\
\hline
\textbf{2} & 0 & 0 & 96 & $+1\,200$ & 3\,720 \\
\hline
\textbf{1} & 0 & 0 & 0 & $+144$ & 1\,152 \\
\hline
\textbf{0} & 0 & 0 & 0 & $+0$ & 186 \\
\hline
\end{tabular}
}
\bigskip
\caption{Number of boundaries with $N=1,2,3$ number of $4 \times 4$ minors which have both positive and negative terms, and the corresponding added contribution to the total number of boundaries, obtained by assuming these minors to be completely independent and setting them to zero. The final column shows the number of boundaries $\boldsymbol{\mathfrak{N}_{M, \text{deformed}}}$ of the deformed $G_+(0,4;3)$.\label{tab:4particle3loopResolved}}
\end{center}
\end{table}

\bigskip

\subsubsection{4-loops}

\label{section_deformed_4-loops}

Let us now consider the deformed $G_+(0,4;4)$. In this case there are six $4 \times 4$ minors $\Delta^{(i,j)}_{1234}$. As usual, the first step is to obtain the $4^{th}$ power of the positroid stratification of $G_{+}(2,4)$. This contains a total of $33^4 = 1\, 185\, 921$ potential boundaries, which are stratified as shown in the first column $\mathbb{N}$ of \tref{tab:4particle4loop}. In agreement with the general result, this has Euler number equal to 1:
\begin{equation}
\sum_{i=0}^{16} (-1)^i \mathbb{N}^{(i)} = 1296-10368+\ldots -16+1=1 \, .
\end{equation}
Many of these boundaries explicitly violate the positivity of some $\Delta^{(i,j)}_I$, as can be easily found using the methods of \sref{extended_positivity_permutations}. Keeping only those boundaries which satisfy extended positivity, we obtain the column labeled $\mathcal{N}_M$ in \tref{tab:4particle4loop}. Interestingly, similarly to the $L=2$ case this again has Euler number equal to 1:
\begin{equation}
\sum_{i=0}^{16} (-1)^i \mathcal{N}_M^{(i)} = 994-6976+\ldots -16+1=1 \, .
\end{equation}

For each of these boundaries it is then necessary to classify which $\Delta^{(i,j)}_I$ may be turned off without turning off any $2\times 2$ minors; this corresponds to step (3) in \sref{sec:methodsummary} and is also easily implemented as in \sref{extended_positivity_permutations}. The additional boundaries which stem from the boundaries in the column $\mathcal{N}_M$ are added assuming that the $\Delta^{(i,j)}_I$ are completely independent. For example, if it is possible to turn off all six $\Delta^{(i,j)}_I$, we see that a large number boundaries are added: $\binom{6}{1}=6$ boundaries of one dimension lower,  $\binom{6}{2}=15$ boundaries of two dimensions lower, and so on; this will add a total of $\sum_{i=1}^6 \binom{6}{i}=63$ boundaries. The result of adding the boundaries from the $\Delta^{(i,j)}_I$ is the deformed $G_+(0,4;4)$, whose boundaries are shown in the right-hand column of \tref{tab:4particle4loop}. Remarkably, there is a total of $4828226$ boundaries, but cancellations are such that the Euler number is again
\begin{equation}
\sum_{i=0}^{16} (-1)^i \mathfrak{N}_{M, \text{deformed}}^{(i)} = 1162-10880+\ldots -22+1=2 \quad .
\end{equation}

\begin{table}
\begin{center}
\bigskip 
{\small
\begin{tabular}{|c|c|c|c|}
\hline
\textbf{Dim} & $\boldsymbol{\mathbb{N}}$ &  $\boldsymbol{\mathcal{N}_M}$ & $\boldsymbol{\mathfrak{N}_{M, \text{deformed}}}$\\
\hline
\textbf{16} & 1 & 1  & 1 \\
\hline
\textbf{15} & 16 & 16 & 22 \\
\hline
\textbf{14} & 136 & 136 & 247 \\
\hline
\textbf{13} & 784 & 784 & 1\,860\\
\hline
\textbf{12} & 3\,376 & 3\,212  & 10\,243\\
\hline
\textbf{11} & 11\,392 & 9\,856 & 42\,846 \\
\hline
\textbf{10} & 30\,928 & 23\,288 & 138\,421 \\
\hline
\textbf{9} & 68\,512 & 43\,616 & 346\,320 \\
\hline
\textbf{8} & 124\,552 & 67\,626 & 666\,654 \\
\hline
\textbf{7} & 185\,664 & 88\,128 & 974\,212 \\
\hline
\textbf{6} & 225\,312 & 96\,496 & 1\,061\,154 \\
\hline
\textbf{5} & 219\,456 & 90\,720 & 843\,992 \\
\hline
\textbf{4} & 167\,616 & 73\,144 & 480\,870 \\
\hline
\textbf{3} & 96\,768 & 47\,744 & 193\,980 \\
\hline
\textbf{2} & 39\,744 & 22\,944 & 55\,362 \\
\hline
\textbf{1} & 10\,368 & 6\,976 & 10\,880 \\
\hline
\textbf{0} & 1\,296 & 994 & 1\,162 \\
\hline
\end{tabular} 
\bigskip
\caption{Stratification of the deformed $G_+(0,4;4)$.\label{tab:4particle4loop}}
}
\end{center}
\end{table}

\bigskip

The explicit examples presented in this section hint that the deformed $G_+(0,n;L)$ might have a remarkably simple geometry. Summarizing our findings for $G_+(0,4;L)$, we obtained $\mathcal{E}=1$ for $L=1$ and $\mathcal{E}=2$ for $2\leq L \leq 4$. If such simplicity is indeed general, it would be interesting to understand how the complicated geometry of $\Gamma_0$ that arises after demanding extended positivity on the $L^{th}$ power of positroid stratification gets ``fixed" by the deformed $\Gamma_1$'s. These questions certainly deserve further study.

\bigskip

\section{Conclusions and Outlook}

\label{section_conclusions}

The amplituhedron is a new geometric formulation of scattering amplitudes in planar ${\cal N}=4$ Super Yang-Mills theory and perhaps it can potentially lead to a completely new, geometric formulation of quantum field theory. In this article we initiated a systematic investigation of the geometry of the amplituhedron. To do so, we introduced a stratification for it and developed a combinatorial implementation based on graphs and hyper perfect matchings. The combinatorial stratification of the amplituhedron considerably generalizes the positroid stratification of the positive Grassmannian and its graphical implementation \cite{2006math09764P,2007arXiv0706.2501P}. Extended positivity plays a central role in the definition of the amplituhedron. Our combinatorial stratification efficiently takes care of it. Furthermore, we explained how extended positivity is beautifully captured by permutations.

We then proceeded to the combinatorial stratification of explicit examples, focusing on $k=0$ and $n=4$. We first considered a {\it mini stratification} which lists boundaries with distinct labels --- lists of vanishing \pl coordinates and non-minimal minors (in this case $4\times4$ determinants). This is an interesting simplification of the structure which follows from the definition of the amplituhedron. To capture all boundaries we have to consider the {\it full stratification} which uses extended labels --- not only listing all vanishing \pl coordinates and non-minimal minors but also additional conditions between \pl coordinates which come from factorizing non-minimal minors. 

We first studied the amplitude at 2-loops. In the mini stratification, it contains \mbox{$1\,232$} boundaries which interplay to produce an extremely simple topology with $\mathcal{E}=2$. We repeated the analysis for the log of the amplitude at 2-loops, which has \mbox{$1\,072$} boundaries and, once again, just $\mathcal{E}=2$. We also discussed how these two objects beautifully combine into the square of the positroid stratification of $G_+(2,4)$. In the full stratification there are \mbox{$1\,434$} boundaries in the amplitude and \mbox{$1\,274$} boundaries in the log and both have $\mathcal{E}=8$, while the gluing region has $\mathcal{E}=7$. This shows that the topology is substantially different from the square of $G_+(2,4)$.

We also performed the mini stratification of the $L=3$ amplitude. Unlike the 2-loop result, we obtained a rather large Euler number (in absolute value), ${\cal E}=-14$ which also shows that the topology is much more involved than $[G_+(2,4)]^3$.
The fact that a relatively complicated topology can in general arise from the simple definition of the amplituhedron is certainly a logical possibility and, perhaps, the most natural expectation. Note that the available Euler numbers for the mini stratification are even Catalan numbers. It would not be surprising if this persists at higher loops, as Catalan numbers play an important role in the positive Grassmannian, so it is tempting to conjecture that for $L=4$ we should get ${\cal E}=132$. We should of course warn that this conjecture is based on extrapolation from very limited data.

We rederived the entire mini stratifications of the $L=2$ amplitude and its log in terms of the integrand. It is important to remark that the computations involved in this approach are completely different from the ones based on the amplituhedron. In particular, this method is based on looking for singularities of a function and makes no reference to positivity. We succeeded in not only reproducing the counting of boundaries at each dimension but also in explicitly verifying that the identities of all boundaries obtained by the two methods match. This is a very important piece of explicit evidence supporting the amplituhedron conjecture and supplements the direct triangulation provided in \cite{Arkani-Hamed:2013kca}.

Finally, we introduced the deformed amplituhedron, which corresponds to deforming the relations between non-minimal minors in order to make them independent. The stratification of this object is considerably simpler than the one for the ordinary amplituhedron. We computed several explicit examples and, quite remarkably, they exhibit an extremely simple topology: $\mathcal{E}=1$ for $L=1$ and $\mathcal{E}=2$ for $2\leq L \leq 4$.

There are several directions worth investigating in the future, among them:

\begin{itemize}

\item One of the main questions we expect to address in future work is how to exploit the combinatorial tools we developed for triangulating the amplituhedron. Different triangulations should correlate with the different forms the integrand can take.

\item Another natural next step is to study how our ideas need to be extended to deal with $k>0$ and $n>4$. In this cases, positivity becomes more involved due to the addition of a tree-level contribution to the matrix $\mathcal{C}$ and the importance of external data, respectively.

\item As a mathematical question, it would be interesting to investigate the geometry of $G_+(0,n;L)$ for $n>4$. Notice that, contrary to the amplituhedron, $G_+(0,n;L)$ does not have additional positivity constraints involving external data for $n>4$. In fact, the mini stratification and its combinatorial implementation can be applied without modifications to this geometry for arbitrary $n$ and $L$ and provide a powerful handle on it.

\item The amplituhedron is just one example inside a large list of spaces which are related to it by relaxing some of the extended positivity conditions \cite{Jara_in_progress}. For example, for $k=0$ and $n=4$ the parent of all these spaces corresponds to the $L^{th}$ power of the positroid stratification of $G_+(2,4)$. Dealing with extended positivity is straightforward in our combinatorial stratification, so our tools can be readily extended for the stratification of these spaces. These geometries are relatively simpler than that of the amplituhedron and it is expected that they can be exploited to constraint or even infer the structure of the integrand \cite{Jara_in_progress}. It would also be interesting to investigate whether the deformed amplituhedron, which similarly results from the relaxation of some relations, can likewise be used for determining the integrand.

\item From a purely mathematical standpoint, it would be interesting to investigate whether the simplicity of the deformed stratification we have observed in explicit examples holds more generally. If so, it would interesting to understand the underlying reason for this. It is important to keep in mind that the general definition of deformation might turn out to be more sophisticated than the one we considered. On a related note, it is possible that the deformations of relations cannot be arbitrary but must obey certain structure in order to preserve a simple geometry. Further exploration of these questions can potentially uncover a rather rich story. It would also be interesting to investigate whether the deformed stratification has any physical significance.

\item Intriguingly, hyper perfect matchings have recently also appeared in the combinatorial interpretation of cluster algebras \cite{2013arXiv1308.2998K}. Generalizing what happens for usual perfect matchings, cluster variables obtained by certain sequences of mutations such as the so-called {\it hexahedron recurrence}, are given by partition functions of hyper perfect matchings. It is interesting to mention that for this application, only hyper perfect matchings satisfying certain asymptotic conditions, called {\it taut conditions}, should be considered. This is, at least superficially, reminiscent of the conditions imposed by extended positivity. It would be interesting to investigate whether there is connection between the amplituhedron and cluster algebras. If it exists, it would be a new addition to the long list of applications of cluster algebras to scattering \cite{ArkaniHamed:2012nw,Franco:2013pg,Golden:2013xva,Amariti:2013ija,Golden:2014xqa,Franco:2014nca,Golden:2014xqf,Paulos:2014dja}.

\item Similarly to the story for 4d $\mathcal{N}=4$ SYM, a connection between scattering amplitudes in the planar ABJM theory in 3d \cite{Aharony:2008ug} and the positive orthogonal Grassmannian has been established in \cite{Huang:2013owa,Huang:2014xza}. It would be interesting to investigate whether something like the amplituhedron exists for this theory and, if so, how our ideas extend to it.

\end{itemize}

\bigskip

\section*{Acknowledgements}

We would like to thank N. Arkani-Hamed for very useful discussions. D. G. would like to thank the Simons Summer Workshop at the Simons Center for Geometry and Physics for hospitality during the completion of this work. The work of S. F. and D. G. is supported by the U.K. Science and Technology Facilities Council (STFC). A.M. acknowledges funding by the Durham International Junior Research Fellowship. J. T. is supported in part by the David and Ellen Lee
Postdoctoral Scholarship and by DOE grant DE-FG03-92-ER40701.

\newpage


\appendix

\section{Two-Loop Boundaries Before Extended Positivity}

\begin{figure}[H]
\begin{center}
\includegraphics[width=15.7cm,angle=90]{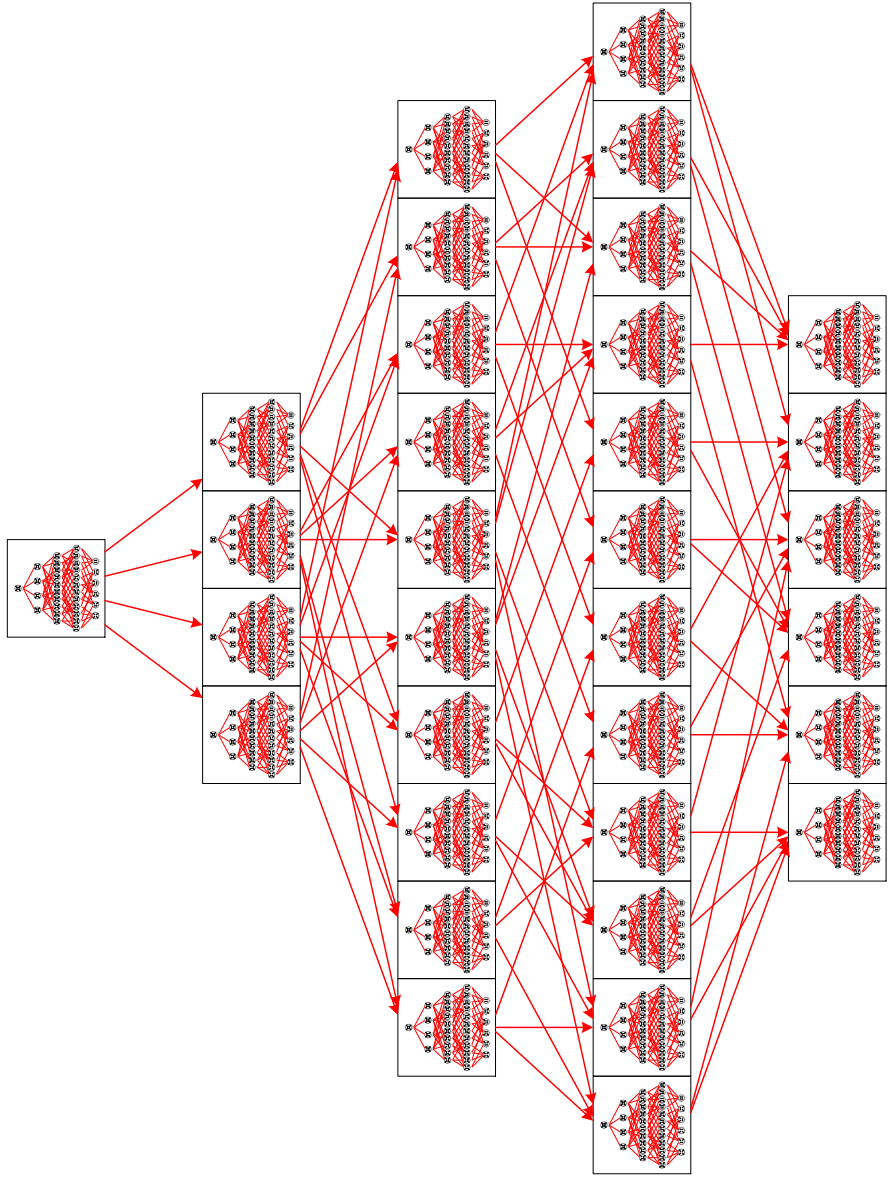}
\caption{Graphical representation of all potential boundaries of the 2-loop $n=4$ amplituhedron, before taking into account extended positivity. Each square corresponds to an element in the the positroid decomposition of the first graph and contains the positroid decomposition of the second graph. The small graphs are $1\, 089$ in total.}
\label{fig:4particle2loop}
\end{center}
\end{figure}

In \fref{fig:4particle2loop} we present a graphical representation of the square of the positroid stratification of $G_+(2,4)$.

\newpage

\section{Geometric Versus Integrand Stratification: Explicit Examples}

\label{appendix_comparison_integrand_geometric}

In \sref{section_integrand_stratification_amplitude}, we obtained the mini stratification of $G_+(0,4;2)$ using the integrand. We have explicitly verified the one-to-one agreement of all boundaries obtained with the stratifications based on the integrand and the amplituhedron. In this appendix we collect several explicit examples of this precise match for illustration purposes. They have been chosen to provide a good representation of all qualitatively different cases that arise. 

Strictly speaking, the language used in this study is the one of labels, i.e. the mini stratification. As explained in \sref{sec:ministrat}, labels really correspond to classes of boundaries. In particular, for every label in which the $4\times 4$ minor vanishes, there can be multiple boundaries, i.e. different integrands. Furthermore, these boundaries in general have different dimensions. For these cases, the table below provides the integrand corresponding to the maximal vanishing of the $4\times 4$ minor. As in the mini stratification, we list this configuration at the highest dimension at which the $4\times 4$ vanishes. All other integrands corresponding to the same labels can be easily constructed.

\bigskip
\paragraph{\textbf{Dimension 8.}} There is only one 8-dimensional boundary, which is the top-dimensional one. It is the integrand \eref{eq:topdimIntegrand}, where the lines $AB$ and $CD$ are completely free. In the table below, we compare the integrand and geometric methods. The same format will be used for all other examples. The first two rows show the integrand and the restrictions on the lines.  The comparison with our other method is seen in the last two rows, where we specify the set of \pl coordinates and hyper perfect matchings present. The hyper perfect matchings contributing to the $4\times 4$ minor $\langle ABCD\rangle$ are highlighted in color, with the ones contributing positively ($P_{23}$, $P_{32}$, $P_{45}$, $P_{54}$) in blue and the ones contributing negatively ($P_{16}$, $P_{61}$) in red. Notice that $\langle ABCD\rangle$ can vanish while some of them are present due to cancellations. However, if none of these perfect matchings are present, $\langle ABCD\rangle$ is forced to automatically vanish.

\bigskip

{\footnotesize
\begin{equation} \label{eq:LOGdim8guy}
\begin{array}{|c|c|}
\hline
\Gape[6pt][6pt]{ \textbf{Integrand} }&  \frac{\langle AB34 \rangle \langle CD12 \rangle+\langle AB23 \rangle \langle CD14 \rangle+\langle AB14 \rangle \langle CD23 \rangle+\langle AB12 \rangle \langle CD34 \rangle}{\langle ABCD \rangle \langle AB12 \rangle \langle AB14 \rangle \langle AB23 \rangle \langle AB34 \rangle \langle CD12 \rangle \langle CD14 \rangle \langle CD23 \rangle \langle CD34 \rangle} \\
\hdashline 
\textbf{Constraints on $AB$ and $CD$} &  \text{Free} \\
\hdashline 
\Gape[6pt][0pt]{ \textbf{\pl coordinates} } &   \Delta^{(1,2)}_{1234}, \Delta^{(1)}_{12},\Delta^{(1)}_{13},\Delta^{(1)}_{14},\Delta^{(1)}_{23},\Delta^{(1)}_{24},\Delta^{(1)}_{34}, \\
  \Gape[0pt][6pt]{ \textbf{turned on} }&  \Delta^{(2)}_{12},\Delta^{(2)}_{13},\Delta^{(2)}_{14},\Delta^{(2)}_{23},\Delta^{(2)}_{24},\Delta^{(2)}_{34}\\
\hdashline 
&  P_{1,1},P_{1,2},P_{1,3},P_{1,4},P_{1,5},\textcolor{red}{\boldsymbol{P_{1,6}}},P_{2,1},P_{2,2},\textcolor{blue}{\boldsymbol{P_{2,3}}},P_{2,4},\\
\textbf{Hyper perfect} &  P_{2,5},P_{2,6}, P_{3,1},\textcolor{blue}{\boldsymbol{P_{3,2}}},P_{3,3},P_{3,4},P_{3,5},P_{3,6},P_{4,1},P_{4,2}, \\
\textbf{matchings present}  &  P_{4,3},P_{4,4},\textcolor{blue}{\boldsymbol{P_{4,5}}},P_{4,6},P_{5,1},P_{5,2},P_{5,3},\textcolor{blue}{\boldsymbol{P_{5,4}}},P_{5,5},P_{5,6}, \\
 & \textcolor{red}{\boldsymbol{P_{6,1}}},P_{6,2}, P_{6,3},P_{6,4},P_{6,5},P_{6,6} \\
\hline
\end{array} \nonumber
\end{equation}}

\ \ \ \ \paragraph{\textbf{Dimension 7.}} There are 9 integrands corresponding to 7-dimensional boundaries. We present all of them below.
\bigskip

\bigskip

\noindent{\scriptsize
\begin{equation} \label{eq:dim7guys}
\begin{array}{|c|c|}
\hline
\Gape[2pt][2pt]{\frac{\langle AB34 \rangle \langle CD12 \rangle+\langle AB23 \rangle \langle CD14 \rangle+\langle AB14 \rangle \langle CD23 \rangle+\langle AB12 \rangle \langle CD34 \rangle}{\langle AB12 \rangle \langle AB14 \rangle \langle AB23 \rangle \langle AB34 \rangle \langle CD12 \rangle \langle CD14 \rangle \langle CD23 \rangle \langle CD34 \rangle}} 
& \frac{\langle AB34 \rangle \langle CD12 \rangle+\langle AB23 \rangle \langle CD14 \rangle+\langle AB14 \rangle \langle CD23 \rangle}{\langle ABCD \rangle \langle AB14 \rangle \langle AB23 \rangle \langle AB34 \rangle \langle CD12 \rangle \langle CD14 \rangle \langle CD23 \rangle \langle CD34 \rangle}   \\
\hdashline 
\langle ABCD \rangle \to  0 
& \langle AB12 \rangle \to  0  \\
\hdashline 
\Gape[3pt][2pt]{\Delta^{(1)}_{12},\Delta^{(1)}_{13},\Delta^{(1)}_{14},\Delta^{(1)}_{23},\Delta^{(1)}_{24},\Delta^{(1)}_{34},}
& \Delta^{(1,2)}_{1234},\Delta^{(1)}_{12},\Delta^{(1)}_{13},\Delta^{(1)}_{14},\Delta^{(1)}_{23},\Delta^{(1)}_{24}, \\
\Gape[0pt][3pt]{\Delta^{(2)}_{12},\Delta^{(2)}_{13},\Delta^{(2)}_{14},\Delta^{(2)}_{23},\Delta^{(2)}_{24},\Delta^{(2)}_{34} }
& \Delta^{(2)}_{12},\Delta^{(2)}_{13},\Delta^{(2)}_{14},\Delta^{(2)}_{23},\Delta^{(2)}_{24},\Delta^{(2)}_{34} \\
 \hdashline 
P_{1,1},P_{1,2},P_{1,3},P_{1,4},P_{1,5},\textcolor{red}{\boldsymbol{P_{1,6}}},P_{2,1},P_{2,2},\textcolor{blue}{\boldsymbol{P_{2,3}}},P_{2,4}, & 
P_{1,1},P_{1,2},P_{1,3},P_{1,4},P_{1,5},\textcolor{red}{\boldsymbol{P_{1,6}}},P_{3,1},\textcolor{blue}{\boldsymbol{P_{3,2}}},P_{3,3},P_{3,4}, \\
P_{2,5},P_{2,6}, P_{3,1},\textcolor{blue}{\boldsymbol{P_{3,2}}},P_{3,3},P_{3,4},P_{3,5},P_{3,6},P_{4,1},P_{4,2}, & 
P_{3,5},P_{3,6},P_{4,1},P_{4,2},P_{4,3},P_{4,4},\textcolor{blue}{\boldsymbol{P_{4,5}}},P_{4,6},P_{5,1},P_{5,2}, \\
\hline
\end{array} \nonumber
\end{equation}
}

\newpage

\paragraph{\textbf{Dimension 6.}} There are 44 integrands corresponding to 6-dimensional boundaries. We present some examples below.

\bigskip

\noindent{\scriptsize
\begin{equation} \label{eq:dim6guys}
\begin{array}{|c|c|}
\hline
 \Gape[2pt][2pt]{\frac{\langle \text{AB34}\rangle  \langle \text{CD12}\rangle +\langle \text{AB23}\rangle  \langle \text{CD14}\rangle +\langle \text{AB14}\rangle  \langle \text{CD23}\rangle }{\langle \text{AB14}\rangle  \langle \text{AB23}\rangle  \langle \text{AB34}\rangle  \langle \text{CD12}\rangle  \langle \text{CD14}\rangle  \langle \text{CD23}\rangle  \langle \text{CD34}\rangle }} & \frac{\langle \text{AB23}\rangle  \langle \text{CD14}\rangle +\langle \text{AB14}\rangle  \langle \text{CD23}\rangle }{\langle \text{AB14}\rangle  \langle \text{AB23}\rangle  \langle \text{ABCD}\rangle  \langle \text{CD12}\rangle  \langle \text{CD14}\rangle  \langle \text{CD23}\rangle  \langle \text{CD34}\rangle } \\
 \hdashline 
\langle \text{ABCD}\rangle \to 0,\langle \text{AB12}\rangle \to 0 
& \langle \text{AB12}\rangle \to 0,\langle \text{AB34}\rangle \to 0 \\
 \hdashline 
\Gape[3pt][2pt]{ \Delta^{(1)}_{12},\Delta^{(1)}_{13},\Delta^{(1)}_{14},\Delta^{(1)}_{23},\Delta^{(1)}_{24} }
& \Delta^{(1,2)}_{1234} ,\Delta^{(1)}_{13},\Delta^{(1)}_{14},\Delta^{(1)}_{23},\Delta^{(1)}_{24} \\
\Gape[0pt][3pt]{ \Delta^{(2)}_{12},\Delta^{(2)}_{13},\Delta^{(2)}_{14},\Delta^{(2)}_{23},\Delta^{(2)}_{24},\Delta^{(2)}_{34} }
& \Delta^{(2)}_{12},\Delta^{(2)}_{13},\Delta^{(2)}_{14},\Delta^{(2)}_{23},\Delta^{(2)}_{24},\Delta^{(2)}_{34} \\
 \hdashline 
P_{1,1},P_{1,2},P_{1,3},P_{1,4},P_{1,5},\textcolor{red}{\boldsymbol{P_{1,6}}},P_{3,1},\textcolor{blue}{\boldsymbol{P_{3,2}}},P_{3,3},P_{3,4}, &
P_{1,1},P_{1,2},P_{1,3},P_{1,4},P_{1,5},\textcolor{red}{\boldsymbol{P_{1,6}}},P_{4,1},P_{4,2},P_{4,3},P_{4,4}, \\
P_{3,5},P_{3,6},P_{4,1},P_{4,2},P_{4,3},P_{4,4},\textcolor{blue}{\boldsymbol{P_{4,5}}},P_{4,6},P_{5,1},P_{5,2}, &
\textcolor{blue}{\boldsymbol{P_{4,5}}},P_{4,6},P_{5,1},P_{5,2},P_{5,3},\textcolor{blue}{\boldsymbol{P_{5,4}}},P_{5,5},P_{5,6},\textcolor{red}{\boldsymbol{P_{6,1}}},P_{6,2}, \\
\hline
\end{array} \nonumber
\end{equation}
}

\paragraph{\textbf{Dimension 5.}} There are 140 integrands corresponding to 5-dimensional boundaries. We present some examples below.

\bigskip

\noindent{\scriptsize
\begin{equation}
\begin{array}{|c|c|}
\hline
 \Gape[2pt][2pt]{\frac{\langle \text{AB23}\rangle  \langle \text{CD14}\rangle +\langle \text{AB14}\rangle  \langle \text{CD23}\rangle }{\langle \text{AB14}\rangle  \langle \text{AB23}\rangle  \langle \text{CD12}\rangle  \langle \text{CD14}\rangle  \langle \text{CD23}\rangle  \langle \text{CD34}\rangle } } & \frac{\langle \text{AB23}\rangle  \langle \text{CD14}\rangle +\langle \text{AB14}\rangle  \langle \text{CD23}\rangle }{\langle \text{AB14}\rangle  \langle \text{AB23}\rangle  \langle \text{AB34}\rangle  \langle \text{CD14}\rangle  \langle \text{CD23}\rangle  \langle \text{CD34}\rangle } \\
 \hdashline \langle \text{ABCD}\rangle \to 0,\langle \text{AB12}\rangle \to 0,\langle \text{AB34}\rangle \to 0 &\langle \text{ABCD}\rangle \to 0,\langle \text{AB12}\rangle \to 0,\langle \text{CD12}\rangle \to 0 \\
 \hdashline \Gape[3pt][2pt]{\Delta^{(1)}_{13},\Delta^{(1)}_{14},\Delta^{(1)}_{23},\Delta^{(1)}_{24} }&\Delta^{(1)}_{12},\Delta^{(1)}_{13},\Delta^{(1)}_{14},\Delta^{(1)}_{23},\Delta^{(1)}_{24} \\
 \Gape[0pt][3pt]{ \Delta^{(2)}_{12},\Delta^{(2)}_{13},\Delta^{(2)}_{14},\Delta^{(2)}_{23},\Delta^{(2)}_{24},\Delta^{(2)}_{34}} &\Delta^{(2)}_{12},\Delta^{(2)}_{13},\Delta^{(2)}_{14},\Delta^{(2)}_{23},\Delta^{(2)}_{24} \\
 \hdashline 
P_{1,1},P_{1,2},P_{1,3},P_{1,4},P_{1,5},\textcolor{red}{\boldsymbol{P_{1,6}}},P_{4,1},P_{4,2},P_{4,3},P_{4,4}, &
P_{1,1},P_{1,3},P_{1,4},P_{1,5},\textcolor{red}{\boldsymbol{P_{1,6}}},P_{3,1},P_{3,3},P_{3,4},P_{3,5},P_{3,6}, \\
\textcolor{blue}{\boldsymbol{P_{4,5}}},P_{4,6},P_{5,1},P_{5,2},P_{5,3},\textcolor{blue}{\boldsymbol{P_{5,4}}},P_{5,5},P_{5,6},\textcolor{red}{\boldsymbol{P_{6,1}}},P_{6,2},
 & P_{4,1},P_{4,3},P_{4,4},\textcolor{blue}{\boldsymbol{P_{4,5}}},P_{4,6},P_{5,1}, P_{5,3},\textcolor{blue}{\boldsymbol{P_{5,4}}},P_{5,5},P_{5,6}, \\
P_{6,3},P_{6,4},P_{6,5},P_{6,6} & \textcolor{red}{\boldsymbol{P_{6,1}}},P_{6,3},P_{6,4},P_{6,5},P_{6,6} \\
\hline
\end{array} \nonumber
\end{equation}
}

\noindent{\scriptsize
\begin{equation} 
\begin{array}{|c|c|}
 \hline \Gape[3pt][2pt]{\frac{1}{\langle \text{AB23}\rangle  \langle \text{AB34}\rangle  \langle \text{CD12}\rangle  \langle \text{CD14}\rangle  \langle \text{CD23}\rangle  \langle \text{CD34}\rangle } } & \frac{1}{\langle \text{ABCD}\rangle  \langle \text{CD12}\rangle  \langle \text{CD23}\rangle  \langle \text{CD34}\rangle } \\
 \hdashline  \langle \text{AB12}\rangle \to 0,\langle \text{AB13}\rangle \to 0,\langle \text{AB14}\rangle \to 0,\langle \text{AB24}\rangle \to 0 & \langle \text{AB12}\rangle \to 0,\langle \text{AB13}\rangle \to 0,\langle \text{AB14}\rangle \to 0,\langle \text{AB34}\rangle \to 0 \\
 \hdashline \Gape[3pt][2pt]{ \Delta^{(1,2)}_{1234} ,\Delta^{(1)}_{12},\Delta^{(1)}_{14} }& \Delta^{(1,2)}_{1234} ,\Delta^{(1)}_{13},\Delta^{(1)}_{14}\\
\Gape[0pt][3pt]{  \Delta^{(2)}_{12},\Delta^{(2)}_{13},\Delta^{(2)}_{14},\Delta^{(2)}_{23},\Delta^{(2)}_{24},\Delta^{(2)}_{34} } & \Delta^{(2)}_{12},\Delta^{(2)}_{13},\Delta^{(2)}_{14},\Delta^{(2)}_{23},\Delta^{(2)}_{24},\Delta^{(2)}_{34} \\
 \hdashline  P_{3,1},\textcolor{blue}{\boldsymbol{P_{3,2}}},P_{3,3},P_{3,4},P_{3,5},P_{3,6},P_{4,1},P_{4,2}, P_{4,3},P_{4,4},
 & P_{1,1},P_{1,2},P_{1,3},P_{1,4},P_{1,5},\textcolor{red}{\boldsymbol{P_{1,6}}},P_{4,1},P_{4,2},P_{4,3},P_{4,4}, \\
\textcolor{blue}{\boldsymbol{P_{4,5}}},P_{4,6} & \textcolor{blue}{\boldsymbol{P_{4,5}}},P_{4,6} \\
\hline 
\hline  

\frac{1}{\langle \text{AB34}\rangle  \langle \text{ABCD}\rangle  \langle \text{CD23}\rangle  \langle \text{CD34}\rangle } & \Gape[2pt][2pt]{\frac{\langle \text{AB23}\rangle  \langle \text{CD14}\rangle +\langle \text{AB34}\rangle  \langle \text{CD12}\rangle }{\langle \text{AB23}\rangle  \langle \text{AB34}\rangle  \langle \text{ABCD}\rangle  \langle \text{CD12}\rangle  \langle \text{CD14}\rangle  \langle \text{CD34}\rangle } } \\
\hline
\end{array} \nonumber
\end{equation}
}

\bigskip

\paragraph{\textbf{Dimension 4.}} There are 274 integrands corresponding to 4-dimensional boundaries. We present some examples below.

\bigskip

\noindent{\scriptsize
\begin{equation}
\begin{array}{|c|c|}
\hline
 \Gape[2pt][2pt]{\frac{\langle \text{AB23}\rangle  \langle \text{CD14}\rangle +\langle \text{AB14}\rangle  \langle \text{CD23}\rangle }{\langle \text{AB14}\rangle  \langle \text{AB23}\rangle  \langle \text{CD14}\rangle  \langle \text{CD23}\rangle  \langle \text{CD34}\rangle } } & \frac{1}{\langle \text{AB23}\rangle  \langle \text{CD12}\rangle  \langle \text{CD34}\rangle } \\
 \hdashline \langle \text{ABCD}\rangle \to 0,\langle \text{AB12}\rangle \to 0,\langle \text{AB34}\rangle \to 0,\langle \text{CD12}\rangle \to 0 &\langle \text{ABCD}\rangle \to 0,\langle \text{AB12}\rangle \to 0,\langle \text{AB34}\rangle \to 0,\langle \text{CD14}\rangle \to 0 \\
 \hdashline \Gape[3pt][2pt]{ \Delta^{(1)}_{13},\Delta^{(1)}_{14},\Delta^{(1)}_{23},\Delta^{(1)}_{24}} &\Delta^{(1)}_{13},\Delta^{(1)}_{14},\Delta^{(1)}_{23},\Delta^{(1)}_{24} \\
 \Gape[0pt][3pt]{ \Delta^{(2)}_{12},\Delta^{(2)}_{13},\Delta^{(2)}_{14},\Delta^{(2)}_{23},\Delta^{(2)}_{24} }&\Delta^{(2)}_{12},\Delta^{(2)}_{13},\Delta^{(2)}_{14},\Delta^{(2)}_{24},\Delta^{(2)}_{34} \\
 \hdashline 
P_{1,1},P_{1,3},P_{1,4},P_{1,5},\textcolor{red}{\boldsymbol{P_{1,6}}},P_{4,1}, P_{4,3},P_{4,4},\textcolor{blue}{\boldsymbol{P_{4,5}}},P_{4,6},
 & P_{1,1},P_{1,2},P_{1,3},P_{1,4},\textcolor{red}{\boldsymbol{P_{1,6}}},P_{4,1}, P_{4,2},P_{4,3},P_{4,4},P_{4,6}, \\
P_{5,1},P_{5,3},\textcolor{blue}{\boldsymbol{P_{5,4}}},P_{5,5},P_{5,6},\textcolor{red}{\boldsymbol{P_{6,1}}},P_{6,3},P_{6,4},P_{6,5},P_{6,6} &
P_{5,1},P_{5,2},P_{5,3},\textcolor{blue}{\boldsymbol{P_{5,4}}},P_{5,6},\textcolor{red}{\boldsymbol{P_{6,1}}},P_{6,2},P_{6,3},P_{6,4},P_{6,6} \\
\hline
\end{array} \nonumber
\end{equation}
}

\newpage

\noindent{\scriptsize
\begin{equation}
\begin{array}{|c|c|}
\hline
  \Gape[3pt][2pt]{\frac{1}{\langle \text{AB23}\rangle  \langle \text{AB34}\rangle  \langle \text{CD14}\rangle  \langle \text{CD23}\rangle  \langle \text{CD34}\rangle } } & \frac{1}{\langle \text{AB34}\rangle  \langle \text{ABCD}\rangle  \langle \text{CD23}\rangle } \\
 \hdashline \langle \text{AB12}\rangle \to 0,\langle \text{AB13}\rangle \to 0,\langle \text{AB14}\rangle \to 0,\langle \text{AB24}\rangle \to 0 & \langle \text{AB12}\rangle \to 0,\langle \text{AB13}\rangle \to 0,\langle \text{AB14}\rangle \to 0,\langle \text{CD12}\rangle \to 0 \\
 \langle \text{CD12}\rangle \to 0 & \langle \text{CD34}\rangle \to 0 \\
 \hdashline \Gape[3pt][2pt]{ \Delta^{(1,2)}_{1234} ,\Delta^{(1)}_{12},\Delta^{(1)}_{14} } & \Delta^{(1,2)}_{1234} ,\Delta^{(1)}_{12},\Delta^{(1)}_{13},\Delta^{(1)}_{14} \\
 \Gape[0pt][3pt]{ \Delta^{(2)}_{12},\Delta^{(2)}_{13},\Delta^{(2)}_{14},\Delta^{(2)}_{23},\Delta^{(2)}_{24}} & \Delta^{(2)}_{13},\Delta^{(2)}_{14},\Delta^{(2)}_{23},\Delta^{(2)}_{24} \\
 \hdashline P_{3,1},P_{3,3},P_{3,4},P_{3,5},P_{3,6},P_{4,1},P_{4,3},P_{4,4},\textcolor{blue}{\boldsymbol{P_{4,5}}},P_{4,6} & 
P_{1,1},P_{1,4},P_{1,5},\textcolor{red}{\boldsymbol{P_{1,6}}},P_{3,1},P_{3,4},P_{3,5},P_{3,6},P_{4,1},P_{4,4}, \\
 & \textcolor{blue}{\boldsymbol{P_{4,5}}},P_{4,6} \\
\hline
\hline 

\Gape[3pt][2pt]{\frac{1}{\langle \text{AB23}\rangle  \langle \text{AB34}\rangle  \langle \text{CD14}\rangle  \langle \text{CD34}\rangle } } & \frac{1}{\langle \text{AB34}\rangle  \langle \text{ABCD}\rangle  \langle \text{CD34}\rangle } \\
\hline
\end{array} \nonumber
\end{equation}
}

\bigskip

\paragraph{\textbf{Dimension 3.}} There are 330 integrands corresponding to 3-dimensional boundaries. We present some examples below.

\bigskip

\noindent{\scriptsize
\begin{equation}
\begin{array}{|c|c|}
\hline
 \Gape[2pt][2pt]{\frac{\langle \text{AB23}\rangle  \langle \text{CD14}\rangle +\langle \text{AB14}\rangle  \langle \text{CD23}\rangle }{\langle \text{AB14}\rangle  \langle \text{AB23}\rangle  \langle \text{CD14}\rangle  \langle \text{CD23}\rangle }} &  \frac{1}{\langle \text{CD23}\rangle  \langle \text{CD34}\rangle } \\
 \hdashline \langle \text{ABCD}\rangle \to 0,\langle \text{AB12}\rangle \to 0,\langle \text{AB34}\rangle \to 0,\langle \text{CD12}\rangle \to 0 & \langle \text{ABCD}\rangle \to 0,\langle \text{AB12}\rangle \to 0,\langle \text{AB13}\rangle \to 0,\langle \text{AB14}\rangle \to 0 \\
 \langle \text{CD34}\rangle \to 0 & \langle \text{AB23}\rangle \to 0,\langle \text{CD14}\rangle \to 0 \\
 \hdashline \Gape[3pt][3pt]{ \Delta^{(1)}_{13},\Delta^{(1)}_{14},\Delta^{(1)}_{23},\Delta^{(1)}_{24},\Delta^{(2)}_{13},\Delta^{(2)}_{14},\Delta^{(2)}_{23},\Delta^{(2)}_{24} }& \Delta^{(1)}_{12},\Delta^{(1)}_{13},\Delta^{(2)}_{12},\Delta^{(2)}_{13},\Delta^{(2)}_{14},\Delta^{(2)}_{24},\Delta^{(2)}_{34} \\
 \hdashline 
P_{1,1},P_{1,4},P_{1,5},\textcolor{red}{\boldsymbol{P_{1,6}}},P_{4,1},P_{4,4},\textcolor{blue}{\boldsymbol{P_{4,5}}},P_{4,6},P_{5,1},\textcolor{blue}{\boldsymbol{P_{5,4}}}, 
 & P_{1,1},P_{1,2},P_{1,3},P_{1,4},\textcolor{red}{\boldsymbol{P_{1,6}}},P_{3,1},\textcolor{blue}{\boldsymbol{P_{3,2}}},P_{3,3},P_{3,4},P_{3,6} \\
P_{5,5},P_{5,6},\textcolor{red}{\boldsymbol{P_{6,1}}},P_{6,4},P_{6,5},P_{6,6} & \\
\hline 
\hline 

\Gape[3pt][2pt]{\frac{1}{\langle \text{AB34}\rangle  \langle \text{CD23}\rangle } } & \frac{1}{\langle \text{AB34}\rangle  \langle \text{CD12}\rangle  \langle \text{CD23}\rangle  \langle \text{CD34}\rangle } \\
 \hdashline \langle \text{ABCD}\rangle \to 0,\langle \text{AB12}\rangle \to 0,\langle \text{AB13}\rangle \to 0,\langle \text{AB14}\rangle \to 0 & \langle \text{AB12}\rangle \to 0,\langle \text{AB13}\rangle \to 0,\langle \text{AB14}\rangle \to 0,\langle \text{AB23}\rangle \to 0 \\
\hline
\end{array} \nonumber
\end{equation}
}

\newpage

\noindent{\scriptsize
\begin{equation}
\begin{array}{|c|c|}
\hline

 \Gape[3pt][2pt]{\frac{1}{\langle \text{AB34}\rangle  \langle \text{CD14}\rangle  \langle \text{CD23}\rangle  \langle \text{CD34}\rangle } } & \frac{1}{\langle \text{AB34}\rangle  \langle \text{CD34}\rangle } \\
 \hdashline \langle \text{ABCD}\rangle \to 0,\langle \text{AB12}\rangle \to 0,\langle \text{AB13}\rangle \to 0,\langle \text{AB14}\rangle \to 0 & \langle \text{ABCD}\rangle \to 0,\langle \text{AB12}\rangle \to 0,\langle \text{AB13}\rangle \to 0,\langle \text{AB14}\rangle \to 0 \\
 \langle \text{AB23}\rangle \to 0,\langle \text{AB24}\rangle \to 0,\langle \text{CD12}\rangle \to 0 & \langle \text{CD12}\rangle \to 0,\langle \text{CD23}\rangle \to 0,\langle \text{CD24}\rangle \to 0  \\
 \hdashline \Gape[3pt][3pt]{ \Delta^{(1)}_{12},\Delta^{(2)}_{12},\Delta^{(2)}_{13},\Delta^{(2)}_{14},\Delta^{(2)}_{23},\Delta^{(2)}_{24}} & \Delta^{(1)}_{12},\Delta^{(1)}_{13},\Delta^{(1)}_{14},\Delta^{(2)}_{12},\Delta^{(2)}_{23},\Delta^{(2)}_{24} \\
 \hdashline P_{3,1},P_{3,3},P_{3,4},P_{3,5},P_{3,6} & P_{1,3},P_{1,5},\textcolor{red}{\boldsymbol{P_{1,6}}},P_{3,3},P_{3,5},P_{3,6},P_{4,3},\textcolor{blue}{\boldsymbol{P_{4,5}}},P_{4,6} \\ 
\hline 
\hline 

\Gape[2pt][2pt]{\frac{\langle \text{AB34}\rangle  \langle \text{CD12}\rangle +\langle \text{AB23}\rangle  \langle \text{CD14}\rangle }{\langle \text{AB23}\rangle  \langle \text{AB34}\rangle  \langle \text{CD12}\rangle  \langle \text{CD14}\rangle } }&  \frac{1}{\langle \text{AB34}\rangle  \langle \text{CD12}\rangle  \langle \text{CD23}\rangle }  \\
 \hdashline \langle \text{ABCD}\rangle \to 0,\langle \text{AB12}\rangle \to 0,\langle \text{AB13}\rangle \to 0,\langle \text{AB14}\rangle \to 0 & \langle \text{AB12}\rangle \to 0,\langle \text{AB13}\rangle \to 0,\langle \text{AB14}\rangle \to 0,\langle \text{AB23}\rangle \to 0 \\
\hline
\end{array} \nonumber
\end{equation}
}

\paragraph{\textbf{Dimension 2.}} There are 264 integrands corresponding to 2-dimensional boundaries. We present some examples below.

\bigskip

\noindent{\scriptsize
\begin{equation}
\begin{array}{|c|c|}
\hline
\Gape[3pt][2pt]{\frac{1}{\langle \text{CD34}\rangle } } & \frac{1}{\langle \text{AB34}\rangle  \langle \text{CD12}\rangle  \langle \text{CD34}\rangle }  \\
 \hdashline \langle \text{ABCD}\rangle \to 0,\langle \text{AB12}\rangle \to 0,\langle \text{AB13}\rangle \to 0,\langle \text{AB14}\rangle \to 0 & \langle \text{AB12}\rangle \to 0,\langle \text{AB13}\rangle \to 0,\langle \text{AB14}\rangle \to 0,\langle \text{AB23}\rangle \to 0 \\
  \langle \text{AB23}\rangle \to 0,\langle \text{CD14}\rangle \to 0,\langle \text{CD23}\rangle \to 0 & \langle \text{AB24}\rangle \to 0,\langle \text{CD14}\rangle \to 0,\langle \text{CD23}\rangle \to 0 \\
 \hdashline \Gape[3pt][3pt]{ \Delta^{(1)}_{12},\Delta^{(1)}_{13}, \Delta^{(2)}_{12},\Delta^{(2)}_{13},\Delta^{(2)}_{24},\Delta^{(2)}_{34}} & \Delta^{(1,2)}_{1234},\Delta^{(1)}_{12},\Delta^{(2)}_{12},\Delta^{(2)}_{13},\Delta^{(2)}_{24},\Delta^{(2)}_{34} \\
 \hdashline P_{1,1},P_{1,2},P_{1,3},\textcolor{red}{\boldsymbol{P_{1,6}}},P_{3,1},\textcolor{blue}{\boldsymbol{P_{3,2}}},P_{3,3},P_{3,6} & P_{3,1},\textcolor{blue}{\boldsymbol{P_{3,2}}},P_{3,3},P_{3,6} \\
\hline 
\hline 

\Gape[3pt][2pt]{\frac{1}{\langle \text{AB34}\rangle  \langle \text{CD14}\rangle  \langle \text{CD23}\rangle } } & \frac{1}{\langle \text{CD23}\rangle } \\
 \hdashline \langle \text{ABCD}\rangle \to 0,\langle \text{AB12}\rangle \to 0,\langle \text{AB13}\rangle \to 0,\langle \text{AB14}\rangle \to 0 & \langle \text{ABCD}\rangle \to 0,\langle \text{AB12}\rangle \to 0,\langle \text{AB13}\rangle \to 0,\langle \text{AB14}\rangle \to 0 \\
  \langle \text{AB23}\rangle \to 0,\langle \text{AB24}\rangle \to 0,\langle \text{CD12}\rangle \to 0,\langle \text{CD34}\rangle \to 0 & \langle \text{AB23}\rangle \to 0,\langle \text{CD14}\rangle \to 0,\langle \text{CD24}\rangle \to 0,\langle \text{CD34}\rangle \to 0 \\
\hline
\end{array} \nonumber
\end{equation}
}

\noindent{\scriptsize
\begin{equation}
\begin{array}{|c|c|}
\hline

 \Gape[3pt][2pt]{\frac{1}{\langle \text{AB23}\rangle  \langle \text{AB34}\rangle  \langle \text{CD14}\rangle  \langle \text{CD34}\rangle } } &  \frac{1}{\langle \text{AB34}\rangle  \langle \text{CD23}\rangle  \langle \text{CD34}\rangle } \\
 \hdashline \langle \text{AB12}\rangle \to 0,\langle \text{AB13}\rangle \to 0,\langle \text{AB14}\rangle \to 0,\langle \text{AB24}\rangle \to 0 & \langle \text{ABCD}\rangle \to 0,\langle \text{AB12}\rangle \to 0,\langle \text{AB13}\rangle \to 0,\langle \text{AB14}\rangle \to 0 \\
  \langle \text{CD12}\rangle \to 0,\langle \text{CD13}\rangle \to 0,\langle \text{CD23}\rangle \to 0,\langle \text{CD24}\rangle \to 0 & \langle \text{AB23}\rangle \to 0,\langle \text{AB24}\rangle \to 0,\langle \text{CD12}\rangle \to 0,\langle \text{CD13}\rangle \to 0 \\
 & \langle \text{CD14}\rangle \to 0 \\
 \hdashline \Gape[3pt][3pt]{ \Delta^{(1,2)}_{1234},\Delta^{(1)}_{12},\Delta^{(1)}_{14},\Delta^{(2)}_{12},\Delta^{(2)}_{23} }& \Delta^{(1)}_{12},\Delta^{(2)}_{12},\Delta^{(2)}_{13},\Delta^{(2)}_{14} \\
 \hdashline P_{3,3},P_{3,5},P_{4,3},\textcolor{blue}{\boldsymbol{P_{4,5}}} & P_{3,1},P_{3,3},P_{3,4} \\
\hline 
\hline 

\Gape[3pt][2pt]{\frac{1}{\langle \text{CD23}\rangle  \langle \text{CD34}\rangle } } & \frac{1}{\langle \text{AB34}\rangle  \langle \text{CD34}\rangle } \\
 \hdashline \langle \text{ABCD}\rangle \to 0,\langle \text{AB12}\rangle \to 0,\langle \text{AB13}\rangle \to 0,\langle \text{AB14}\rangle \to 0 & \langle \text{ABCD}\rangle \to 0,\langle \text{AB12}\rangle \to 0,\langle \text{AB13}\rangle \to 0,\langle \text{AB14}\rangle \to 0 \\
\hline
\end{array} \nonumber
\end{equation}
}

\bigskip

\paragraph{\textbf{Dimension 1.}} There are 136 integrands corresponding to 1-dimensional boundaries. We present some examples below.

\bigskip

\noindent{\scriptsize
\begin{equation}
\begin{array}{|c|c|}
\hline
  1 & \Gape[3pt][2pt]{\frac{1}{\langle \text{AB34}\rangle  \langle \text{CD12}\rangle }}  \\
 \hdashline \langle \text{ABCD}\rangle \to 0,\langle \text{AB12}\rangle \to 0,\langle \text{AB13}\rangle \to 0,\langle \text{AB14}\rangle \to 0 & \langle \text{AB12}\rangle \to 0,\langle \text{AB13}\rangle \to 0,\langle \text{AB14}\rangle \to 0,\langle \text{AB23}\rangle \to 0 \\
  \langle \text{AB23}\rangle \to 0,\langle \text{CD14}\rangle \to 0,\langle \text{CD23}\rangle \to 0,\langle \text{CD24}\rangle \to 0 & \langle \text{AB24}\rangle \to 0,\langle \text{CD13}\rangle \to 0,\langle \text{CD14}\rangle \to 0,\langle \text{CD23}\rangle \to 0 \\
 \langle \text{CD34}\rangle \to 0  & \langle \text{CD34}\rangle \to 0  \\
 \hdashline \Gape[3pt][3pt]{ \Delta^{(1)}_{12},\Delta^{(1)}_{13},\Delta^{(2)}_{24},\Delta^{(2)}_{34} }& \Delta^{(1,2)}_{1234},\Delta^{(1)}_{12},\Delta^{(2)}_{13},\Delta^{(2)}_{34} \\
 \hdashline P_{1,2},\textcolor{red}{\boldsymbol{P_{1,6}}},\textcolor{blue}{\boldsymbol{P_{3,2}}},P_{3,6} & P_{3,1},\textcolor{blue}{\boldsymbol{P_{3,2}}}  \\
\hline 
\hline  

\Gape[3pt][2pt]{\frac{1}{\langle \text{AB34}\rangle  \langle \text{CD12}\rangle  \langle \text{CD23}\rangle } }& \frac{1}{\langle \text{AB34}\rangle  \langle \text{CD34}\rangle } \\
 \hdashline \langle \text{AB12}\rangle \to 0,\langle \text{AB13}\rangle \to 0,\langle \text{AB14}\rangle \to 0,\langle \text{AB23}\rangle \to 0 & \langle \text{ABCD}\rangle \to 0,\langle \text{AB12}\rangle \to 0,\langle \text{AB13}\rangle \to 0,\langle \text{AB14}\rangle \to 0 \\
\hline
\end{array} \nonumber
\end{equation}
}

\newpage

\paragraph{\textbf{Dimension 0.}} There are 34 integrands corresponding to 0-dimensional boundaries. We present some examples below.

\bigskip

\noindent{\scriptsize
\begin{equation}
\begin{array}{|c|c|}
\hline
 \Gape[3pt][2pt]{\frac{1}{\langle \text{AB34}\rangle  \langle \text{CD12}\rangle } }& \frac{1}{\langle \text{AB34}\rangle  \langle \text{CD34}\rangle }  \\
 \hdashline \langle \text{AB12}\rangle \to 0,\langle \text{AB13}\rangle \to 0,\langle \text{AB14}\rangle \to 0,\langle \text{AB23}\rangle \to 0 & \langle \text{ABCD}\rangle \to 0,\langle \text{AB12}\rangle \to 0,\langle \text{AB13}\rangle \to 0,\langle \text{AB14}\rangle \to 0 \\
  \langle \text{AB24}\rangle \to 0,\langle \text{CD13}\rangle \to 0,\langle \text{CD14}\rangle \to 0,\langle \text{CD23}\rangle \to 0 & \langle \text{AB23}\rangle \to 0,\langle \text{AB24}\rangle \to 0,\langle \text{CD12}\rangle \to 0,\langle \text{CD13}\rangle \to 0 \\
 \langle \text{CD24}\rangle \to 0,\langle \text{CD34}\rangle \to 0  & \langle \text{CD14}\rangle \to 0,\langle \text{CD23}\rangle \to 0,\langle \text{CD24}\rangle \to 0  \\
 \hdashline \Gape[3pt][3pt]{ \Delta^{(1,2)}_{1234},\Delta^{(1)}_{12},\Delta^{(2)}_{34} }& \Delta^{(1)}_{12},\Delta^{(2)}_{12} \\
 \hdashline \textcolor{blue}{\boldsymbol{P_{3,2}}} & P_{3,3} \\
\hline 
\hline 

\Gape[3pt][2pt]{\frac{1}{\langle \text{AB34}\rangle }} & 1 \\
 \hdashline  \langle \text{ABCD}\rangle \to 0,\langle \text{AB12}\rangle \to 0,\langle \text{AB13}\rangle \to 0,\langle \text{AB14}\rangle \to 0 & \langle \text{ABCD}\rangle \to 0,\langle \text{AB12}\rangle \to 0,\langle \text{AB13}\rangle \to 0,\langle \text{AB14}\rangle \to 0 \\
\hline
\end{array} \nonumber
\end{equation}
}


\bibliographystyle{JHEP}
\bibliography{amplituhedronNotes}
\end{document}